\documentclass[format=acmsmall, review=false, 11pt]{acmart} 
\usepackage{acm-ec-26}
\usepackage{booktabs} 
\startPage{1}

\setcitestyle{acmnumeric}
\usepackage{cleveref}
\usepackage[most]{tcolorbox} 
\usepackage{threeparttable}
\usepackage{tabularx, booktabs} 
  
 \usepackage{pdfpages}
 \usepackage{mathtools}
\usetikzlibrary{positioning}

\usepackage[normalem]{ulem}

\usepackage{threeparttable}
\usepackage{tabularx, booktabs} 
\usepackage{multirow}
\usepackage{color, colortbl}
\definecolor{Gray}{gray}{0.9}

\makeatletter
\newcommand{\thickhline}{%
    \noalign {\ifnum 0=`}\fi \hrule height 1pt
    \futurelet \reserved@a \@xhline
}
\newcolumntype{"}{@{\hskip\tabcolsep\vrule width 1pt\hskip\tabcolsep}}
\makeatother

\tcbset{
    colback=blue!3,
    colframe=blue!60!black,
    coltitle=white,
    colbacktitle=blue!70!black,
    fonttitle=\bfseries
}

\usepackage{macros_pan}
\author{Pan Xu}
\authornote{\texttt{{\color{blue}tcs.enthusiast@gmail.com}}}

\title{Certifying Capacity and Gaps of the Rank-Dependent Gain-Sharing Framework for Ranking}

\begin{abstract}
\normalsize

The randomized primal--dual framework of Devanur, Jain,
and Kleinberg (SODA 2013) distributes each realized unit matching
gain between the endpoint dual variables according to the matched
offline vertex's rank. A gain-sharing function specifies this split
but does not affect Ranking's decisions. We distinguish the choice of this function from the
underlying protocol: the rank-dependent splitting rule and its
expected edgewise dual-feasibility criterion. Optimizing the
function over a prescribed class yields a functional optimization
problem whose value is the protocol's certifying capacity---the
strongest uniform guarantee obtainable within that protocol.
Its shortfall from Ranking's actual competitiveness is the
certifying gap. This separates optimizing the gain-sharing function from identifying
the certifying limitations of the underlying protocol.

We study unweighted online bipartite matching under adversarial
arrivals on two instance classes: graphs with online degree at
most $d$ and $d$-regular graphs. The optimized protocol is tight
on both degree-one classes and recovers the exact $1-1/e$
competitiveness in the unrestricted setting. At degree two,
however, both classes exhibit a strictly positive certifying gap,
even when all measurable functions specifying nonnegative
complementary shares are allowed. In both degree-two settings,
we determine Ranking's exact competitiveness and the framework's
exact certifying capacity over nondecreasing gain-sharing
functions. These comparisons distinguish the loss from imposing
monotonicity from the barrier that persists over the full
measurable class.

For general $d$, we focus on nondecreasing, right-continuous
gain-sharing functions $f:[0,1]\to[0,1]$ with $f(1)=1$.
Their representation by probability measures provides structure
for the functional optimization. Even over this restricted class,
the certifying capacity and Ranking's actual competitiveness
both converge to $1-1/e$ in both instance families as
$d\to\infty$. For online degree at most $d$, the certifying gap
is at most $1/(2ed)+O(d^{-2})$. On $d$-regular graphs, the gap
tends to zero but is at least $(1/(2e)-o(1))/d$.

Our results quantify the distinction between function-class
restrictions and limitations of a fixed gain-sharing protocol.
The broader goal is to use these barriers to guide the design
of randomized primal--dual protocols with greater certifying
capacity, ideally matching the exact competitiveness of the
algorithm being analyzed.
\end{abstract}

\begin{document}

%
%
%
%
%


\pagenumbering{roman}

\maketitle

\setcounter{tocdepth}{1}
\tableofcontents

\clearpage

\pagenumbering{arabic}


\section{Introduction}
\label{sec:intro}

Randomized primal--dual methods are a standard tool for the design and
analysis of online algorithms. A canonical example is the framework of
Devanur, Jain, and Kleinberg~\cite{DJK13} for analyzing \rank{} in
online bipartite matching under adversarial arrivals. By requiring dual
constraints to hold in expectation rather than in every execution,
their framework gives a short proof of the optimal competitiveness
$1-1/e$ of \rank{}. We briefly review the setting and the analysis
before formulating the questions studied in this paper.

\xhdr{Online matching and \rank{}.} Karp, Vazirani, and
Vazirani~\cite{KVV90} introduced the classical online bipartite
matching problem and the algorithm \rank{}. The offline vertex set
$I$ of a bipartite graph $G=(I,J,E)$ is known in advance, while the
online vertices in $J$ arrive sequentially, each revealing its incident
edges. Each arrival must be matched irrevocably to an unmatched
neighbor or left unmatched. The objective is to maximize the matching
cardinality. Before the arrivals begin, \rank{} assigns independent
uniform ranks in $[0,1]$ to the offline vertices. It then matches each
arrival, whenever possible, to its unmatched neighbor of minimum rank.

\xhdr{The randomized primal--dual analysis of~\cite{DJK13}.}
The analysis distributes each realized unit matching gain between the
dual variables of the matched edge's endpoints. A \emph{gain-sharing
function} $f$ specifies the split: when an offline vertex of rank $y$
is matched, it receives $f(y)$, and its online partner receives
$1-f(y)$. The total dual charge therefore equals the matching
cardinality in every execution. A uniform lower bound on the expected
sum of endpoint charges over all graph edges then yields a
competitiveness guarantee.

\xhdr{From unrestricted to degree-restricted matching.} For
unrestricted online matching, the choice $f(y)=e^{y-1}$
in~\cite{DJK13} certifies the exact competitiveness $1-1/e$ of
\rank{}. Degree restrictions provide a natural setting in which to
ask whether this agreement between the certificate and the actual
competitiveness persists.

Cohen and Wajc~\cite{cohen2018randomized} showed that, on $d$-regular
bipartite graphs under adversarial arrivals, the competitiveness of
\rank{} remains bounded away from one as $d\to\infty$, whereas
algorithms designed to exploit regularity achieve competitiveness
approaching one. Feng et al.~\cite{feng2025degree} subsequently
established stronger upper bounds on the competitiveness of \rank{}
on degree-bounded and regular instances. These works show that
\rank{} can be strictly suboptimal on such classes, while leaving
its exact worst-case competitiveness largely unresolved.

Algorithmic suboptimality, however, does not imply a limitation of the
analysis: a suboptimal algorithm may still admit a tight certificate.
This raises a separate question about the analysis of \rank{}:
\emph{Does the exponential gain-sharing function still certify its
exact competitiveness under degree restrictions? If not, what is the
best competitiveness certifiable within the framework
of~\cite{DJK13}?}

\xhdr{Certifying capacity through functional optimization.} To make
``best'' precise, we distinguish the \emph{gain-sharing rule} from
the \emph{gain-sharing function}. The rule prescribes the structure
of the dual updates: each unit matching gain is split into
nonnegative complementary endpoint charges that depend only on the
matched offline rank. This rule, together with the expected edgewise
dual-feasibility criterion, defines the underlying primal--dual
\emph{protocol}. The function $f$ specifies the particular split
within that fixed structure.

In this protocol, the role of $f$ is purely analytical: changing it
changes the dual certificate but does not affect any matching decision
of \rank{}.\footnote{In \emph{algorithm-defining} randomized
primal--dual frameworks, the gain-sharing function also enters the
allocation rule and is therefore part of the algorithm design.
Appendix~\ref{sec:other-related-work} reviews such frameworks and
extends the notions of certifying capacity and certifying gap to
this setting.} We can therefore optimize $f$ over a prescribed admissible function
class while keeping both the algorithm and the protocol fixed. One
function must provide a guarantee uniformly over the entire instance
class. The supremum of these guarantees is the protocol's
\emph{certifying capacity}, and computing it is a functional
optimization problem.

This optimization is distinct from seeking the strongest guarantee
among all possible randomized primal--dual protocols for \rank{}.
Our paper addresses the former question: it optimizes the gain-sharing
function within the fixed protocol of~\cite{DJK13}, rather than
optimizing the dual-update structure or the feasibility criterion.
This distinction separates the task of choosing the best function
within an analysis from the broader task of designing a better
primal--dual protocol.

\xhdr{The certifying gap as a limitation of the protocol.} The
difference between the actual competitiveness of \rank{} and the
optimized certifying capacity is the \emph{certifying gap}. A positive
gap cannot be closed by choosing another function within the prescribed
class. If it persists even over the full class of measurable,
nonnegative rank-only splits permitted by the protocol, it identifies
a limitation of the protocol itself, rather than a suboptimal
function choice or a regularity restriction on that function. This
is a barrier for the specified gain-sharing rule and expected edgewise
certificate, not for randomized primal--dual analysis in general.
Quantifying this barrier may guide the design of more effective
randomized primal--dual protocols with greater certifying capacity.

\subsection{Certifying Capacity as a Functional Optimization Problem}
\label{subsec:intro-certifying-capacity}

We now formalize these questions for the rank-dependent randomized
primal--dual framework of~\cite{DJK13}, keeping \rank{} fixed
throughout.

\xhdr{Competitiveness of \rank{}.} Let $\cG$ be a collection of
finite online bipartite matching instances containing at least one
instance with positive optimum. Each instance $G=(I,J,E)$ includes an
online arrival order fixed before the offline ranks are sampled.
Write $\rank(G)$ for the random matching cardinality produced by
\rank{} and $\operatorname{OPT}(G)$ for the maximum matching
cardinality in the full graph. The competitiveness of \rank{} on
$\cG$ is
\begin{equation}
\label{eq:intro-algorithm-competitiveness}
\gamma(\cG):=\inf_{\substack{G\in\cG\\\operatorname{OPT}(G)>0}}\frac{\mathbb E[\rank(G)]}{\operatorname{OPT}(G)},
\end{equation}
where the expectation is over the independent offline ranks.

\xhdr{The function-parameterized primal--dual protocol.} Denote the
fixed protocol by $\mathfrak P$ and its instantiation with a
gain-sharing function $f$ by $\mathfrak P(f)$. The standard LP dual
of maximum bipartite matching is
\begin{equation}
\label{eq:matching-dual-LP}
\min\quad\sum_{i\in I}z_i+\sum_{j\in J}z_j
\qquad\text{s.t.}\quad z_i+z_j\geq1\ \ \forall(i,j)\in E,\qquad z_i,z_j\geq0.
\end{equation}
Integrality of the bipartite matching LP and strong duality imply
that its optimum equals $\operatorname{OPT}(G)$.

The protocol initializes all dual variables to zero. Whenever
\rank{} matches an online vertex $j$ to an offline vertex $i$ of
rank $Y_i$, it updates
\[
Z_i\gets f(Y_i),\qquad Z_j\gets1-f(Y_i).
\]
Here $f:[0,1]\to[0,1]$ is Borel measurable and satisfies
$f(1)=1$.\footnote{A nonnegative complementary split of a fixed
finite gain $\rho>0$ has the form $(g(y),\rho-g(y))$ with
$0\leq g\leq\rho$. Dividing all dual charges by $\rho$ and
setting $f=g/\rho$ scales both the expected edge scores and the
total dual gain by the same factor, preserving the certified
competitiveness. This normalization concerns complementary splits
of a fixed total gain. Changing $f(1)$ to one affects no expected
charge, because the sampled offline ranks equal one with probability
zero.}
The allocation rule remains independent of $f$.

For each offline vertex $i$ and online vertex $j$, let $X_i$ and
$X_j$ be their respective indicators of being matched in the final
matching. If $j$ is matched, let $Y_j$ denote the rank of its offline
partner; otherwise set $Y_j:=1$. Thus
\[
Z_i=f(Y_i)X_i,\qquad Z_j=(1-f(Y_j))X_j.
\]
For $G\in\cG$ and $(i,j)\in E(G)$, define the expected endpoint
score
\begin{equation}
\label{eq:Phi-G-i-j-f-preliminaries}
\Phi_G(i,j;f):=\mathbb E[f(Y_i)X_i]+\mathbb E[(1-f(Y_j))X_j]=\mathbb E[Z_i+Z_j].
\end{equation}
These scores are the quantities controlled by the protocol's expected
edgewise dual-feasibility criterion.

\xhdr{Best certifiable competitiveness.} The worst-case edgewise
guarantee associated with $f$ is
\begin{equation}
\label{eq:chi-Gamma-definitions-preliminaries}
\chi(f,\cG):=\inf_{\substack{G\in\cG\\(i,j)\in E(G)}}\Phi_G(i,j;f).
\end{equation}
For a nonempty admissible function class $\cS$, define the
\emph{certifying capacity} of $\mathfrak P(\cS)$ on $\cG$ by
\begin{equation}
\label{eq:intro-certifying-capacity}
\Gamma_{\mathfrak P(\cS)}(\cG):=\sup_{f\in\cS}\chi(f,\cG)
=\sup_{f\in\cS}\inf_{\substack{G\in\cG\\(i,j)\in E(G)}}\Phi_G(i,j;f).
\end{equation}
The order of optimization is essential: one function must provide a
uniform guarantee over the entire instance class. This optimization
varies the gain-sharing function, not the splitting protocol or its
expected edgewise feasibility criterion.

The normalized protocol permits every Borel-measurable $[0,1]$-valued
function with the endpoint normalization. Neither monotonicity nor
continuity is needed for the following soundness statement, which
also defines the full admissible class $\cA$.

\begin{proposition}[Soundness of the gain-sharing framework;
Appendix~\ref{app:intro-gain-sharing-soundness}]
\label{prop:intro-gain-sharing-soundness}
For every instance class $\cG$ as above and every nonempty function
class
\[
\cS\subseteq\cA:=\{f:[0,1]\to[0,1]:f\text{ is Borel measurable},\ f(1)=1\},
\]
one has
\begin{equation}
\label{eq:intro-soundness}
\chi(f,\cG)\leq\gamma(\cG)\qquad\text{for every }f\in\cS,
\end{equation}
and consequently
\begin{equation}
\label{eq:intro-capacity-soundness}
\Gamma_{\mathfrak P(\cS)}(\cG)\leq\gamma(\cG).
\end{equation}
\end{proposition}


\xhdr{Certifying gap and tightness.} We define the
\emph{certifying gap} by
\begin{equation}
\label{eq:intro-general-certifying-gap}
\Del_{\mathfrak P(\cS)}(\cG):=\gamma(\cG)-\Gamma_{\mathfrak P(\cS)}(\cG)\geq0.
\end{equation}
We say that $\mathfrak P(\cS)$ is \emph{tight} for \rank{} on
$\cG$ if its certifying gap is zero, or equivalently, if its
certifying capacity equals the actual competitiveness of \rank{}
on $\cG$.\footnote{Tightness concerns equality of the optimal
values; it does not require the certifying supremum to be attained within the prescribed function space.}
A positive gap means that the framework is not tight, even after
optimizing over all $f\in\cS$. Tightness is therefore relative to
both the admissible function class and the instance class.

For unrestricted online matching, the protocol of~\cite{DJK13} is
tight even over the continuous nondecreasing functions, because
$f(y)=e^{y-1}$ certifies the exact competitiveness $1-1/e$.
Our motivating question is  \emph{whether this tightness persists when
the instance class is restricted by degree}.

\subsection{Instance Classes and the Role of Gain-Sharing Functions}
\label{subsec:intro-instance-and-function-classes}

\xhdr{Two instance classes: online degree at most $d$ and
$d$-regular graphs.} Motivated by the degree-restricted settings
studied in~\cite{cohen2018randomized,feng2025degree}, we consider
two families of instances.  For an integer $d\geq1$, let $\mathcal B_{\leq d}^{\mathrm{on}}$
denote the collection of finite bipartite matching instances whose
online vertices have degree at most $d$, with no restriction on
offline degrees. Let $\mathcal B_\infty^{\mathrm{on}}$ denote all
finite instances, without a degree restriction on either side. We
also consider the class $\mathcal B_d$ of nonempty finite
$d$-regular bipartite instances, in which every offline and online
vertex has degree exactly $d$. Each instance includes an arbitrary
online arrival order fixed before the ranks are sampled.

For finite $d$, specialize the preceding definitions as follows:
\[
\begin{aligned}
\chi(f;d)&:=\chi(f,\mathcal B_{\leq d}^{\mathrm{on}}),
&\gamma(d)&:=\gamma(\mathcal B_{\leq d}^{\mathrm{on}}),\\
\tilde\chi(f;d)&:=\chi(f,\mathcal B_d),
&\tilde\gamma(d)&:=\gamma(\mathcal B_d).
\end{aligned}
\]
For a nonempty $\cS\subseteq\cA$, abbreviate
$\Gamma_{\cS}(\cG):=\Gamma_{\mathfrak P(\cS)}(\cG)$ and write
\[
\Gamma_{\cS}(d):=\sup_{f\in\cS}\chi(f;d),\qquad
\tilde\Gamma_{\cS}(d):=\sup_{f\in\cS}\tilde\chi(f;d).
\]
Untilded quantities concern an upper bound on online degree, whereas
tilded quantities concern exact regularity on both sides. At
$d=\infty$, the untilded definitions use
$\mathcal B_\infty^{\mathrm{on}}$ directly.

\xhdr{A strict certifying gap already at degree two.}  The following result
answers the tightness question negatively: on both degree-two
instance classes, the protocol falls short even when the gain-sharing
function ranges over the full measurable class $\cA$.

\begin{proposition}[Strict certifying gaps at degree two;
Appendix~\ref{app:gain-sharing-function-classes}]
\label{prop:intro-degree-two-incompleteness}
One has
\begin{equation}
\label{eq:intro-degree-two-protocol-barrier}
\Gamma_{\cA}(2)<\frac{715361}{10^6}<\gamma(2),\qquad
\tilde\Gamma_{\cA}(2)=\frac34<\frac{119}{144}=\tilde\gamma(2).
\end{equation}
Thus the rank-dependent protocol $\mathfrak P$ is not tight for \rank{} on either
$\mathcal B_{\leq2}^{\mathrm{on}}$ or $\mathcal B_2$, even when the
gain-sharing function ranges over the full admissible class $\cA$.
\end{proposition}

These gaps cannot be attributed to a suboptimal function choice or to
imposing continuity or monotonicity. They identify limitations of the
fixed rank-dependent splitting rule and its expected edgewise
certificate, not of randomized primal--dual analysis in general.
For two-regular graphs, the gap over $\cA$ is exactly $11/144$.

\xhdr{Our principal gain-sharing class.} To study certifying capacity
across degrees, we focus on
\begin{equation}\label{def:cF}
\mathcal F:=\{f\in\cA:f\text{ is nondecreasing and right-continuous on }[0,1)\}.
\end{equation}
Its structure makes the outer functional optimization tractable:
there is a canonical affine bijection between $\cF$ and
$\cP([0,1])$, the Borel probability measures on $[0,1]$, given by
\[
f(y)=\pi_f([0,y])=\int_{[0,1]}\mathbf1\{y\geq s\}\,\pi_f(ds),\qquad y\in[0,1].
\]
This representation enables optimization through convex mixtures of
threshold certificates. Moreover, $\cF$ is compact in
$L^1([0,1])$, and the inner certifying values are $L^1$-continuous
at each fixed finite degree. Both certifying suprema are therefore
attained; the details are given in
Appendix~\ref{app:gain-sharing-function-classes}.

Unless a function-class subscript is displayed, we work over $\cF$:
\[
\begin{aligned}
\Gamma(d)&:=\Gamma_{\cF}(d),
&\Del(d)&:=\gamma(d)-\Gamma(d),\\
\tilde\Gamma(d)&:=\tilde\Gamma_{\cF}(d),
&\tilde\Del(d)&:=\tilde\gamma(d)-\tilde\Gamma(d).
\end{aligned}
\]
For $\tilde\Gamma(d)$, $\tilde\gamma(d)$, and $\tilde\Del(d)$,
entries at $d=\infty$ denote the corresponding limits as
$d\to\infty$, whose existence is established later. They do not
refer to an $\infty$-regular instance class.

\xhdr{Continuity versus monotonicity.} Within nondecreasing profiles,
right-continuity on $[0,1)$ can be imposed by changing values on a
null set, leaving every expected edge score unchanged.\footnote{A
nondecreasing real-valued function has at most countably many
discontinuities, and every interior discontinuity is a jump.
Replacing its value at each $t<1$ by its right-hand limit changes
the function only at discontinuities. Here a null set means a set
of Lebesgue measure zero.}
More generally, imposing full continuity preserves all certifying
suprema studied here, with or without monotonicity; see
Proposition~\ref{prop:continuity-preserves-certifying-values} in
Appendix~\ref{app:gain-sharing-function-classes}.
Monotonicity, however, can strictly reduce certifying capacity.
For example, on $\mathcal B_{\leq2}^{\mathrm{on}}$,
\[
\Gamma(2)<\Gamma_{\cA}(2),
\]
so restricting the gain-sharing function from $\cA$ to $\cF$
incurs a strict loss in certifying capacity. 
The quantitative comparison is proved in the same appendix. We retain right-continuity rather than full continuity
so that optimal step profiles remain admissible and attainment
and uniqueness can be stated directly.


\section{Main Contributions}
\label{sec:main-contributions}

\xhdr{Conceptual contribution.} We formulate the best guarantee
obtainable from a fixed randomized primal--dual protocol as a
functional optimization problem. Its \emph{certifying capacity}
is the strongest uniform guarantee over a prescribed gain-sharing
function class, while its \emph{certifying gap} compares this
optimized guarantee with the
actual competitiveness of the target algorithm. A zero gap means
that the framework is \emph{tight}. This formulation separates
optimizing a function within a protocol from designing a better
protocol, making both function-class restrictions and protocol
limitations explicit.\footnote{In Appendix~\ref{sec:other-related-work}, we extend this perspective to algorithm-defining frameworks where gain-sharing functions enter the allocation rule directly as part of algorithm design.}


\xhdr{Quantitative results and their interpretation.}
We use the instance classes and notation introduced in
Subsection~\ref{subsec:intro-instance-and-function-classes}.
The certifying values $\Gamma(d)$ and $\tilde\Gamma(d)$ optimize
over $\cF$ from~\eqref{def:cF}, whereas $\gamma(d)$ and
$\tilde\gamma(d)$ denote the actual competitiveness of \rank{}.
The certifying gap separates into two nonnegative components:
\[
\Del(d)=\underbrace{\Gamma_{\cA}(d)-\Gamma(d)}_{\text{loss due to monotonicity}}
+\underbrace{\gamma(d)-\Gamma_{\cA}(d)}_{\text{gap of }\mathfrak P(\cA)}.
\]
The first measures the loss due to restricting the gain-sharing
function to $\cF$; the second measures the gap that persists over
the full admissible class $\cA$. The analogous decomposition holds
for regular instances. At degree two, both components are positive
under an online-degree bound, whereas only the second is positive
on regular graphs.

Table~\ref{tab:main-results} summarizes our results. Decimal
expansions of exact constants are truncated after four decimal
places; numerical lower and upper bounds are rounded downward and
upward, respectively, to four decimal places. All asymptotic
statements concern integer $d\to\infty$, and all implicit
constants are absolute.

\begin{table}[!htbp]
\centering
\caption{Summary of the certifying values over $\cF$, the
competitiveness of \rank{}, and the corresponding certifying gaps.}
\label{tab:main-results}
\small
\setlength{\tabcolsep}{4pt}
\renewcommand{\arraystretch}{1.3}
\renewcommand{\tabularxcolumn}[1]{m{#1}}
\begin{tabularx}{\linewidth}{@{}l c c c >{\centering\arraybackslash}X@{}}
\toprule
Quantity
& $d=1$
& $d=2$\textsuperscript{\bluee{$\ast$}}
& $d=\infty$\textsuperscript{\bluee{\S}}
& Large $d$
\\
\specialrule{0.6pt}{5pt}{8pt}
$\Gamma(d)$
& $1$
& $0.7149$
& $1-\frac1e$
& $\displaystyle\begin{gathered}
\geq 1-\frac1e+\frac{1}{3e\,d(d+2)}\\[5pt]
\leq 1-\frac1e+O\!\left(\frac{\ln d}{d^2}\right)
\end{gathered}$
\\
\specialrule{0.4pt}{8pt}{8pt}
$\gamma(d)$
& $1$
& $0.7177$
& $1-\frac1e$
& $\displaystyle\begin{gathered}
\geq 1-\frac1e+\frac{1}{ed^2}-O(d^{-9/4})\\[5pt]
\leq 1-\frac1e+\frac{1}{2ed}+O(d^{-2})
\end{gathered}$
\\
\specialrule{0.4pt}{8pt}{8pt}
$\Del(d)$
& $0$
& $0.0027$
& $0$
& $\displaystyle\begin{gathered}
\text{Positive lower rate: ?\textsuperscript{\bluee{$\dagger$}}}\\[5pt]
\leq \frac{1}{2ed}+O(d^{-2})
\end{gathered}$
\\
\specialrule{1.1pt}{12pt}{12pt}
$\tilde\Gamma(d)$
& $1$
& $\frac34$
& $1-\frac1e$
& $\displaystyle 1-\frac1e+\left(\frac1{2e}+o(1)\right)\frac{(\ln d)^2}{d^2}$
\\
\specialrule{0.4pt}{8pt}{8pt}
$\tilde\gamma(d)$
& $1$
& $\displaystyle\frac{119}{144}$
& $1-\frac1e$
& $\displaystyle\begin{gathered}
\geq 1-\left(1-\frac1d\right)^d\\[5pt]
=1-\frac1e+o(1);\quad\text{upper rate: ?\textsuperscript{\bluee{$\ddagger$}}}
\end{gathered}$
\\
\specialrule{0.4pt}{8pt}{8pt}
$\tilde\Del(d)$
& $0$
& $\displaystyle\frac{11}{144}$
& $0$
& $\displaystyle\begin{gathered}
\liminf_{d\to\infty}d\,\tilde\Del(d)\geq\frac1{2e}\\[5pt]
\tilde\Del(d)=o(1);\quad\text{upper rate: ?\textsuperscript{\bluee{$\ddagger$}}}
\end{gathered}$
\\
\specialrule{0.8pt}{8pt}{0pt}
\end{tabularx}

\par\medskip
\parbox{\linewidth}{\footnotesize
\textsuperscript{\bluee{$\ast$}} The three decimal entries are
truncated after four decimal places.
\par\smallskip
\textsuperscript{\bluee{\S}} Untilded entries at $d=\infty$ refer
to the unrestricted class of finite instances. Tilded entries denote
limits as $d\to\infty$, not values for an $\infty$-regular graph class.
\par\smallskip
\textsuperscript{\bluee{$\dagger$}} No positive asymptotic lower
rate for $\Del(d)$ beyond $\Del(d)\geq0$ is established here.
\par\smallskip
\textsuperscript{\bluee{$\ddagger$}} Quantitative upper convergence
rates for $\tilde\gamma(d)-(1-1/e)$ and $\tilde\Del(d)$ are not
established here. Convergence to zero does not assert monotonicity.}
\end{table}

\subsection{Bipartite Graphs with Bounded Online Degree}
\label{subsec:main-contributions-bounded-online-degree}

We begin with the boundary cases, where the optimized certificate
is tight, before turning to the strict degree-two gap and the
large-degree bounds. For $f\in\cF$, define
\begin{equation}
\label{eq:F-H-f-definition}
F_f(t):=\int_0^t f(y)\,dy,\qquad H_f(t):=F_f(t)+1-f(t),\qquad t\in[0,1].
\end{equation}
Since $F_f$ is continuous and $f$ is upper semicontinuous,
$H_f$ is lower semicontinuous and attains its minimum on $[0,1]$.

\begin{theorem}[Unrestricted online degree;
Section~\ref{sec:unrestricted-degree} and
Appendix~\ref{app:unrestricted-degree-characterization}]
\label{thm:Gamma-infinity}
For every $f\in\cF$,
\begin{equation}
\label{eq:Gamma-infinity}
\chi(f;\infty)=\min_{t\in[0,1]}H_f(t),\qquad
\Gamma(\infty)=\gamma(\infty)=1-\frac1e,\qquad \Del(\infty)=0.
\end{equation}
The unique optimizer for $\Gamma(\infty)$ is
\begin{equation}
\label{eq:optimal-f-infinity}
f_\infty^\star(t)=e^{t-1},\qquad t\in[0,1].
\end{equation}
\end{theorem}

\xhdr{Relation to the classical analysis of~\cite{DJK13}.} Theorem~\ref{thm:Gamma-infinity}
recovers the classical exponential certificate and its agreement with
the unrestricted competitiveness of \rank{}. It additionally
determines the exact inner certifying value for every $f\in\cF$
and proves uniqueness of the exponential optimizer in this class.

\begin{theorem}[Online degree at most one;
Appendix~\ref{sec:degree-one}]
\label{thm:Gamma-degree-one}
For every $f\in\cF$,
\begin{equation}
\label{eq:Gamma-degree-one}
\chi(f;1)=H_f(1)=\int_0^1f(y)\,dy,\qquad
\Gamma(1)=\gamma(1)=1,\qquad \Del(1)=0.
\end{equation}
The unique optimizer for $\Gamma(1)$ is
\begin{equation}
\label{eq:optimal-f-degree-one}
f_1^\star\equiv1.
\end{equation}
\end{theorem}

Already at online degree at most two, the optimized certificate
falls strictly below the actual competitiveness. We determine both
values and their difference exactly.

\begin{theorem}[Online degree at most two: exact values and gap;
Sections~\ref{sec:proof-degree-two-certifying-value}
and~\ref{sec:proof-degree-two-competitiveness}]
\label{thm:Gamma-two}
Let $s^\star$ be the unique root in $(0,1)$ of
\begin{equation}
\label{eq:s-two-star-polynomial}
2s^4+2s^3-8s^2-33s+7=0.
\end{equation}
Then
\begin{align}
\Gamma(2)&=\frac{2}{3-s^\star}=0.7149\ldots,
\label{eq:Gamma-two-exact-value}\\
\gamma(2)&=1-\sum_{k=1}^{\infty}2^{-(2^k+k-1)}=0.7177\ldots,
\label{eq:gamma-two-exact-value}\\
\Del(2)&=1-\sum_{k=1}^{\infty}2^{-(2^k+k-1)}-\frac{2}{3-s^\star}=0.0027\ldots.
\label{eq:Delta-two-exact-value}
\end{align}
The unique optimizer for $\Gamma(2)$ is
\begin{equation}
\label{eq:optimal-step-profile-degree-two}
f_2^\star(y)=
\begin{cases}
0,&0\leq y<s^\star,\\[1mm]
w^\star,&s^\star\leq y<1,\\[1mm]
1,&y=1,
\end{cases}
\qquad w^\star:=\frac{2}{(1-s^\star)(3-s^\star)}.
\end{equation}
Its Stieltjes measure is
\[
\pi_{f_2^\star}=w^\star\delta_{s^\star}+(1-w^\star)\delta_1,
\]
where $\delta_s$ denotes the Dirac probability measure at $s$.
\end{theorem}

Numerically, $s^\star=0.2027\ldots$ and $w^\star=0.8968\ldots$.

\xhdr{\rank{} is optimal at online degree two.} Bhangale,
Chakraborty, and Harsha~\cite{BCH25} proved that \textsc{Half-Half}
achieves the value in~\eqref{eq:gamma-two-exact-value} and that no
randomized integral online matching algorithm can do better on
$\mathcal B_{\leq2}^{\mathrm{on}}$. Theorem~\ref{thm:Gamma-two}
shows that \rank{} attains the same value and is therefore another
optimal algorithm for this class. Our proof combines a
state-dependent lower certificate with a matching binary layered
construction.

\xhdr{Two sources of certifying loss.} The bounds in
Appendix~\ref{app:gain-sharing-function-classes} give
\[
\Gamma(2)<\frac{17877}{25000}\leq\Gamma_{\cA}(2)<\frac{715361}{10^6}<\gamma(2).
\]
Thus both components of $\Del(2)$ are strictly positive:
monotonicity reduces certifying capacity, but removing it does not
make the rank-dependent protocol tight.

\xhdr{How state-dependent certification closes the gap.} The
stronger proof replaces the fixed rank-only split by dual updates
based on the evolving safety states of offline vertices.\footnote{A
safety state records a vertex's probability of remaining unmatched
after a fixed arrival prefix and the distribution of its rank
conditional on this event. The state describes the randomized
execution in distribution; it does not modify the matching decisions
of \rank{}.}
The total increase in dual variables at each arrival is bounded by
that arrival's expected matching gain, rather than obtained by
splitting each realized unit gain according to one common function
of the matched rank. The proof retains edgewise dual feasibility
and certifies the exact value $\gamma(2)$. Thus a more flexible
dual-update rule suffices to remove the protocol barrier, without
changing either \rank{} or the underlying matching LP.

For larger online-degree bounds, we quantify the corrections to the
unrestricted value $1-1/e$.

\begin{theorem}[Large online-degree bound;
Sections~\ref{sec:proof-large-degree-certifying-value}
and~\ref{sec:proof-large-degree-competitiveness}]
\label{thm:Gamma-large-degree}
For all sufficiently large integers $d$,
\begin{align}
\frac{1}{3e\,d(d+2)}
&\leq\Gamma(d)-\left(1-\frac1e\right)
\leq O\!\left(\frac{\ln d}{d^2}\right),
\label{eq:Gamma-large-degree-bounds}\\
\frac{1}{ed^2}-O(d^{-9/4})
&\leq\gamma(d)-\left(1-\frac1e\right)
\leq\frac{1}{2ed}+O(d^{-2}),
\label{eq:gamma-large-degree-bounds}\\
\Del(d)&\leq\frac{1}{2ed}+O(d^{-2}).
\label{eq:Delta-large-degree-upper}
\end{align}
The lower bound on $\Gamma(d)$ holds for every integer $d\geq2$.
\end{theorem}

\xhdr{Relation to the algorithmic upper bound of~\cite{azar2017online}.}
The upper bound on $\gamma(d)$ in~\eqref{eq:gamma-large-degree-bounds}
already follows from the algorithm-independent bound of Azar, Cohen,
and Roytman~\cite{azar2017online}. We retain our layered
construction in
Appendix~\ref{app:large-degree-competitiveness-proofs} to provide
a direct, self-contained analysis of \rank{} on explicit finite
instances. This algorithmic bound gives only an $O(1/d)$ upper
correction for $\Gamma(d)$ through soundness. The sharper
$O(\ln d/d^2)$ certifying correction
in~\eqref{eq:Gamma-large-degree-bounds} requires a separate
mixture of graph-edge kernels; that construction also supplies
the reduced-threshold laws used in the regular-degree analysis.

\xhdr{A vanishing gap even over $\cF$.} Both $\Gamma(d)$ and
$\gamma(d)$ converge to $1-1/e$, and $\Del(d)\to0$. Thus
$\mathfrak P(\cF)$ is asymptotically tight in the additive sense,
without enlarging the function class to $\cA$. This does not imply
tightness at any fixed finite degree or agreement of the leading
corrections.

\subsection{Regular Bipartite Graphs}
\label{subsec:main-contributions-regular-graphs}

We next restrict the adversary to the regular instance class
$\mathcal B_d$ defined in the introduction. For $d=1$, every
instance is a disjoint union of isolated edges, so
\[
\tilde\chi(f;1)=\tilde\Gamma(1)=\tilde\gamma(1)=1,\qquad
\tilde\Del(1)=0\qquad(\forall f\in\cF).
\]
Unlike the online-degree-at-most-one class, every admissible
profile is optimal here. Degree two again exhibits a strict gap,
which we determine exactly.

\begin{theorem}[Two-regular graphs: exact values and gap;
Sections~\ref{sec:proof-Gamma-two-regular}
and~\ref{sec:proof-regular-two-competitiveness}]
\label{coro:Gamma-two-regular}
One has
\begin{equation}
\label{eq:Gamma-two-regular-value}
\tilde\Gamma(2)=\frac34,\qquad
\tilde\gamma(2)=\frac{119}{144},\qquad
\tilde\Del(2)=\frac{11}{144}.
\end{equation}
The unique optimizer for $\tilde\Gamma(2)$ is
\begin{equation}
\label{eq:optimal-profile-two-regular}
\tilde f_2^\star(y)=
\begin{cases}
\frac34,&0\leq y<1,\\[1mm]
1,&y=1,
\end{cases}
\qquad
\pi_{\tilde f_2^\star}=\frac34\delta_0+\frac14\delta_1.
\end{equation}
\end{theorem}

\xhdr{Three implications.} \emph{(i) Matching the known upper bound.}
Feng et al.~\cite[Lemma~3.7]{feng2025degree} proved
$\tilde\gamma(2)\leq119/144$.
Theorem~\ref{coro:Gamma-two-regular} establishes equality: a global
cycle decomposition into path blocks supplies the matching lower
bound.

\emph{(ii) The entire gap persists over $\cA$.} By
Appendix~\ref{app:gain-sharing-function-classes},
$\tilde\Gamma_{\cA}(2)=\tilde\Gamma(2)=3/4$. Consequently,
\[
\tilde\Gamma_{\cA}(2)-\tilde\Gamma(2)=0,\qquad
\tilde\gamma(2)-\tilde\Gamma_{\cA}(2)=\tilde\Del(2)=\frac{11}{144}.
\]
There is no loss due to monotonicity: the entire gap is due to the
failure of $\mathfrak P(\cA)$ to be tight. It separates the
protocol's uniform edgewise certificate from the global matching
performance of \rank{}.

\emph{(iii) Comparison with fractional matching.} The value
$\tilde\Gamma(2)=3/4$ also equals the optimal competitiveness of
fractional online matching on $\mathcal B_{\leq2}^{\mathrm{on}}$,
attained by \textsc{Water-Level}; see~\cite[Theorem~2]{BCH25}.
This is an equality of numerical values across different instance
classes and solution models, not an equivalence of the two analyses.

For large regular degree, the certifying value and the competitiveness
have the same limit, but their improvements over that limit occur
at different scales.

\begin{theorem}[Large regular degree;
Sections~\ref{app:Gamma-regular-large-degree}
and~\ref{sec:proof-regular-large-degree-competitiveness}]
\label{thm:Gamma-regular-large-degree}
For every integer $d\geq2$,
\begin{equation}
\label{eq:gamma-regular-finite-degree-lower}
\tilde\gamma(d)\geq1-\left(1-\frac1d\right)^d.
\end{equation}
As $d\to\infty$,
\begin{align}
\tilde\Gamma(d)&=1-\frac1e+\left(\frac1{2e}+o(1)\right)\frac{(\ln d)^2}{d^2},
\label{eq:Gamma-regular-asymptotic-value}\\
\tilde\gamma(d)&=1-\frac1e+o(1),
\label{eq:gamma-regular-limit}\\
\tilde\Del(d)&=o(1),\qquad
\liminf_{d\to\infty}d\,\tilde\Del(d)\geq\frac1{2e}.
\label{eq:Delta-regular-asymptotic-bounds}
\end{align}
Moreover, every sequence $f_d\in\cF$ satisfying
$\tilde\Gamma(d)-\tilde\chi(f_d;d)\to0$ also satisfies
\begin{equation}
\label{eq:regular-optimal-profiles-converge}
\|f_d-f_\infty^\star\|_1\longrightarrow0,\qquad f_\infty^\star(y)=e^{y-1}.
\end{equation}
\end{theorem}

\xhdr{Comparison with the previous upper bound by~\cite{feng2025degree}.} The lower bound
in~\eqref{eq:gamma-regular-finite-degree-lower} has the expansion
\[
1-\left(1-\frac1d\right)^d=1-\frac1e+\frac1{2ed}+O(d^{-2}).
\]
In contrast, the upper bound of Feng et
al.~\cite[Theorem~3.1]{feng2025degree} is
\[
\tilde\gamma(d)\leq1-\frac{d-1}{2d-1}\left(1-\frac1d\right)^d
=1-\frac1{2e}+\frac1{2ed}+O(d^{-2}).
\]
Theorem~\ref{thm:Gamma-regular-large-degree} identifies the exact
limiting competitiveness as $1-1/e$, improving the limiting upper
bound $1-1/(2e)$. Our lower bound uses a global threshold-cut
argument, while the limiting upper bound uses exactly regular
instances that approximate triangular batch behavior. The latter
holds for all sufficiently large integers $d$, not merely a
subsequence, but does not supply a quantitative upper convergence rate.

\xhdr{A vanishing gap, but a vanishing fraction of the improvement.}
Since $\tilde\Del(d)\to0$, the protocol $\mathfrak P(\cF)$ is
asymptotically tight in the additive sense on regular graphs as well.
Nevertheless, the gap is at least of order $1/d$, whereas the
certifiable improvement over $1-1/e$ is of order
$(\ln d)^2/d^2$. Consequently,
\[
\frac{\tilde\Gamma(d)-(1-1/e)}{\tilde\gamma(d)-(1-1/e)}\longrightarrow0.
\]
Thus even though the certifying gap vanishes, optimization over
$\cF$ captures a vanishing fraction of the improvement guaranteed
by regularity. Equality of limiting constants is therefore weaker
than tightness at the scale of the finite-degree correction. This
separation of scales concerns the rank-dependent protocol over
$\cF$; unlike the degree-two result, it does not establish the
same separation over $\cA$.

\section{Main Challenges and Our Approach}
\label{sec:main-challenges-and-approach}

The certifying gap compares two different worst-case problems. The
certifying value is determined by a minimum expected edge score,
optimized over gain-sharing functions; the competitiveness depends on
the total matching size of an entire instance. A graph with one poor
edge score need not be a hard instance for \rank{}. We combine
functional optimization with state-dependent and global analyses,
retaining structural and distributional information that simpler
bounds discard.

\xhdr{AI-assisted methodology.}
The author used ChatGPT (OpenAI) to explore proof strategies,
develop and refine mathematical arguments, and assist with
symbolic and numerical calculations. This assistance concerned
parts of the degree-two and large-degree analyses and the
gain-sharing function-class comparisons; further details appear
in the AI methodology statement at the end of the paper.

\subsection{Functional Optimization and Positive Kernel Mixtures}
\label{subsec:approach-large-degree-kernel-mixtures}

An upper bound on certifying capacity must hold for every admissible
function, although the worst graph-edge pair may depend on that
function. For $h_s(y):=\mathbf1\{y\geq s\}$, the Stieltjes
representation and Fubini's theorem give
\[
\Xi_{G,i,j}(s):=\Phi_G(i,j;h_s),\qquad
\Phi_G(i,j;f)=\int_{[0,1]}\Xi_{G,i,j}(s)\,\pi_f(ds),\qquad f\in\cF.
\]
This identity uses the independence of the execution from $f$ and
the unit mass of $\pi_f$, which preserves the constant in the
online charge.

Fix finitely many graph-edge pairs $(G_r,i_r,j_r)$ in the instance
class $\cG$ and nonnegative weights $\lambda_r$ summing to one.
Their mixed kernel $\breve K(s):=\sum_r\lambda_r\Xi_{G_r,i_r,j_r}(s)$
satisfies
\[
\Gamma_{\cF}(\cG)\leq\sup_{\pi\in\cP([0,1])}\int_{[0,1]}\breve K(s)\,\pi(ds)=\sup_{s\in[0,1]}\breve K(s).
\]
The inequality replaces the adversary's infimum by a convex
combination; the equality uses the admissibility of every Dirac
measure. No minimax interchange is needed. A candidate function,
verified against all graph-edge pairs, supplies the opposite bound.
When these bounds agree, equality in the mixture constrains the
support of an optimal Stieltjes measure to the contact set of
$\breve K$, which can also determine the optimizer uniquely.

\xhdr{The unrestricted equalizing measure.}
The characterization $\chi(f;\infty)=\min_t H_f(t)$ provides a
target for finite-degree upper constructions. Since
$H_{h_s}(t)=\mathbf1\{t<s\}+(t-s)_+$, the probability measure
\begin{equation}
\label{eq:large-degree-exponential-dual-measure}
\mu_\infty(dt):=e^{-t}\,dt+\frac1e\delta_1(dt)
\end{equation}
equalizes every threshold response:
\begin{equation}
\label{eq:large-degree-exponential-mixture}
\int_0^1e^{-t}H_{h_s}(t)\,dt+\frac1eH_{h_s}(1)=1-\frac1e,\qquad s\in[0,1].
\end{equation}
Integrating against $\pi_f$ yields the same identity for every
$f\in\cF$. The finite-degree challenge is to realize an accurate
positive approximation to this measure by actual graph-edge scores,
not merely by an abstract distribution of thresholds.

\subsection{Degree Two: Exact Structure and Stronger Certificates}
\label{subsec:approach-degree-two-structure}

\xhdr{An exact reduction of local scores.}
Conditioning on survival biases offline ranks and can correlate
vertices that had independent initial ranks. At online degree two,
rooted partial orders capture this dependence exactly. After any
fixed arrival prefix, each positive-probability safety event
$\mathcal E_i$ has a monomial profile
$\Pr[\mathcal E_i\mid Y_i=y]=a_i y^{p_i}$. Two such events either
have disjoint rank-witness sets and are independent, or are mutually
exclusive. This dichotomy reduces every distinguished-edge score to
\[
\chi(f;2)=\min\left\{H_f(1),\inf_{p,q\in\mathbb Z_{\geq0}}\Psi_{p,q}(f)\right\},
\]
where $\Psi_{p,q}$ is realized by two roots subjected to $p$ and
$q$ private filters before a final common arrival. This is an exact
characterization, not a sufficient relaxation of the dual constraints.

Section~\ref{sec:proof-degree-two-certifying-value} uses a positive
mixture of $H_f(1)$, $\Psi_{0,0}(f)$, and $\Psi_{2,1}(f)$ whose
threshold kernel is maximal exactly at $s^\star$ and $1$.
The contact conditions identify the candidate step profile; verifying
all remaining filtered-root constraints proves the matching lower
bound, and the contact set gives uniqueness. To establish barriers
over $\cA$, where the Stieltjes representation is unavailable, we
instead average the affine score formulas directly and use
$\int_0^1 k(y)f(y)\,dy\leq\int_0^1(k(y))_+\,dy$.
This separates the protocol barrier from the loss due to monotonicity;
see Appendix~\ref{app:gain-sharing-function-classes}.

\xhdr{A state-dependent certificate for the same algorithm.}
The same safety structure supports a stronger analysis of \rank{}.
A state records the witness size and the survival probability, and
thus the conditional rank distribution. We assign budgets to these
states so that each arrival's increase in dual mass is at most its
matching probability, while maintaining edgewise dual feasibility.
These updates need not split each realized gain according to one
common rank function. Equal-state merges determine a dyadic budget
recursion and the matching binary layered obstruction; the remaining
budget inequalities are checked by finite rational calculations and
an analytic tail bound. This yields the exact competitiveness in
Section~\ref{sec:proof-degree-two-competitiveness} without changing
the allocation rule.

\xhdr{Global accounting on two-regular graphs.}
Regularity requires preserving the complete instance rather than
truncating after a distinguished arrival. For competitiveness, only
local maxima of the arrival order on a cycle can remain unmatched.
Removing these arrivals leaves path blocks. An endpoint of chain
length $r$ survives with probability $1/r!$, while a block with
endpoint lengths $r,s$ contains $r+s-2$ offline vertices. Charging
failure probabilities to these supporting vertices bounds the total
loss and yields $119/144$, attained by the cycle construction
of~\cite{feng2025degree}. This global accounting differs from
minimizing a single edge score, which gives the certifying value
$3/4$; see Sections~\ref{sec:proof-Gamma-two-regular}
and~\ref{sec:proof-regular-two-competitiveness}.

\subsection{Finite Online Degree: Local Slack and Global Surplus}
\label{subsec:approach-finite-online-degree}

\xhdr{Retaining two critical ranks.}
The single-threshold DJK bound discards information needed to exploit
finite degree. Delete the distinguished offline vertex and retain
both its preprocessing threshold $\alpha$, the largest reduced
partner rank among its earlier neighbors, and the final arrival's
reduced partner rank $\tau$. Insertion monotonicity gives
$\Phi_G(i^\star,j^\star;f)\geq\mathbb E[\mathcal R_f(\alpha,\tau)]$,
where
\[
\mathcal R_f(a,t):=1+F_f(a)-\bigl(1-(t-a)_+\bigr)f(t).
\]
For the exponential profile, the slack above $1-1/e$ admits a
quadratic lower bound in $t-a$ when $t\geq a$ and a linear lower
bound in $a-t$ otherwise. A rank-raising argument
preserves the earlier execution and yields conditional moment bounds
for $\tau$ given $\alpha$. We then choose a perturbation of the
exponential function that cancels the threshold dependence of the
expected-slack bound, giving a uniform improvement of order $d^{-2}$;
see Section~\ref{sec:proof-large-degree-certifying-value}.

\xhdr{Realizing the equalizing mixture at finite degree.}
For the upper bound, a corrected Bernstein approximation represents
\eqref{eq:large-degree-exponential-dual-measure} by a positive mixture
of order-statistic laws with error $O(d^{-2})$. Realizing these laws
by graphs creates additional edge-score contributions, which must be
controlled at the same scale. The common-pool construction preserves
correlations between consecutive order statistics, while a private
singleton phase removes the availability discrepancy caused by
inserting the distinguished root. Except for an exponentially rare
ordering of pool labels, the excess is confined between consecutive
shared-pool ranks. Together with complete bipartite and singleton
instances, this gives a positive graph mixture with uniform error
$O(N/(d-1)^2)+2^{-N}$, where $N$ indexes the relevant shared-pool
order statistic. Taking $N=\Theta(\ln d)$ with a sufficiently large
constant yields the $O(\ln d/d^2)$ certifying correction.
Subsection~\ref{subsec:large-degree-upper-gadgets} states the
precise construction.

\xhdr{Aggregating slack for competitiveness.}
A separate argument is needed for the stronger lower bound on
$\gamma(d)$. After a reduction to forward instances with a prescribed
perfect matching, an equivalent execution in increasing offline-rank
order expresses the improvement over $1-1/e$ as an integrated global
threshold surplus. Loss accounting connects this surplus to two
mechanisms: self-matches with positive replacement-rank slack, and
the time for which vertices remain active after losing their
prescribed partners. Binomial concentration bounds the number of
competitors, yielding the $1/(ed^2)-O(d^{-9/4})$ correction in
Section~\ref{sec:proof-large-degree-competitiveness}. The layered
upper construction instead joins roots with independent histories;
it does not assume that surviving ranks remain uniform.

\subsection{Regularity: Sharp Local Bounds and Global Constructions}
\label{subsec:approach-regular-large-degree}

\xhdr{Full-history thresholds and a sharp local correction.}
For regular instances, truncation can destroy regularity and discard
matching opportunities. We retain each offline vertex's full-history
threshold $\beta_i$, the maximum reduced partner rank over all its
incident arrivals. Sequential exposure gives
\[
\Pr[\beta_i\leq y]\leq\bigl(1-(1-y)^{d-1}\bigr)^d,\qquad 0\leq y<1.
\]
This envelope makes small thresholds unlikely. Flattening the
exponential profile below a carefully chosen cutoff just below
$(\ln d)/d$ sacrifices little at those thresholds while improving
the response elsewhere, producing the leading lower correction
$(\ln d)^2/(2ed^2)$.

The matching upper bound requires more than the scalar estimate on
$\Gamma(d)$. We complete graph-edge obstructions whose degrees are
bounded on both sides to exactly regular graphs, controlling the
increase in the distinguished score. The common-pool mixture also
controls the reduced-threshold distribution functions; this supplies
an extra factor $y$ in the regularization cost, reducing its scale
from $\ln d/d$ to $(\ln d)^2/d^2$. Compactness and uniqueness of
the unrestricted optimizer force asymptotically optimal profiles to
converge in $L^1$ to $e^{y-1}$. Monotonicity then bounds their
contribution near zero through values at fixed interior points;
letting those points tend to zero identifies the factor $1/e$ in
the sharp constant. Section~\ref{app:Gamma-regular-large-degree}
combines these distributional and functional estimates without
interchanging a limit and the outer supremum.

\xhdr{Degree balance and globally hard regular instances.}
For the algorithmic lower bound, let $A_G(t)$ be the expected fraction
of offline vertices matched with rank at most $t$. Exact degree
balance across a rank-threshold cut gives
\[
\left(1-\frac td\right)\frac{dA_G(t)}{dt}+A_G(t)\geq1,\qquad A_G(0)=0,
\]
with the differential inequality holding almost everywhere on
$(0,1)$. Integration yields $A_G(1)\geq1-(1-1/d)^d$.

The limiting upper bound cannot be obtained merely by regularizing a
poor edge score. We construct entire regular instances with triangular
block neighborhoods, using geometric block sizes to balance degrees
on both sides. Finite-field incidence graphs and vertex blow-ups
supply exactly biregular block pairs with no large empty bipartite
rectangles. This property prevents a batch from missing many
low-ranked available vertices and yields a deterministic comparison
with complete triangular batches for every fixed rank realization.
Finally, a separate degree-completion argument reaches every
sufficiently large integer degree, not merely a subsequence, since
the regular classes are not nested. This proves convergence to
$1-1/e$ but supplies no quantitative upper rate; see
Section~\ref{sec:proof-regular-large-degree-competitiveness}.

\section{Proof of the Unrestricted-Degree Assertions in
Theorem~\ref{thm:Gamma-infinity}}
\label{sec:unrestricted-degree}

In this section, we prove all assertions of
Theorem~\ref{thm:Gamma-infinity}. We first characterize the exact
inner certifying value in
Subsection~\ref{subsec:unrestricted-inner-characterization}. We then
optimize this characterization and prove uniqueness of the exponential
profile in Subsection~\ref{subsec:proof-Gamma-infinity}. Finally,
Subsection~\ref{subsec:unrestricted-competitiveness} identifies the
actual competitiveness and the certifying gap.

We first record the insertion monotonicity used throughout the
unrestricted-degree analysis. For a fixed realization of the offline
ranks, let $R_G(j)$ denote the rank of the offline node matched to
$j$ by \rank{} on $G$, with $R_G(j):=1$ when $j$ is unmatched.

\begin{lemma}[Offline-vertex insertion monotonicity]
\label{lem:offline-insertion-boundary}
Let $G=(I,J,E)$, and let
\[
\widetilde G=(I\cup\{i^\star\},J,\widetilde E)
\]
be obtained by adding one offline node $i^\star$ and an arbitrary set
of edges incident to it, leaving the original graph and online arrival
order unchanged. Fix distinct offline ranks on $I\cup\{i^\star\}$,
and couple the two executions using the same ranks on $I$. Then
\[
R_{\widetilde G}(j)\leq R_G(j)
\qquad
\text{for every }j\in J.
\]
This follows from the monotonicity argument
in~\cite[proof of Lemma~2.3]{DJK13}.
\end{lemma}

\subsection{Exact Inner Characterization}
\label{subsec:unrestricted-inner-characterization}

Recall that $F_f$ is continuous. Since every $f\in\mathcal F$ is
nondecreasing and right-continuous on $[0,1)$, with $f(1)=1$, it is
upper semicontinuous on $[0,1]$. Hence
$H_f=F_f+1-f$ is lower semicontinuous and attains its minimum.

\begin{lemma}[Exact unrestricted-degree characterization]
\label{thm:chi-infinity-characterization}
For every $f\in\mathcal F$,
\begin{equation}
\label{eq:chi-infinity-characterization}
\chi(f;\infty)
=
\min_{t\in[0,1]}H_f(t).
\end{equation}
\end{lemma}

The lower bound follows by deleting the distinguished offline node,
conditioning on the reduced partner rank of the distinguished online
node, and then reinserting the offline node. For the reverse bound,
complete bipartite instances realize every continuity point of $f$
asymptotically through an order-statistic quantile. Right-continuity
extends the bound to discontinuity points in $[0,1)$, while a
two-arrival singleton instance realizes the endpoint $t=1$ exactly.
The full proof is given in
Appendix~\ref{app:unrestricted-degree-characterization}.

\subsection{Optimal Certifying Value and Uniqueness}
\label{subsec:proof-Gamma-infinity}

\xhdr{Optimal certifying value.}
Fix $f\in\mathcal F$ and write
\[
c:=\min_{t\in[0,1]}H_f(t)=\chi(f;\infty),
\]
where the equality follows from
Lemma~\ref{thm:chi-infinity-characterization}. Since
$H_f(t)\geq c$,
\[
f(t)\leq F_f(t)+1-c
\qquad
\text{for every }t\in[0,1].
\]
Define
\[
U(t):=F_f(t)+1-c.
\]
Then $U$ is absolutely continuous, $U(0)=1-c$, and
\[
\frac{d U(t)}{dt}=f(t)\leq U(t)
\qquad
\text{for Lebesgue-almost every }t\in(0,1).
\]
Gronwall's inequality gives
\[
U(t)\leq(1-c)e^t.
\]
Using $f(1)=1$,
\[
c
\leq H_f(1)
=
F_f(1)
=
U(1)-(1-c)
\leq
(1-c)(e-1),
\]
and therefore
\[
c\leq1-\frac1e.
\]
Taking the supremum over $f\in\mathcal F$ gives
\[
\Gamma(\infty)\leq1-\frac1e.
\]

For
\[
f_\infty^\star(t):=e^{t-1},
\]
we have
\[
F_{f_\infty^\star}(t)=e^{t-1}-\frac1e,
\qquad
H_{f_\infty^\star}(t)=1-\frac1e
\quad
\text{for every }t\in[0,1].
\]
Hence
\[
\Gamma(\infty)=1-\frac1e,
\]
and $f_\infty^\star$ attains this value.

\xhdr{Uniqueness.}
Suppose $f\in\mathcal F$ attains $\Gamma(\infty)=1-1/e$. The
preceding inequalities must then be equalities, and in particular
\[
F_f(1)=1-\frac1e.
\]
Define
\[
U(t):=F_f(t)+\frac1e.
\]
Then
\[
U(0)=\frac1e,
\qquad
U(1)=1,
\qquad
\frac{d U(t)}{dt}=f(t)\leq U(t)
\quad\text{a.e.}
\]
Therefore,
\[
\frac{d}{dt}\bigl(e^{-t}U(t)\bigr)
=
e^{-t}\Bigl(\frac{d U(t)}{dt}-U(t)\Bigr)
\leq0
\quad\text{a.e.}
\]
Since $e^{-t}U(t)$ is absolutely continuous, it is nonincreasing.
Its values at both endpoints equal $1/e$, so it is constant.
Consequently,
\[
U(t)=e^{t-1}
\qquad
\text{for every }t\in[0,1],
\]
and hence
\[
f(t)=e^{t-1}
\qquad
\text{for Lebesgue-almost every }t\in[0,1].
\]
For each $t<1$, choose a sequence $t_k\downarrow t$ outside this
null set. Right-continuity of $f$ then gives
$f(t)=e^{t-1}$; the endpoint follows from $f(1)=1$. Thus
$f_\infty^\star$ is the unique optimizer in $\mathcal F$.

\subsection{Competitiveness and Gap}
\label{subsec:unrestricted-competitiveness}

Soundness gives
\[
\gamma(\infty)\geq\Gamma(\infty)=1-\frac1e.
\]
Conversely, the classical hard-instance construction
of~\cite{KVV90} implies that, for every $\varepsilon>0$, there is a
finite instance $G\in\mathcal B_\infty^{\mathrm{on}}$ with
$\operatorname{OPT}(G)>0$ such that
\[
\frac{\mathbb E[\rank(G)]}{\operatorname{OPT}(G)}
\leq
1-\frac1e+\varepsilon.
\]
Taking the infimum over finite instances and then letting
$\varepsilon\downarrow0$ gives
\[
\gamma(\infty)=1-\frac1e.
\]
Consequently,
\[
\gamma(\infty)=\Gamma(\infty)=1-\frac1e,
\qquad
\Del(\infty)=0,
\]
which completes the proof of Theorem~\ref{thm:Gamma-infinity}.

\section{Proof of the Online-Degree-Two Certifying Assertions in
Theorem~\ref{thm:Gamma-two}}
\label{sec:proof-degree-two-certifying-value}

In this section, we prove the assertions concerning $\Gamma(2)$ and
its unique optimizer $f_2^\star$ in
Theorem~\ref{thm:Gamma-two}.

We first characterize degree-two safety events by rooted partial orders
in Subsection~\ref{subsec:degree-two-order-states-Gamma}. Using this
structure, Subsection~\ref{subsec:filtered-root-functionals} reduces
the score of an arbitrary distinguished edge to a discrete family of
filtered-root functionals. We then establish the upper bound on
$\Gamma(2)$ in Subsection~\ref{subsec:three-constraint-upper-bound}
by means of a three-constraint kernel mixture, and prove the matching
lower bound in
Subsection~\ref{subsec:matching-lower-bound-degree-two} by verifying
the candidate step profile against the full family of filtered-root
constraints. Finally,
Subsection~\ref{subsec:uniqueness-degree-two-closure} proves that this
profile is the unique optimizer in $\mathcal F$.

\subsection{Degree-Two Order States}
\label{subsec:degree-two-order-states-Gamma}

An \emph{arrival prefix} consists of the first $t$ online arrivals,
for some $t\in\{0,\ldots,|J|\}$, in their prescribed order. Fix such
a prefix in an instance $G\in\mathcal B_{\leq2}^{\mathrm{on}}$.
For an offline node $i$, define
\[
\mathcal E_i
:=
\{i\text{ is safe after \rank{} processes the fixed prefix}\}.
\]
This is a random event determined by the offline ranks
$\mathbf Y=(Y_i)_{i\in I}$. We call $i$ \emph{positive} after the
prefix if $\Pr[\mathcal E_i]>0$, and \emph{dead} otherwise.
When analyzing a distinguished edge $(i^\star,j^\star)$, we will
truncate the instance immediately after $j^\star$ and take the prefix
to consist of all arrivals strictly before $j^\star$.

Let $\mathcal W\subseteq I$ be finite, and let $\prec$ be a strict
partial order on $\mathcal W$. Define the associated \emph{order event}
by
\begin{equation}
\label{eq:order-event-definition}
\mathcal O(\mathcal W,\prec)
:=
\left\{
Y_{i_1}<Y_{i_2}
\text{ whenever }i_1,i_2\in\mathcal W\text{ and }i_1\prec i_2
\right\}.
\end{equation}
A \emph{rooted order state} for $i$ is a triple
$(\mathcal W,\prec;i)$ such that $i\in\mathcal W$ is the unique
maximal element of the finite poset $(\mathcal W,\prec)$.
The set $\mathcal W$ is its \emph{witness set}. We use strict rank
comparisons throughout; the offline ranks are distinct with
probability one.

\begin{lemma}[Rooted order-state representation]
\label{lem:rooted-order-state-representation}
Fix an arrival prefix in an online-degree-two instance. Every positive
offline node $i$ admits a rooted order state
$(\mathcal W_i,\prec_i;i)$ such that, on the event of distinct
offline ranks,
\begin{equation}
\label{eq:safety-event-order-representation}
\mathcal E_i=\mathcal O(\mathcal W_i,\prec_i).
\end{equation}
\end{lemma}

The proof is by induction over the arrivals and is deferred to
Appendix~\ref{app:degree-two-state-characterization}. Initially,
$\mathcal W_i=\{i\}$ and the partial order is empty. When an arrival
joins two positive states with disjoint witness sets, their orders are
merged and one additional rank comparison is imposed.

\begin{corollary}[Monomial safety profile]
\label{cor:monomial-safety-profile-Gamma}
For every offline node $i$, there exist $p_i\in\mathbb Z_{\geq0}$
and $a_i\in[0,1]$ such that
\begin{equation}
\label{eq:degree-two-monomial-profile-Gamma}
\Pr[\mathcal E_i\mid Y_i=y]=a_i y^{p_i}
\qquad\text{for every }y\in[0,1].
\end{equation}
Moreover, $a_i=0$ if and only if $\Pr[\mathcal E_i]=0$.
The conditional probabilities are taken in the versions furnished by
the order-state representations: specified ranks are fixed, and the
remaining ranks are sampled independently and uniformly on $[0,1]$.
\end{corollary}

\begin{proof}
If $i$ is dead, take $a_i=0$ and $p_i=0$. Otherwise, let
$(\mathcal W_i,\prec_i;i)$ be its rooted order state, set
$p_i:=|\mathcal W_i|-1$, and let $\ell_i$ be the number of linear
extensions of the induced order on $\mathcal W_i\setminus\{i\}$.

Because $i$ is the unique maximal element, every other witness
precedes $i$. Conditional on $Y_i=y$, all $p_i$ remaining witness
ranks must therefore lie below $y$ and respect the induced order.
For $y>0$, each prescribed strict ordering has probability
$y^{p_i}/p_i!$. Distinct linear extensions give disjoint events up
to rank ties, so
\[
\Pr[\mathcal E_i\mid Y_i=y]
=
\frac{\ell_i}{p_i!}y^{p_i}.
\]
The formula also holds at $y=0$: the probability is zero if $p_i>0$
and one if $p_i=0$, with the convention $y^0=1$.
Since $1\leq\ell_i\leq p_i!$, setting $a_i:=\ell_i/p_i!$ gives
$a_i\in(0,1]$. Finally,
\[
\Pr[\mathcal E_i]=\int_0^1a_i y^{p_i}\,dy
=\frac{a_i}{p_i+1},
\]
which proves the last assertion.
\end{proof}

The recursively constructed order states satisfy the following
pairwise dichotomy.

\begin{lemma}[Independent or mutually exclusive safety events]
\label{lem:independent-or-exclusive-safety-events}
Let $i_1,i_2$ be distinct positive offline nodes after the same
arrival prefix, with rooted order states
$(\mathcal W_1,\prec_1;i_1)$ and $(\mathcal W_2,\prec_2;i_2)$
from the recursive construction above. Exactly one of the following
cases holds.
\begin{enumerate}
    \item The witness sets are disjoint:
    $\mathcal W_1\cap\mathcal W_2=\varnothing$.
    The safety events are then independent. In particular, if their
    monomial parameters are $(a_1,p_1)$ and $(a_2,p_2)$, then
    \begin{equation}
    \label{eq:independent-profile-factorization-Gamma}
    \Pr\!\left[
    \mathcal E_{i_1}\cap\mathcal E_{i_2}
    \,\middle|\,
    Y_{i_1}=x,\,Y_{i_2}=y
    \right]
    =a_1a_2x^{p_1}y^{p_2}
    \qquad(x,y\in[0,1]),
    \end{equation}
    using the conditional versions induced by the order events.

    \item The safety events are mutually exclusive:
    \begin{equation}
    \label{eq:mutually-exclusive-safety-events-Gamma}
    \Pr[\mathcal E_{i_1}\cap\mathcal E_{i_2}]=0.
    \end{equation}
\end{enumerate}
\end{lemma}

The distinction is between disjoint \emph{witness sets}, which give
independent safety events, and mutually exclusive \emph{safety events}.
The proof is also deferred to
Appendix~\ref{app:degree-two-state-characterization}, where these
properties are established jointly with the rooted order-state
representation.

\subsection{Filtered-Root Functionals}
\label{subsec:filtered-root-functionals}

\begin{figure}[t]
\centering
\begin{tikzpicture}[x=1cm,y=1cm,scale=0.9]
\tikzset{
  off/.style={circle,draw,inner sep=1.2pt,minimum size=16pt},
  on/.style={circle,draw,inner sep=1.2pt,minimum size=16pt}
}

\node[off] (istar) at (0,0) {$i^\star$};
\node[off] (itilde) at (5.6,0) {$\tilde i^\star$};
\node[on] (jstar) at (2.8,-1.6) {$j^\star$};

\node[on] (j1) at (-2.0,1.8) {$j_1$};
\node[on] (j2) at (-2.0,0.7) {$j_2$};
\node (jdots) at (-2.0,-0.35) {$\vdots$};
\node[on] (jp) at (-2.0,-1.4) {$j_p$};

\node[off] (i1) at (-4.0,1.8) {$i_1$};
\node[off] (i2) at (-4.0,0.7) {$i_2$};
\node (idots) at (-4.0,-0.35) {$\vdots$};
\node[off] (ip) at (-4.0,-1.4) {$i_p$};

\node[on] (jt1) at (7.6,1.8) {$\tilde j_1$};
\node[on] (jt2) at (7.6,0.7) {$\tilde j_2$};
\node (jtdots) at (7.6,-0.35) {$\vdots$};
\node[on] (jtq) at (7.6,-1.4) {$\tilde j_q$};

\node[off] (it1) at (9.6,1.8) {$\tilde i_1$};
\node[off] (it2) at (9.6,0.7) {$\tilde i_2$};
\node (itdots) at (9.6,-0.35) {$\vdots$};
\node[off] (itq) at (9.6,-1.4) {$\tilde i_q$};

\draw (j1)--(istar) (j2)--(istar) (jp)--(istar);
\draw (j1)--(i1) (j2)--(i2) (jp)--(ip);

\draw (jt1)--(itilde) (jt2)--(itilde) (jtq)--(itilde);
\draw (jt1)--(it1) (jt2)--(it2) (jtq)--(itq);

\draw[line width=1.0pt] (istar)--(jstar);
\draw (itilde)--(jstar);

\node at (0,2.55) {$p$ private filters for $i^\star$};
\node at (7.6,2.55) {$q$ private filters for $\tilde i^\star$};
\end{tikzpicture}

\caption{The $(p,q)$ power-filter gadget $G_{p,q}$. The final online
node $j^\star$ has neighborhood $\{i^\star,\tilde i^\star\}$,
and the bold edge $(i^\star,j^\star)$ is distinguished. Before
$j^\star$, each $j_r$, $r\in[p]$, arrives with neighborhood
$\{i^\star,i_r\}$, and each $\tilde j_s$, $s\in[q]$, arrives
with neighborhood $\{\tilde i^\star,\tilde i_s\}$. All private
offline nodes are distinct, and each has only its corresponding filter
as an online neighbor; hence it is certainly safe when that filter
arrives. Conditional on $Y_{i^\star}=x$, the root $i^\star$
survives all its filters exactly when $Y_{i_r}<x$ for every $r\in[p]$,
with probability $x^p$. The other root similarly survives with
probability $y^q$ conditional on $Y_{\tilde i^\star}=y$.
The two filter sequences may be interleaved arbitrarily. If $p=0$
or $q=0$, the corresponding side has no filters.}
\label{fig:power-filter-gadget}
\end{figure}
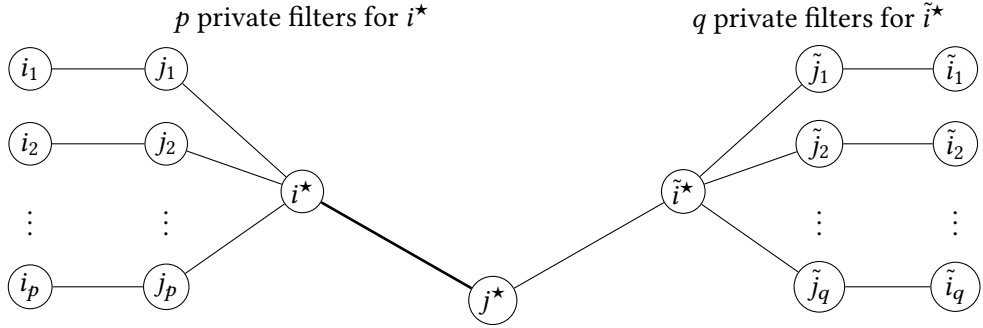

For $p,q\in\mathbb Z_{\geq0}$, let $G_{p,q}$ be the
\emph{$(p,q)$ power-filter gadget} in
Figure~\ref{fig:power-filter-gadget}. Its two offline roots
$i^\star,\tilde i^\star$ undergo $p$ and $q$ independent private
filters, respectively, before the final arrival $j^\star$ with
neighborhood $\{i^\star,\tilde i^\star\}$. We distinguish the
edge $(i^\star,j^\star)$.

\begin{lemma}[Score of the power-filter gadget]
\label{lem:power-filter-gadget-score}
For every $p,q\in\mathbb Z_{\geq0}$ and $f\in\mathcal F$,
\begin{equation}
\label{eq:power-filter-gadget-score}
\Phi_{G_{p,q}}(i^\star,j^\star;f)
=\Psi_{p,q}(f)
:=\frac{p+q+1}{(p+1)(q+1)}
+\int_0^1
\left(1-y^p-y^q+\frac{y^{p+q+1}}{p+1}\right)f(y)\,dy.
\end{equation}
\end{lemma}

The ordered pair $(p,q)$ is not symmetric: $p$ corresponds to the
distinguished offline endpoint $i^\star$, whereas $q$ corresponds
to the other root $\tilde i^\star$. The direct calculation is given
in Appendix~\ref{app:power-filter-gadget-score}.

We now reduce every degree-two edge score to the family consisting of
$H_f(1)$ and these filtered-root functionals.

\begin{lemma}[Exact degree-two edge reduction]
\label{lem:exact-degree-two-edge-reduction}
For every $f\in\mathcal F$,
\begin{equation}
\label{eq:exact-degree-two-inner-characterization}
\chi(f;2)
=\min\left\{
H_f(1),\,
\inf_{p,q\in\mathbb Z_{\geq0}}\Psi_{p,q}(f)
\right\}.
\end{equation}
\end{lemma}

\begin{proof}
Fix $G\in\mathcal B_{\leq2}^{\mathrm{on}}$ and
$(i^\star,j^\star)\in E(G)$. Truncating immediately after
$j^\star$ leaves its online contribution unchanged and can only
decrease the offline contribution of $i^\star$. For the lower
bound, it therefore suffices to consider the truncated instance,
which we continue to denote by $G$.

Let $\mathcal E_{i^\star}$ be the event that $i^\star$ is safe
immediately before $j^\star$. If $j^\star$ has a second neighbor,
denote it by $\tilde i^\star$ and define
$\mathcal E_{\tilde i^\star}$ analogously.

\xhdr{The easy cases.} Suppose $j^\star$ has only one neighbor,
one of the two safety events has probability zero, or the two events
are mutually exclusive. Then $i^\star$ is matched by the end,
almost surely: either it was matched earlier, or it is the only safe
neighbor when $j^\star$ arrives. Since the online contribution is
nonnegative,
\begin{equation}
\label{eq:easy-degree-two-edge-lower-bound}
\Phi_G(i^\star,j^\star;f)
\geq\mathbb E[f(Y_{i^\star})]
=\int_0^1f(y)\,dy
=H_f(1).
\end{equation}

\xhdr{Two independent positive safety events.} By
Lemma~\ref{lem:independent-or-exclusive-safety-events}, the remaining
case has two positive states with disjoint witness sets. Their
conditional safety probabilities factor, and
Corollary~\ref{cor:monomial-safety-profile-Gamma} gives
\[
\Pr[\mathcal E_{i^\star}\mid Y_{i^\star}=x]=ax^p,
\qquad
\Pr[\mathcal E_{\tilde i^\star}\mid Y_{\tilde i^\star}=y]=by^q,
\]
for $a,b\in(0,1]$ and $p,q\in\mathbb Z_{\geq0}$.
For $r\geq0$, write $M_r(f):=\int_0^1y^r f(y)\,dy$.

The final arrival is matched exactly when at least one root is safe.
Independence yields
\begin{equation}
\label{eq:final-match-probability-independent}
\Pr[X_{j^\star}=1]
=\frac{a}{p+1}+\frac{b}{q+1}
-\frac{ab}{(p+1)(q+1)}.
\end{equation}
The distinguished offline node remains unmatched exactly when both
roots are safe and $Y_{\tilde i^\star}<Y_{i^\star}$. Hence
\begin{equation}
\label{eq:offline-f-contribution-independent}
\mathbb E[f(Y_{i^\star})X_{i^\star}]
=H_f(1)-\frac{ab}{q+1}M_{p+q+1}(f).
\end{equation}
Conditioning on the selected root similarly gives
\begin{equation}
\label{eq:partner-f-contribution-independent}
\mathbb E[f(Y_{j^\star})X_{j^\star}]
=aM_p(f)+bM_q(f)
-ab\left(\frac1{p+1}+\frac1{q+1}\right)M_{p+q+1}(f).
\end{equation}
Since the online contribution is
$\Pr[X_{j^\star}=1]-\mathbb E[f(Y_{j^\star})X_{j^\star}]$,
combining these identities gives
\begin{equation}
\label{eq:Theta-p-q-a-b}
\begin{aligned}
\Phi_G(i^\star,j^\star;f)
={}&\frac{a}{p+1}+\frac{b}{q+1}
-\frac{ab}{(p+1)(q+1)}\\
&+\int_0^1
\left(1-ay^p-by^q+\frac{ab}{p+1}y^{p+q+1}\right)f(y)\,dy.
\end{aligned}
\end{equation}

For fixed $p,q,f$, the right-hand side is affine in each of $a,b$
separately, so its minimum over $[0,1]^2$ is attained at a corner.
The four values are
\begin{align*}
(0,0):\quad& H_f(1),\\
(1,0):\quad& H_f(1)+\frac1{p+1}-M_p(f),\\
(0,1):\quad& H_f(1)+\frac1{q+1}-M_q(f),\\
(1,1):\quad& \Psi_{p,q}(f).
\end{align*}
Because $0\leq f\leq1$, one has
$M_p(f)\leq1/(p+1)$ and $M_q(f)\leq1/(q+1)$. Therefore,
\[
\Phi_G(i^\star,j^\star;f)
\geq\min\{H_f(1),\Psi_{p,q}(f)\}.
\]
Together with the easy cases, this proves
\begin{equation}
\label{eq:degree-two-edge-reduction-lower}
\chi(f;2)
\geq\min\left\{
H_f(1),\,\inf_{p,q\in\mathbb Z_{\geq0}}\Psi_{p,q}(f)
\right\}.
\end{equation}

\xhdr{Tightness.} The value $H_f(1)$ is realized by one offline
node and two singleton online neighbors, with the later edge
distinguished. For every $p,q\geq0$, the gadget $G_{p,q}$ belongs
to $\mathcal B_{\leq2}^{\mathrm{on}}$, and
Lemma~\ref{lem:power-filter-gadget-score} gives its distinguished-edge
score as $\Psi_{p,q}(f)$. Hence the reverse inequality holds in
\eqref{eq:degree-two-edge-reduction-lower}, proving the claim.
\end{proof}

\xhdr{Remarks on the admissible function class.} The order-state
results are independent of $f$. The score calculation and edge
reduction use only Borel measurability, the bounds $0\leq f\leq1$,
and the normalization $f(1)=1$; they require neither continuity nor
monotonicity. Thus both lemmas remain valid for every such measurable
profile. The upper bound $f\leq1$ is essential to the proof above:
it ensures nonnegative online contributions and
$M_r(f)\leq1/(r+1)$. This extension concerns the inner
characterization only. \emph{The outer optimization below remains over
$\mathcal F$, whose monotonicity and right-continuity provide the
probability-measure representation.}

\subsection{A Three-Constraint Upper Bound}
\label{subsec:three-constraint-upper-bound}

For the remainder of the certifying-value analysis, write
\[
r^\star:=\frac{2}{3-s^\star},
\]
where $s^\star$ is defined in~\eqref{eq:s-two-star-polynomial}.
We first prove the upper bound on $\Gamma(2)$.

\begin{lemma}[Three-constraint upper bound]
\label{lem:three-constraint-upper-bound}
One has
\begin{equation}
\label{eq:Gamma-two-upper-bound}
\Gamma(2)\leq r^\star.
\end{equation}
\end{lemma}

\xhdr{Choice of the three active constraints.} Lemma~\ref{lem:exact-degree-two-edge-reduction}
expresses $\chi(f;2)$ as the minimum of $H_f(1)$ and the family
$\{\Psi_{p,q}(f)\}_{p,q\geq0}$. Retaining any subfamily gives an
upper bound. We use
\[
H_f(1),\qquad\Psi_{0,0}(f),\qquad\Psi_{2,1}(f).
\]
The first is the singleton obstruction; the second is the simplest
two-root constraint, with no preprocessing.

To motivate this choice, consider the admissible step profile
$f_{s,w}$ with Stieltjes measure
\[
\pi_{s,w}=w\delta_s+(1-w)\delta_1,
\qquad 0<s<\frac14,\quad0<w<1.
\]
Equivalently, $f_{s,w}$ is zero below $s$, equals $w$ on $[s,1)$,
and equals one at $1$. Equalizing the first two constraints gives
\[
w(1-s)=1-\frac{w(1-s)^2}{2}.
\]
Thus the common value and the required mass are
\begin{equation}
\label{eq:heuristic-eta-omega-degree-two}
H_{f_{s,w}}(1)=\Psi_{0,0}(f_{s,w})=\frac{2}{3-s},
\qquad
w=\frac{2}{(1-s)(3-s)}.
\end{equation}
For $0<s<1/4$, this choice indeed satisfies $0<w<1$.

The first two constraints leave one free parameter $s$, and their
common value increases with $s$. The constraint $\Psi_{2,1}$ becomes
tight at the candidate optimum.\footnote{Along this family,
$\Psi_{2,1}$ becomes tight at $s^\star=0.2027\ldots$, before
$\Psi_{1,1}$, whose first positive equality point is
$0.2050\ldots$. Lemma~\ref{lem:all-filtered-root-constraints}
verifies that no remaining constraint is smaller at $s^\star$.}
Indeed, substituting~\eqref{eq:heuristic-eta-omega-degree-two} gives
\begin{equation}
\label{eq:quartic-origin-degree-two}
\Psi_{2,1}(f_{s,w})-\frac{2}{3-s}
=
\frac{2s^4+2s^3-8s^2-33s+7}{15(3-s)}.
\end{equation}
Requiring this difference to vanish yields exactly
\eqref{eq:s-two-star-polynomial}, and then
$w=w^\star=r^\star/(1-s^\star)$.
This restricted family serves only to identify a candidate profile
and active constraints. The proof below bounds every $f\in\mathcal F$;
it does not assume that an optimizer has this step form.

\begin{proof}[\textbf{Proof of Lemma~\ref{lem:three-constraint-upper-bound}}]
For $x\in[0,1]$, let $h_x(y):=\mathbf1\{y\geq x\}$. Each
$h_x$ belongs to $\mathcal F$, and every $f\in\mathcal F$ has the
representation
\begin{equation}
\label{eq:stieltjes-threshold-representation}
f(y)=\pi_f([0,y])
=\int_{[0,1]}h_x(y)\,\pi_f(dx).
\end{equation}
Each of the three selected functionals is a constant plus an integral
of $f$ against a bounded kernel. Since $\pi_f$ is a probability
measure, Fubini's theorem therefore gives
\[
L(f)=\int_{[0,1]}L(h_x)\,\pi_f(dx)
\]
for each of these functionals $L$. The probability normalization
preserves its constant term, and boundedness ensures absolute
integrability.

The corresponding threshold kernels are
\begin{align}
K_H(x)
&:=H_{h_x}(1)=1-x,
\label{eq:K-H-degree-two}\\
K_{0,0}(x)
&:=\Psi_{0,0}(h_x)=\frac12+x-\frac{x^2}{2},
\label{eq:K-00-degree-two}\\
K_{2,1}(x)
&:=\Psi_{2,1}(h_x)
=\frac9{10}-x+\frac{x^2}{2}+\frac{x^3}{3}-\frac{x^5}{15}.
\label{eq:K-21-degree-two}
\end{align}
Consequently,
\begin{equation}
\label{eq:threshold-representation-three-constraints}
\begin{aligned}
H_f(1)&=\int_{[0,1]}K_H(x)\,\pi_f(dx),\\
\Psi_{0,0}(f)&=\int_{[0,1]}K_{0,0}(x)\,\pi_f(dx),\\
\Psi_{2,1}(f)&=\int_{[0,1]}K_{2,1}(x)\,\pi_f(dx).
\end{aligned}
\end{equation}

\xhdr{A dual mixture of the three constraints.} We seek a convex
combination $\breve K$ that is bounded above by $r^\star$ and
attains this value at the candidate support points $s^\star$ and
$1$. The interior contact at $s^\star$ also suggests tangency:
\[
\breve K(s^\star)=\breve K(1)=r^\star,
\qquad
\left.\frac{d\breve K(x)}{dx}\right|_{x=s^\star}=0.
\]
Solving these contact conditions together with normalization gives
the weights below.

The defining quartic is positive at $1/5$ and negative at $1/4$, so
$1/5<s^\star<1/4$. Define
\[
u^\star
:=3(s^\star)^4-10(s^\star)^3-13(s^\star)^2+24s^\star+46
\]
and
\begin{align}
\lambda_H
&:=\frac{3(s^\star)^4-2(s^\star)^3-7(s^\star)^2+8s^\star+8}
{u^\star},
\label{eq:lambda-H-degree-two}\\
\lambda_{0,0}
&:=\frac{-8(s^\star)^3-6(s^\star)^2+16s^\star+23}
{u^\star},
\label{eq:lambda-00-degree-two}\\
\lambda_{2,1}
&:=\frac{15}{u^\star}.
\label{eq:lambda-21-degree-two}
\end{align}
For $0<s^\star<1/4$, all three numerators are positive, and their
sum is $u^\star$. Hence the weights are positive and sum to
one.

Set
\[
\breve K(x)
:=\lambda_HK_H(x)+\lambda_{0,0}K_{0,0}(x)
+\lambda_{2,1}K_{2,1}(x).
\]
Using~\eqref{eq:s-two-star-polynomial}, direct simplification gives
\begin{equation}
\label{eq:mixed-kernel-factorization-degree-two}
r^\star-\breve K(x)
=
\frac{(x-s^\star)^2(1-x)
\bigl(4-3(s^\star)^2-2s^\star x-2s^\star-x^2-x\bigr)}
{u^\star}.
\end{equation}
The remaining factor is strictly positive on $[0,1]$, since
\[
4-3(s^\star)^2-2s^\star x-2s^\star-x^2-x
\geq2-4s^\star-3(s^\star)^2>0.
\]
Therefore,
\begin{equation}
\label{eq:mixed-kernel-upper-bound}
\breve K(x)\leq r^\star
\qquad\text{for every }x\in[0,1],
\end{equation}
with equality exactly at $x=s^\star$ and $x=1$.

For every $f\in\mathcal F$,
Lemma~\ref{lem:exact-degree-two-edge-reduction} now gives
\[
\begin{aligned}
\chi(f;2)
&\leq\min\{H_f(1),\Psi_{0,0}(f),\Psi_{2,1}(f)\}\\
&\leq\lambda_HH_f(1)
+\lambda_{0,0}\Psi_{0,0}(f)
+\lambda_{2,1}\Psi_{2,1}(f)=\int_{[0,1]}\breve K(x)\,\pi_f(dx)
\leq r^\star.
\end{aligned}
\]
Taking the supremum over $f\in\mathcal F$ proves the lemma.
\end{proof}

\subsection{The Matching Lower Bound}
\label{subsec:matching-lower-bound-degree-two}

We next show that the candidate profile attains $r^\star$.

\begin{lemma}[Matching lower bound]
\label{lem:degree-two-matching-lower-bound}
One has
\begin{equation}
\label{eq:Gamma-two-lower-bound}
\Gamma(2)\geq r^\star.
\end{equation}
\end{lemma}

The dual kernel is tight only at $s^\star$ and $1$, suggesting a
Stieltjes measure supported at those points. Balancing the first two
constraints fixes its mass at $s^\star$ to be $w^\star$. Thus
we use the profile already defined in
\eqref{eq:optimal-step-profile-degree-two}:
\[
f_2^\star(y)=
\begin{cases}
0,&0\leq y<s^\star,\\
w^\star,&s^\star\leq y<1,\\
1,&y=1,
\end{cases}
\qquad
w^\star=\frac{r^\star}{1-s^\star}.
\]
It belongs to $\mathcal F$, and its Stieltjes measure is
$\pi_{f_2^\star}=w^\star\delta_{s^\star}+(1-w^\star)\delta_1$.

The three selected constraints are tight:
\begin{equation}
\label{eq:first-two-active-degree-two}
H_{f_2^\star}(1)=w^\star(1-s^\star)=r^\star,
\qquad
\Psi_{0,0}(f_2^\star)
=1-\frac{w^\star(1-s^\star)^2}{2}=r^\star,
\end{equation}
where the second identity uses $r^\star=2/(3-s^\star)$.
Similarly,
\begin{equation}
\label{eq:third-active-degree-two}
\Psi_{2,1}(f_2^\star)-r^\star
=
\frac{2(s^\star)^4+2(s^\star)^3-8(s^\star)^2-33s^\star+7}
{15(3-s^\star)}
=0.
\end{equation}
It remains to verify that none of the other filtered-root constraints
falls below this value.

\begin{lemma}[All filtered-root constraints]
\label{lem:all-filtered-root-constraints}
For every $p,q\in\mathbb Z_{\geq0}$,
\begin{equation}
\label{eq:all-filtered-root-constraints}
\Psi_{p,q}(f_2^\star)\geq r^\star.
\end{equation}
Equality holds exactly for $(p,q)=(0,0)$ and $(p,q)=(2,1)$.
\end{lemma}

\begin{proof}
Evaluating~\eqref{eq:power-filter-gadget-score} at $f_2^\star$ and
using
\[
\frac{1-w^\star}{w^\star}
=\frac{(s^\star)^2-4s^\star+1}{2}
\]
gives
\begin{equation}
\label{eq:Psi-p-q-minus-gamma-Q}
\Psi_{p,q}(f_2^\star)-r^\star
=
\frac{w^\star}{(p+1)(q+1)(p+q+2)}Q_{p,q}(s^\star),
\end{equation}
where
\begin{equation}
\label{eq:Q-p-q-definition}
\begin{aligned}
Q_{p,q}(s^\star)
:={}&(q+1)(p+q+2)(s^\star)^{p+1}
+(p+1)(p+q+2)(s^\star)^{q+1}\\
&-(q+1)(s^\star)^{p+q+2}-(p+1)+\frac{(s^\star)^2-4s^\star+1}{2}
(p+q+2)(p+q+1).
\end{aligned}
\end{equation}
Since $w^\star>0$, it suffices to show $Q_{p,q}(s^\star)\geq0$.

The quartic in~\eqref{eq:s-two-star-polynomial} is strictly decreasing
on $[0,1]$. Evaluating it at $1/5$ and $203/1000$ gives
\begin{equation}
\label{eq:coarse-s-ratio-bounds}
\frac15<s^\star<\frac{203}{1000},
\qquad
\frac{(s^\star)^2-4s^\star+1}{2}>\frac{57}{500}.
\end{equation}

\xhdr{The case $p+q+2\geq10$.} Because $0<s^\star<1$, the first
and third monomial terms in~\eqref{eq:Q-p-q-definition} have
nonnegative sum:
\[
(q+1)(s^\star)^{p+1}
\bigl(p+q+2-(s^\star)^{q+1}\bigr)\geq0.
\]
The second monomial term is also nonnegative. Hence
\[
\begin{aligned}
Q_{p,q}(s^\star)
&\geq
\frac{(s^\star)^2-4s^\star+1}{2}
(p+q+2)(p+q+1)-(p+1)\\
&\geq
(p+q+1)\left(\frac{57}{500}(p+q+2)-1\right)
>0.
\end{aligned}
\]

\xhdr{The case $p+q+2\leq9$.} Using
\eqref{eq:s-two-star-polynomial}, every $Q_{p,q}(s^\star)$ reduces
to a polynomial in $s^\star$ of degree at most three. The exact
rational verification in Appendix~\ref{app:finite-filter-verification}
gives
\[
Q_{p,q}(s^\star)\geq0
\qquad
\text{for all }p,q\in\mathbb Z_{\geq0}\text{ with }p+q+2\leq9,
\]
with equality exactly at $(p,q)=(0,0)$ and $(p,q)=(2,1)$.
Together with~\eqref{eq:Psi-p-q-minus-gamma-Q}, this proves the lemma.
\end{proof}

\begin{proof}[\textbf{Proof of Lemma~\ref{lem:degree-two-matching-lower-bound}}]
The exact reduction in
Lemma~\ref{lem:exact-degree-two-edge-reduction}, the identity
$H_{f_2^\star}(1)=r^\star$, and
Lemma~\ref{lem:all-filtered-root-constraints} imply
\begin{equation}
\label{eq:step-profile-certificate-degree-two}
\chi(f_2^\star;2)=r^\star.
\end{equation}
Since $f_2^\star\in\mathcal F$,
\[
\Gamma(2)=\sup_{f\in\mathcal F}\chi(f;2)
\geq\chi(f_2^\star;2)=r^\star.
\]
Combined with Lemma~\ref{lem:three-constraint-upper-bound}, this gives
$\Gamma(2)=r^\star$ and shows that the supremum is attained.
\end{proof}


\subsection{Uniqueness of the Optimal Profile}
\label{subsec:uniqueness-degree-two-closure}

Let $g\in\mathcal F$ attain $\Gamma(2)=r^\star$, and let
$\pi_g$ be its Stieltjes probability measure. The upper-bound proof
gives
\[
r^\star=\chi(g;2)
\leq\int_{[0,1]}\breve K(x)\,\pi_g(dx)
\leq r^\star.
\]
Equality therefore holds throughout. By
\eqref{eq:mixed-kernel-factorization-degree-two},
$\breve K(x)<r^\star$ away from $s^\star$ and $1$. Hence
\begin{equation}
\label{eq:optimal-measure-support-degree-two}
\operatorname{supp}(\pi_g)\subseteq\{s^\star,1\}.
\end{equation}
Write
\[
\pi_g=\theta\delta_{s^\star}+(1-\theta)\delta_1,
\qquad\theta\in[0,1].
\]
Since $\chi(g;2)=r^\star$, the exact edge reduction implies
$H_g(1)\geq r^\star$ and $\Psi_{0,0}(g)\geq r^\star$.
Equivalently,
\[
\theta(1-s^\star)\geq r^\star,
\qquad
1-\frac{\theta(1-s^\star)^2}{2}\geq r^\star.
\]
Using $r^\star=2/(3-s^\star)$, these yield
\[
\theta\geq\frac{r^\star}{1-s^\star}=w^\star,
\qquad
\theta\leq\frac{2(1-r^\star)}{(1-s^\star)^2}
=\frac{r^\star}{1-s^\star}=w^\star.
\]
Thus $\theta=w^\star$ and $\pi_g=\pi_{f_2^\star}$.
The bijection between $\mathcal F$ and $\mathcal P([0,1])$ now gives
$g=f_2^\star$ at every point of $[0,1]$. This proves uniqueness in
$\mathcal F$.


\section{Proof of the Online-Degree-Two Competitiveness Assertion in
Theorem~\ref{thm:Gamma-two}}
\label{sec:proof-degree-two-competitiveness}

In this section, we prove
Equation~\eqref{eq:gamma-two-exact-value} in
Theorem~\ref{thm:Gamma-two}. We first encode the monomial safety
profiles from the preceding section by a two-parameter state and
formulate a state-dependent dual-budget criterion in
Subsection~\ref{subsec:degree-two-state-budget-criterion}. We then
construct an explicit feasible budget in
Subsection~\ref{subsec:degree-two-feasible-budget}; its technical
feasibility verification is deferred to
Appendix~\ref{app:degree-two-competitiveness-auxiliary}. Finally,
Subsection~\ref{subsec:degree-two-binary-upper-construction} gives a
matching binary layered construction, proving that the lower guarantee
is exact.

Throughout this section, write
\[
c:=1-\sum_{k=1}^{\infty}2^{-(2^k+k-1)}.
\]
We prove that every instance in $\mathcal B_{\leq2}^{\mathrm{on}}$
has expected matching size at least $c\operatorname{OPT}(G)$, and
then construct instances whose competitiveness converges to $c$.

\subsection{Safety States and a Dual-Budget Criterion}
\label{subsec:degree-two-state-budget-criterion}

For a positive offline node, write its monomial safety profile as
\[
\Pr[\mathcal E_i\mid Y_i=y]=m_iq_i y^{m_i-1},
\qquad
m_i:=p_i+1,
\qquad
q_i:=\Pr[\mathcal E_i]=\frac{a_i}{p_i+1}.
\]
The state is determined by the fixed arrival prefix, not by the
realized matching. Write $\mathsf s_i=(m_i,q_i)$, where
$m_i\geq1$ and $0<q_i\leq1/m_i$. A fresh node has state $(1,1)$;
write $\bot$ for a dead state and set $q(\bot)=0$.

When an arrival joins two positive states with disjoint witness sets,
the matching probability and the two output states are
\begin{equation}
\label{eq:degree-two-budget-transition}
\begin{aligned}
P(\mathsf s_1,\mathsf s_2)&=q_1+q_2-q_1q_2,\\
\mathsf s_1^+&=\left(m_1+m_2,\frac{m_1q_1q_2}{m_1+m_2}\right),\\
\mathsf s_2^+&=\left(m_1+m_2,\frac{m_2q_1q_2}{m_1+m_2}\right).
\end{aligned}
\end{equation}
Indeed, the post-arrival conditional safety probability of the first
root is
\[
m_1q_1y^{m_1-1}\int_0^y m_2q_2z^{m_2-1}\,dz
=m_1q_1q_2y^{m_1+m_2-1},
\]
and the other expression is symmetric. For mutually exclusive safety
events, both outputs are dead and the matching probability is
$q_1+q_2$. A singleton arrival makes its neighbor dead and matches
with probability $q_i$.

Let $\mathcal S_2$ be the smallest set containing $(1,1)$ and closed
under the two outputs in~\eqref{eq:degree-two-budget-transition}.
The order-state characterization implies that every positive state
arising in an online-degree-two instance belongs to $\mathcal S_2$.

\begin{lemma}[Degree-two dual-budget criterion]
\label{lem:degree-two-dual-budget-criterion}
Suppose $b$ assigns a budget to each state in
$\mathcal S_2\cup\{\bot\}$ such that
\[
b(\bot)=0,
\qquad b(1,1)=c,
\qquad 0\leq b(\mathsf s)\leq\min\{c,q(\mathsf s)\}.
\]
If every pair of positive input states satisfies
\begin{equation}
\label{eq:degree-two-dual-budget-inequalities}
b(\mathsf s_1)+b(\mathsf s_2)-b(\mathsf s_k^+)
\leq P(\mathsf s_1,\mathsf s_2),
\qquad k\in\{1,2\},
\end{equation}
for the independent transition in~\eqref{eq:degree-two-budget-transition},
then $\gamma(2)\geq c$.
\end{lemma}

The proof is given in
Appendix~\ref{app:degree-two-dual-budget-criterion-proof}. It assigns
state-dependent dual targets to offline nodes and shows that the total
expected increase in dual mass at each arrival is at most that
arrival's matching probability.

\subsection{A Feasible State-Dependent Budget}
\label{subsec:degree-two-feasible-budget}

Define
\begin{equation}
\label{eq:degree-two-symmetric-budget}
b_{\mathrm{sym}}(q)
:=q-\sum_{\ell=1}^{\infty}
\frac{q^{2^\ell}}{2^{2^\ell+\ell-1}},
\qquad q\in[0,1].
\end{equation}
The series converges uniformly and gives
\begin{equation}
\label{eq:degree-two-symmetric-budget-identities}
\begin{aligned}
b_{\mathrm{sym}}(1)&=c,\\
2b_{\mathrm{sym}}(q)-b_{\mathrm{sym}}(q^2/2)&=2q-q^2,\\
0\leq b_{\mathrm{sym}}(q)&\leq q-\frac{q^2}{4}.
\end{aligned}
\end{equation}
The second identity follows by shifting the summation index. For
nonnegativity, use $q^{2^\ell}\leq q$ and
$\sum_{\ell\geq1}2^{-(2^\ell+\ell-1)}=1-c<1$.

Call a state \emph{dyadic} if it is
\[
\left(2^k,2^{-(2^k-1)}\right),
\qquad k\in\mathbb Z_{\geq0}.
\]
Assign it the budget $b_{\mathrm{sym}}(2^{-(2^k-1)})$.
For non-dyadic states, assign the seven positive budgets in
Table~\ref{tab:degree-two-state-budgets}, and set every other budget
to zero, including $b(\bot)$. The table also displays the first four
dyadic states. These two prescriptions agree on those four entries
by~\eqref{eq:degree-two-symmetric-budget-identities}.

\begin{table}[t]
\centering
\caption{Small positive-budget states. All dyadic states receive the
budget in~\eqref{eq:degree-two-symmetric-budget}; all non-dyadic
states not listed here receive budget zero.}
\label{tab:degree-two-state-budgets}
\renewcommand{\arraystretch}{1.15}
\begin{tabular}{c c c}
\toprule
Index & State $(m,q)$ & Budget $b(m,q)$\\
\midrule
1 & $(1,1)$ & $c$\\
2 & $(2,1/2)$ & $2c-1$\\
3 & $(3,1/6)$ & $3c-2$\\
4 & $(3,1/3)$ & $3c-2$\\
5 & $(4,1/8)$ & $4c-11/4$\\
6 & $(5,1/30)$ & $5c-43/12$\\
7 & $(5,1/20)$ & $5c-43/12$\\
8 & $(6,1/72)$ & $6c-155/36$\\
9 & $(7,1/112)$ & $7c-241/48$\\
10 & $(7,1/84)$ & $7c-241/48$\\
11 & $(8,1/128)$ & $8c-367/64$\\
\bottomrule
\end{tabular}
\end{table}

The dyadic budgets make equal-state merging transitions tight. The
seven additional states cover the remaining small transitions that
would otherwise violate the dual-budget inequalities.

\begin{lemma}[Feasibility of the state-dependent budget]
\label{lem:degree-two-state-budget-feasible}
The budget above satisfies
Lemma~\ref{lem:degree-two-dual-budget-criterion}.
\end{lemma}

The proof is given in
Appendix~\ref{app:degree-two-state-budget-feasibility-proof}. It
reduces the finite part to exact rational inequalities among the
small positive-budget states and handles the infinite dyadic tail
analytically.

Lemmas~\ref{lem:degree-two-dual-budget-criterion}
and~\ref{lem:degree-two-state-budget-feasible} establish
\[
\gamma(2)\geq c.
\]

\subsection{A Matching Binary Layered Construction}
\label{subsec:degree-two-binary-upper-construction}

We use the binary layered obstruction underlying the upper bound
in~\cite{BCH25}, and evaluate its performance under \rank{} directly.
Fix an integer $L\geq1$ and start with $2^L$ fresh offline roots.
At each level, pair disjoint components and add one online arrival
adjacent to their two roots. Predetermine one root to be retained for
the next level; the other receives no further edges and is called
\emph{retired}. After $L$ levels, append a singleton arrival adjacent
to the last retained root. Every choice is fixed independently of the
ranks, and every online degree is at most two.

This instance has a perfect matching: assign each merge arrival to
its retired root and the final singleton to the last root. Therefore
its optimum is $2^L$.

The two input states at every merge are independent and identical.
Let $q_k$ be the safety probability of either output root after level
$k$, with $q_0=1$. By~\eqref{eq:degree-two-budget-transition},
\[
q_{k+1}=\frac{q_k^2}{2},
\qquad
q_k=2^{-(2^k-1)}.
\]
Exactly $2^{L-k}$ roots retire at level $k$. A retired root remains
unmatched precisely when it is safe immediately after that merge,
and the last retained root is matched by its singleton if it has not
been matched earlier. Linearity of expectation therefore gives
\begin{equation}
\label{eq:degree-two-binary-ratio}
\frac{\mathbb E[\rank(G_L)]}{\operatorname{OPT}(G_L)}
=1-\sum_{k=1}^{L}2^{-k}q_k
=1-\sum_{k=1}^{L}2^{-(2^k+k-1)}.
\end{equation}
Letting $L\to\infty$ proves $\gamma(2)\leq c$. Together with the
lower bound, this proves Equation~\eqref{eq:gamma-two-exact-value}
in Theorem~\ref{thm:Gamma-two}.

\xhdr{The certifying gap.}
Combining the values of $\Gamma(2)$ and $\gamma(2)$ established above
gives
\[
\Del(2)=\gamma(2)-\Gamma(2)
=1-\sum_{k=1}^{\infty}2^{-(2^k+k-1)}
-\frac{2}{3-s^\star}
=0.0027\ldots>0.
\]
For comparison, strict positivity already follows from the coarse
bounds
\[
\Gamma(2)<\frac{2000}{2797}<\frac{71777}{100000}<c,
\]
using~\eqref{eq:coarse-s-ratio-bounds} and
\eqref{eq:degree-two-competitiveness-rational-bracket}. Thus the
strict separation itself does not require the exact evaluation of
$\gamma(2)$. This completes the proof of
Theorem~\ref{thm:Gamma-two}.

\section{Proof of the Certifying Bounds for Large Online Degree in
Theorem~\ref{thm:Gamma-large-degree}}
\label{sec:proof-large-degree-certifying-value}

In this section, we prove
Inequality~\eqref{eq:Gamma-large-degree-bounds}
in Theorem~\ref{thm:Gamma-large-degree}. For the lower bound, we retain the preprocessing threshold of the
distinguished offline node and the reduced partner rank of the final
arrival. An expected-slack estimate for these two ranks determines a
perturbation of the exponential profile. For the upper bound, we use
common-pool and complete bipartite instances to form a positive
mixture of exact edge-score kernels. Throughout this section, write
\[
D:=d-1.
\]

\subsection{Two Critical Ranks}
\label{subsec:large-degree-critical-ranks}

Fix $G\in\mathcal B_{\leq d}^{\mathrm{on}}$ and a distinguished edge
$(i^\star,j^\star)$. Truncating immediately after $j^\star$ leaves
its online contribution unchanged and can only decrease the offline
contribution. For the lower-bound argument, we therefore assume that
$j^\star$ is the final arrival.

Run \rank{} on $G^{-i^\star}:=G-i^\star$. If
$j_1\prec\cdots\prec j_r\prec j^\star$ are the earlier online
neighbors of $i^\star$, define
\begin{equation}
\label{eq:large-degree-alpha-tau}
\alpha:=\max_{\ell\in[r]}R_{G^{-i^\star}}(j_\ell),
\qquad
\tau:=R_{G^{-i^\star}}(j^\star),
\end{equation}
with $\max\varnothing:=0$ and partner rank one for an unmatched
online node. Both variables are independent of $Y_{i^\star}$.
For $a,t\in[0,1]$, define
\begin{equation}
\label{eq:large-degree-response}
\mathscr R_f(a,t)
:=1+F_f(a)-\bigl(1-(t-a)_+\bigr)f(t),
\end{equation}
where $(t-a)_+:=\max\{t-a,0\}$.

\begin{lemma}[Two-threshold response bound]
\label{lem:large-degree-response-bound}
For every $f\in\mathcal F$,
\begin{equation}
\label{eq:large-degree-response-bound}
\Phi_G(i^\star,j^\star;f)
\geq\mathbb E[\mathscr R_f(\alpha,\tau)].
\end{equation}
\end{lemma}

\begin{proof}
Condition on all ranks other than $Y_{i^\star}$. Inserting $i^\star$
with rank $y$ matches it before $j^\star$ exactly when $y<\alpha$,
apart from ties. Until $i^\star$ is selected, the original and
reduced executions agree; the first earlier neighbor whose reduced
partner rank exceeds $y$ selects $i^\star$. If there is no such
neighbor, $i^\star$ remains safe until $j^\star$.

Thus $i^\star$ is matched by the end exactly when
$y<\max\{\alpha,\tau\}$, and its conditional expected contribution
is $F_f(\max\{\alpha,\tau\})$. By
Lemma~\ref{lem:offline-insertion-boundary} and monotonicity of $f$,
the online contribution is at least $1-f(\tau)$. This remains true
when $\tau=1$, since $f(1)=1$. When $\alpha<y<\tau$, the final
arrival selects $i^\star$, so its contribution is $1-f(y)$.
The conditional edge score is therefore at least
\[
F_f(\max\{\alpha,\tau\})+1-f(\tau)
+\mathbf1\{\alpha<\tau\}
\int_\alpha^\tau\bigl(f(\tau)-f(y)\bigr)\,dy.
\]
In the two cases $\alpha\leq\tau$ and $\alpha>\tau$, this simplifies
to $\mathscr R_f(\alpha,\tau)$. Taking expectations proves the
lemma. The argument uses boundedness and monotonicity, not continuity
of $f$.
\end{proof}

For $f_\infty^\star(t)=e^{t-1}$, define the response slack
\begin{equation}
\label{eq:large-degree-exponential-slack}
\mathscr S(a,t)
:=\mathscr R_{f_\infty^\star}(a,t)-\left(1-\frac1e\right).
\end{equation}
Direct calculation gives
\begin{equation}
\label{eq:large-degree-slack-formula}
\mathscr S(a,t)=
\begin{cases}
\displaystyle e^{a-1}\int_0^{t-a}u e^u\,du,&t\geq a,\\[2mm]
e^{a-1}-e^{t-1},&t<a.
\end{cases}
\end{equation}
In particular,
\begin{equation}
\label{eq:large-degree-asymmetric-slack}
\mathscr S(a,t)\geq
\begin{cases}
\frac12e^{a-1}(t-a)^2,&t\geq a,\\[1mm]
\frac1e(a-t),&t<a.
\end{cases}
\end{equation}
The slack is quadratic above the diagonal and linear below it. The
following estimate quantifies the contribution forced by bounded
online degree.

\begin{lemma}[Expected exponential slack]
\label{lem:large-degree-conditional-slack}
For the critical ranks in~\eqref{eq:large-degree-alpha-tau},
\begin{equation}
\label{eq:large-degree-conditional-slack}
\mathbb E[\mathscr S(\alpha,\tau)]
\geq\frac{1}{D^2+4D+2}
\mathbb E\!\left[e^{\alpha-1}(1-\alpha)^2\right].
\end{equation}
\end{lemma}

\begin{proof}
The rank-raising estimates in
Appendix~\ref{app:large-degree-auxiliary-estimates} give, for
$\mathcal L(\alpha)$-almost every $a\in[0,1)$,\footnote{Here
$\mathcal L(\alpha)$ is the probability law of $\alpha$; the
statement holds outside a Borel set $S$ with
$\mathcal L(\alpha)(S)=0$.}
\begin{align}
\mathbb E[(\tau-a)_+^2\mid\alpha=a]
&\geq\frac{2(1-a)^2}{(D+1)(D+2)}(1-p_a),
\label{eq:large-degree-upper-gap-moment}\\
\mathbb E[(a-\tau)_+\mid\alpha=a]
&\geq\frac{(1-a)p_a^2}{2D},
\label{eq:large-degree-lower-gap-moment}
\end{align}
where $p_a:=\Pr[\tau<a\mid\alpha=a]$. Combining these estimates
with~\eqref{eq:large-degree-asymmetric-slack} yields
\[
\mathbb E[\mathscr S(\alpha,\tau)\mid\alpha=a]
\geq\frac{1-a}{2eD}p_a^2
+\frac{e^{a-1}(1-a)^2}{(D+1)(D+2)}(1-p_a).
\]
Completing the square gives
\[
\mathbb E[\mathscr S(\alpha,\tau)\mid\alpha=a]
\geq\frac{e^{a-1}(1-a)^2}{(D+1)(D+2)}
\left(1-\frac{De^a(1-a)}{2(D+1)(D+2)}\right).
\]
Since $e^a(1-a)\leq1$ on $[0,1]$ and
\[
\frac1{(D+1)(D+2)}
-\frac{D}{2(D+1)^2(D+2)^2}
\geq\frac1{D^2+4D+2},
\]
this proves the required conditional lower bound for almost every
$a<1$ with respect to $\mathcal L(\alpha)$. At $a=1$, its
right-hand side is zero and $\mathscr S\geq0$. Taking expectations
proves~\eqref{eq:large-degree-conditional-slack}.
\end{proof}

\subsection{An Explicit Improved Gain-Sharing Function}
\label{subsec:large-degree-lower-bound}

\begin{lemma}[Explicit lower bound]
\label{lem:large-degree-lower-bound}
For every integer $d\geq2$,
\begin{equation}
\label{eq:large-degree-explicit-lower-bound}
\Gamma(d)\geq1-\frac1e+\frac1{3e\,d(d+2)}.
\end{equation}
\end{lemma}

The slack estimate depends on the preprocessing threshold through
$e^{a-1}(1-a)^2$. We choose a perturbation whose diagonal response
cancels that dependence. Define
\begin{equation}
\label{eq:large-degree-perturbation}
g(t):=\frac{t+2}{3}e^{t-1}(1-t)^2,
\qquad
f_d^{\mathsf{lb}}(t):=e^{t-1}+\frac{g(t)}{d(d+2)}.
\end{equation}
The superscript indicates a lower-bound construction; no optimality
of this profile is asserted.

\begin{proof}[\textbf{Proof of Lemma~\ref{lem:large-degree-lower-bound}}]
The function $g$ is nonnegative, $g(1)=0$, and
\[
\frac{dg(t)}{dt}
=-\frac{(1-t)(t^2+4t+1)}3e^{t-1}.
\]
For $t\in[0,1]$,
\[
0\leq(1-t)(t^2+4t+1)
=1+3t(1-t)-t^3\leq\frac74.
\]
Thus $-e^{t-1}\leq dg(t)/dt\leq0$, and $f_d^{\mathsf{lb}}$ is
strictly increasing, nonnegative, and continuous, with terminal
value one. In particular, $f_d^{\mathsf{lb}}\in\mathcal F$.

Direct integration gives
\begin{equation}
\label{eq:large-degree-diagonal-cancellation}
\int_0^a g(y)\,dy-g(a)
=\frac1{3e}-e^{a-1}(1-a)^2.
\end{equation}
If $t\geq a$, nonnegativity and nonincreasingness of $g$ give
\[
g(a)-\bigl(1-(t-a)\bigr)g(t)
=g(a)-g(t)+(t-a)g(t)\geq0.
\]
If $t<a$, the derivative bound gives
\[
g(a)-g(t)\geq-\int_t^a e^{y-1}\,dy=-\mathscr S(a,t).
\]
Substituting~\eqref{eq:large-degree-diagonal-cancellation} into the
response functional therefore yields, in both cases,
\begin{equation}
\label{eq:large-degree-perturbed-response}
\begin{aligned}
\mathscr R_{f_d^{\mathsf{lb}}}(a,t)
\geq{}&1-\frac1e
+\left(1-\frac1{d(d+2)}\right)\mathscr S(a,t)+\frac1{d(d+2)}
\left(\frac1{3e}-e^{a-1}(1-a)^2\right).
\end{aligned}
\end{equation}

Since $D^2+4D+2=d(d+2)-1$, the threshold-dependent terms cancel
upon applying Lemma~\ref{lem:large-degree-conditional-slack}:
\[
\left(1-\frac1{d(d+2)}\right)\frac1{d(d+2)-1}
=\frac1{d(d+2)}.
\]
Consequently,
\[
\mathbb E[\mathscr R_{f_d^{\mathsf{lb}}}(\alpha,\tau)]
\geq1-\frac1e+\frac1{3e\,d(d+2)}.
\]
Lemma~\ref{lem:large-degree-response-bound} gives this lower bound
for every edge score. Taking the infimum over graph-edge pairs and
using $f_d^{\mathsf{lb}}\in\mathcal F$ proves the claim.
\end{proof}

\subsection{Finite Graphs for the Upper Bound}
\label{subsec:large-degree-upper-gadgets}

\begin{lemma}[Logarithmic second-order upper bound]
\label{lem:large-degree-upper-bound}
For all sufficiently large $D$,
\begin{equation}
\label{eq:large-degree-logarithmic-upper-bound}
\Gamma(D+1)\leq1-\frac1e+O\!\left(\frac{\ln D}{D^2}\right),
\end{equation}
with an absolute implicit constant.
\end{lemma}

For $h_s(y):=\mathbf1\{y\geq s\}$, write
$\Xi_{G,i,j}(s):=\Phi_G(i,j;h_s)$. The Stieltjes representation
and Fubini's theorem give
\[
\Phi_G(i,j;f)=\int_{[0,1]}\Xi_{G,i,j}(s)\,\pi_f(ds),
\qquad f\in\mathcal F.
\]
For any finite convex combination $\breve K_D$ of kernels from
$\mathcal B_{\leq D+1}^{\mathrm{on}}$, it follows that
\begin{equation}
\label{eq:large-degree-mixture-upper-bound}
\Gamma(D+1)
\leq\sup_{f\in\mathcal F}
\int_{[0,1]}\breve K_D(s)\,\pi_f(ds)
=\sup_{s\in[0,1]}\breve K_D(s).
\end{equation}
The inequality bounds the adversary's infimum by the chosen convex
combination; the equality uses $h_s\in\mathcal F$. The dual-mixture
interpretation and the equalizing exponential measure are discussed
in Subsection~\ref{subsec:approach-large-degree-kernel-mixtures}.

\xhdr{The common-pool graph.} Partition $D$ offline nodes into a
shared pool $\mathcal U$ of size $\lfloor D/2\rfloor$ and a private
pool $\mathcal P$ of the remaining nodes, and add $i^\star$.
Fix integers $2\leq N\leq D$ and $0\leq q\leq D-N$. Send $q$ arrivals
adjacent to $\mathcal P\cup\mathcal U$, followed by one arrival
adjacent to $\{i^\star\}\cup\mathcal P$ and then one singleton
arrival for each node in $\mathcal P$. Next send $N-2$ arrivals
adjacent to $\mathcal U$, followed by two arrivals adjacent to
$\{i^\star\}\cup\mathcal U$. The latter is $j^\star$, and
$(i^\star,j^\star)$ is distinguished. Every online degree is at
most $D+1$.

After the first $q$ arrivals, let $B_q$ be the first remaining private
rank, and let $S_q,T_q$ be the $(N-1)$st and $N$th remaining shared
ranks, respectively. A nonexistent rank is assigned value one.
Denote this graph's exact threshold-score kernel by $\Xi_{D,q,N}$.
The private singleton phase removes any discrepancy in private-node
availability caused by inserting $i^\star$.

\xhdr{Residual graphs.} We also use the complete bipartite graphs
$K_{D+1,k}$,\footnote{Here $K_{D+1,k}$ denotes the complete bipartite
graph with $D+1$ offline nodes and $k$ online nodes; the online nodes
arrive in any fixed order.} $k\in[D]$, with
an edge incident to the final arrival distinguished; write
$\Xi_{D,k}^{\mathsf{cmp}}$ for their kernels. The two-arrival
singleton graph contributes the kernel $H_{h_s}(1)$.

\begin{lemma}[Positive mixed-kernel construction]
\label{lem:large-degree-positive-mixture}
There is an absolute constant $A>0$ such that, for sufficiently large
$D$ and integers $A\ln D\leq N\leq D^{1/3}$, there are nonnegative
weights
\[
\omega_q\quad(0\leq q\leq D-N),
\qquad \delta_k\quad(k\in[D]),
\qquad \delta_\infty,
\]
whose total mass is one, such that
\begin{equation}
\label{eq:large-degree-mixed-kernel}
\breve K_{D,N}(s)
:=\sum_{q=0}^{D-N}\omega_q\Xi_{D,q,N}(s)
+\sum_{k=1}^{D}\delta_k\Xi_{D,k}^{\mathsf{cmp}}(s)
+\delta_\infty H_{h_s}(1)
\end{equation}
satisfies
\begin{equation}
\label{eq:large-degree-mixed-kernel-bound}
\sup_{s\in[0,1]}\breve K_{D,N}(s)
\leq1-\frac1e+O\!\left(\frac{N}{D^2}\right)+2^{-N},
\end{equation}
with an absolute implicit constant.
\end{lemma}

The proof is given in
Appendix~\ref{app:large-degree-auxiliary-estimates}. A corrected
Bernstein mixture approximates the exponential dual measure
in~\eqref{eq:large-degree-exponential-dual-measure} to order
$D^{-2}$. Common-pool graphs realize most coefficients, and complete
bipartite graphs realize the residual coefficients. Their aggregate
score corrections are $O(N/D^2)$. The exceptional ordering
$B_q>T_q$ requires the first $N$ remaining pool labels to be shared
and has probability at most $2^{-N}$; the other corrections are
localized to $(S_q,T_q]$.

\begin{proof}[\textbf{Proof of Lemma~\ref{lem:large-degree-upper-bound}}]
Choose
\[
N=\left\lceil\left(A+\frac3{\ln2}\right)\ln D\right\rceil.
\]
For sufficiently large $D$, the conditions of
Lemma~\ref{lem:large-degree-positive-mixture} hold and
$2^{-N}\leq D^{-3}$. Every component is the exact kernel of a finite
instance in $\mathcal B_{\leq D+1}^{\mathrm{on}}$. Hence
\eqref{eq:large-degree-mixture-upper-bound} gives
\[
\Gamma(D+1)\leq\sup_{s\in[0,1]}\breve K_{D,N}(s)
\leq1-\frac1e+O\!\left(\frac{\ln D}{D^2}\right).
\]
\end{proof}

\xhdr{Completion of the proof.}
Lemma~\ref{lem:large-degree-lower-bound} establishes the lower bound
in~\eqref{eq:Gamma-large-degree-bounds} for every integer $d\geq2$.
For the upper bound, Lemma~\ref{lem:large-degree-upper-bound}, with
$D=d-1$, yields
\[
\Gamma(d)-\left(1-\frac1e\right)
\leq
O\!\left(\frac{\ln d}{d^2}\right)
\]
for all sufficiently large $d$. This proves
Inequality~\eqref{eq:Gamma-large-degree-bounds}.


\section{Proof of the Competitiveness Bounds for Large Online Degree in
Theorem~\ref{thm:Gamma-large-degree}}
\label{sec:proof-large-degree-competitiveness}

In this section, we prove
Inequality~\eqref{eq:gamma-large-degree-bounds} in Theorem~\ref{thm:Gamma-large-degree}. We first reduce to instances with a prescribed perfect matching and
no backward edges. Exposing \rank{} in increasing rank order then
expresses its improvement over $1-1/e$ as an integrated global
threshold surplus. A loss-accounting estimate gives the lower bound,
while a layered family with independent input histories gives the
upper bound. The auxiliary proofs are collected in
Appendix~\ref{app:large-degree-competitiveness-proofs}.

\subsection{Forward Instances and the Global Surplus}
\label{subsubsec:large-degree-global-surplus}

\begin{lemma}[Forward reduction and rank-order equivalence]
\label{lem:large-degree-forward-reduction}
The infimum defining $\gamma(d)$ may be restricted to instances with
\[
I=\{i_1,\ldots,i_n\},\qquad
J=\{j_1,\ldots,j_n\},\qquad j_1\prec\cdots\prec j_n,
\]
whose diagonal edges form a perfect matching and satisfy
\begin{equation}
\label{eq:large-degree-forward-form}
(i_k,j_r)\in E(G)\quad\Longrightarrow\quad k\geq r.
\end{equation}
For any fixed distinct ranks, \rank{} produces the same matching by
processing offline nodes in increasing rank order and assigning each
to its earliest currently unmatched online neighbor, if one exists.
\end{lemma}

Call an instance satisfying~\eqref{eq:large-degree-forward-form}
\emph{forward}. Its optimum is $n$, and $j_r$ is the last incident
arrival of $i_r$. Set
\[
r(G):=\frac{\mathbb E[\rank(G)]}{n},
\qquad
\delta_G:=r(G)-\left(1-\frac1e\right).
\]
For $r\in[n]$, define
\begin{equation}
\label{eq:large-degree-global-critical-ranks}
\beta_r:=\max_{j:(i_r,j)\in E(G)}R_{G-i_r}(j),
\qquad
\tau_r:=R_{G-i_r}(j_r),
\qquad
R_r:=R_G(j_r).
\end{equation}
Insertion monotonicity implies $R_r\leq\tau_r\leq\beta_r$.
For $0\leq t<1$, define
\begin{equation}
\label{eq:large-degree-global-profiles}
\begin{aligned}
q_G(t)&:=\frac1n\sum_{r=1}^n\Pr[\beta_r>t],\\
A_G(t)&:=\frac1n\mathbb E\!\left[
\sum_{i\in I}X_i\mathbf1\{Y_i\leq t\}\right],\\
\mathcal D_G(t)&:=q_G(t)+A_G(t)-1.
\end{aligned}
\end{equation}
We also use $A_G(1)=r(G)$. The function $q_G$ is nonincreasing and,
for a fixed finite graph, is a polynomial on $[0,1)$: conditional on
a strict rank ordering, $\beta_r$ is an order statistic or the
constant one.

\begin{lemma}[Exact global surplus]
\label{lem:large-degree-global-surplus}
Every balanced perfect-matching instance satisfies
\begin{equation}
\label{eq:large-degree-global-surplus-identities}
A_G(t)=\int_0^tq_G(y)\,dy,
\qquad
\mathcal D_G(t)=\frac1n\sum_{r=1}^n
\Pr[R_r\leq t<\beta_r]\geq0.
\end{equation}
Moreover,
\begin{align}
\delta_G
&=\int_0^1e^{t-1}\mathcal D_G(t)\,dt
=\frac1n\sum_{r=1}^n
\mathbb E[e^{\beta_r-1}-e^{R_r-1}],
\label{eq:large-degree-global-surplus-integral}\\
0&\leq A_G(t)-(1-e^{-t})\leq e\delta_G
\qquad(0\leq t\leq1).
\label{eq:large-degree-global-profile-control}
\end{align}
\end{lemma}

Thus the algorithmic lower bound is a bound on the \emph{aggregate}
slack of a complete instance, rather than on its minimum edge score.

\subsubsection{A Quantitative Lower Bound}
\label{subsubsec:large-degree-competitiveness-lower}

The next estimate converts the decrease of $q_G$ into integrated
surplus. Its proof distinguishes two ways an unprocessed offline
node can lose its final unmatched neighbor. If that neighbor is its
diagonal partner, the same competition also creates self-matches with
positive replacement-rank slack. Otherwise, the node previously lost
its diagonal partner while retaining another neighbor, and the
resulting displacement already contributes surplus. Concentration
of the number of ranks above $t$ controls the cost of these two
mechanisms.

\begin{lemma}[Finite-degree surplus estimate]
\label{lem:large-degree-finite-surplus-estimate}
Let $G$ be a forward instance of maximum online degree at most $d$.
For $T,\eta\in(0,1)$ with $d(1-T)\geq e$, define
\[
\epsilon_{d,T,\eta}
:=\exp\!\left(-\frac{\eta^2(1-T)(d-1)}{2+\eta}\right).
\]
Then
\begin{equation}
\label{eq:large-degree-finite-surplus-estimate}
\delta_G
\geq
\frac{\displaystyle\int_0^T e^{t-1}
\left(-\frac{dq_G(t)}{dt}\right)\,dt}
{(1+\eta)d\bigl((1+\eta)d+(1-T)^{-1}\bigr)}
-\frac{\epsilon_{d,T,\eta}}{d}\ln\frac1{1-T}.
\end{equation}
\end{lemma}

\begin{proof}[\textbf{Lower-bound direction of Inequality~\eqref{eq:gamma-large-degree-bounds} in Theorem~\ref{thm:Gamma-large-degree}}]
By Lemma~\ref{lem:large-degree-forward-reduction}, it suffices to
bound $r(G)$ for forward instances. If $\delta_G>1/d$, the claimed
lower bound is immediate for sufficiently large $d$. Assume henceforth
that $0\leq\delta_G\leq1/d$, and choose
\[
T=1-d^{-1/4},\qquad \eta=d^{-1/4}.
\]
These parameters satisfy $d(1-T)\geq e$ for sufficiently large $d$.
By~\eqref{eq:large-degree-global-profile-control}, $A_G$ is uniformly
within $e/d$ of $1-e^{-t}$. Monotonicity of $q_G$ and difference
quotients over intervals of length $d^{-1/2}$ on each side of $T$
therefore give
\begin{equation}
\label{eq:large-degree-q-at-cutoff}
q_G(T)=e^{-T}+O(d^{-1/2}).
\end{equation}
Indeed, the two averages of $q_G$ bound $q_G(T)$, the corresponding
averages of $e^{-t}$ differ from $e^{-T}$ by $O(d^{-1/2})$, and the
uniform error in $A_G$ contributes another $O(d^{-1/2})$.

Integrating by parts and again using
\eqref{eq:large-degree-global-profile-control} yields
\begin{equation}
\label{eq:large-degree-integrated-q-decrease}
\begin{aligned}
\int_0^T e^{t-1}\left(-\frac{dq_G(t)}{dt}\right)\,dt
&=\frac1e-e^{T-1}q_G(T)
+\int_0^T e^{t-1}q_G(t)\,dt\\
&=\frac{T}{e}+O(d^{-1/2})
=\frac1e-O(d^{-1/4}).
\end{aligned}
\end{equation}
For the integral in the first line, a second integration by parts
bounds its difference from $T/e$ by $O(\delta_G)$.
The denominator in~\eqref{eq:large-degree-finite-surplus-estimate}
is $d^2(1+O(d^{-1/4}))$, and $\epsilon_{d,T,\eta}$ is exponentially
small in $d^{1/4}$. Lemma~\ref{lem:large-degree-finite-surplus-estimate}
therefore gives, uniformly over these instances,
\[
r(G)\geq1-\frac1e+\frac1{ed^2}-O(d^{-9/4}).
\]
Taking the infimum proves the lower bound on $\gamma(d)$.
\end{proof}

\subsubsection{A Layered Upper-Bound Construction}
\label{subsubsec:large-degree-competitiveness-upper}

The upper-bound family repeatedly joins $d$ roots with independent,
identically distributed histories and then retires one predetermined
root. The remaining roots are combined in later layers. This yields
a scalar recurrence for their safety probabilities without assuming
that surviving ranks are fresh uniforms.

\begin{lemma}[Layered upper bound]
\label{lem:large-degree-layered-upper}
For $d\geq2$, let
\begin{equation}
\label{eq:large-degree-layered-recurrence}
u_0:=1,
\qquad
u_{k+1}:=u_k-\frac{1-(1-u_k)^d}{d},
\qquad k\geq0.
\end{equation}
Then
\begin{equation}
\label{eq:large-degree-layered-upper}
\gamma(d)
\leq 1-\frac1d\sum_{k=0}^{\infty}
\left(1-\frac1d\right)^ku_{k+1}
=1-\frac1e+\frac1{2ed}+O(d^{-2}).
\end{equation}
More precisely, the middle expression lies strictly between
$1-(1-1/d)^d-2/d^2$ and $1-(1-1/d)^d$.
\end{lemma}

The finite construction and the recurrence estimate are proved in
Appendix~\ref{app:large-degree-competitiveness-proofs}. Consequently,
Lemma~\ref{lem:large-degree-layered-upper} establishes the upper bound
in~\eqref{eq:gamma-large-degree-bounds}. Together with the lower bound
proved above, this completes the proof of
Inequality~\eqref{eq:gamma-large-degree-bounds}.

\xhdr{The certifying gap.}
Combining soundness with the competitiveness upper bound gives
\[
0\leq
\Del(d)
=
\gamma(d)-\Gamma(d)
\leq
\gamma(d)-\left(1-\frac1e\right)
\leq
\frac1{2ed}+O(d^{-2}),
\]
which proves the remaining assertion of
Theorem~\ref{thm:Gamma-large-degree}.

\section{Proof of the Two-Regular Certifying Assertions in
Theorem~\ref{coro:Gamma-two-regular}}
\label{sec:proof-Gamma-two-regular}

In this section, we prove that $\tilde\Gamma(2)=3/4$ and that the
profile $\tilde f_2^\star$ in~\eqref{eq:optimal-profile-two-regular}
is its unique optimizer in $\mathcal F$. Subsection~\ref{subsec:regular-two-upper-bound}
establishes the upper bound using two regular-instance obstructions
and a two-constraint kernel mixture.
Subsection~\ref{subsec:regular-two-lower-bound} proves the matching
lower bound with the constant gain split, and
Subsection~\ref{subsec:regular-two-uniqueness} establishes uniqueness.
The auxiliary graph constructions and probability estimates are proved
in Appendix~\ref{app:regular-two-certifying-auxiliary}.

Throughout, $G\in\mathcal B_2$: every offline and online node has
degree exactly two, and the online arrival order is fixed in advance.
We evaluate each edge score after the complete execution. In
particular, we do not truncate after the distinguished arrival,
since truncation need not preserve 2-regularity.

\subsection{Two Regular-Instance Obstructions}
\label{subsec:regular-two-upper-bound}

The upper bound uses a sequence of cycles with a nearly always
unmatched final arrival and a six-cycle realizing the filtered-root
functional $\Psi_{1,0}$.

\begin{lemma}[Upper bound on 2-regular instances]
\label{lem:regular-two-upper-bound}
For every $f\in\mathcal F$,
\begin{equation}
\label{eq:regular-two-upper-obstructions}
\tilde\chi(f;2)\leq\min\{H_f(1),\Psi_{1,0}(f)\}.
\end{equation}
Consequently,
\begin{equation}
\label{eq:regular-two-Gamma-upper-bound}
\tilde\Gamma(2)\leq\frac34.
\end{equation}
\end{lemma}

\begin{proof}
The constructions in Appendix~\ref{app:regular-two-obstruction-proofs}
establish~\eqref{eq:regular-two-upper-obstructions}. We derive the
consequence for the outer optimization.

Let $\pi_f$ be the Stieltjes probability measure of $f$, and recall
$h_x(y):=\mathbf1\{y\geq x\}$. The two threshold kernels are
\[
H_{h_x}(1)=1-x,
\qquad
\Psi_{1,0}(h_x)=\frac23+\frac{x^2}{2}-\frac{x^3}{6}.
\]
Their convex combination satisfies
\begin{equation}
\label{eq:regular-two-kernel-identity}
\frac14H_{h_x}(1)+\frac34\Psi_{1,0}(h_x)
=\frac34-\frac{x(1-x)(2-x)}8.
\end{equation}
Both functionals are constants plus bounded integrals of $f$.
Substituting $f(y)=\int_{[0,1]}h_x(y)\,\pi_f(dx)$ and applying
Fubini's theorem therefore gives
\[
\begin{aligned}
\tilde\chi(f;2)
&\leq\frac14H_f(1)+\frac34\Psi_{1,0}(f)
=\frac34-\frac18\int_{[0,1]}x(1-x)(2-x)\,\pi_f(dx)
\leq\frac34.
\end{aligned}
\]
The last inequality uses nonnegativity of the integrand on $[0,1]$.
Taking the supremum over $f\in\mathcal F$ proves
\eqref{eq:regular-two-Gamma-upper-bound}.
\end{proof}

\subsection{The Constant Gain-Sharing Certificate}
\label{subsec:regular-two-lower-bound}

The mixture in~\eqref{eq:regular-two-kernel-identity} equals $3/4$
exactly at $x=0$ and $x=1$. This suggests a Stieltjes measure supported
on those endpoints, whose distribution function is constant on
$[0,1)$. If the constant is $\theta$, the obstruction values are
$\theta$ and $1-\theta/3$. Equalizing them gives $\theta=3/4$,
namely the profile $\tilde f_2^\star$ already defined in
\eqref{eq:optimal-profile-two-regular}.

\begin{lemma}[Constant gain-sharing certificate]
\label{lem:regular-two-constant-certificate}
For the admissible profile $\tilde f_2^\star\in\mathcal F$,
\begin{equation}
\label{eq:regular-two-step-certificate}
\tilde\chi(\tilde f_2^\star;2)=\frac34.
\end{equation}
\end{lemma}

\begin{proof}
Fix $G\in\mathcal B_2$ and $(i^\star,j^\star)\in E(G)$. Offline
ranks equal one with probability zero, so
\[
\Phi_G(i^\star,j^\star;\tilde f_2^\star)
=\frac34\Pr[X_{i^\star}=1]+\frac14\Pr[X_{j^\star}=1].
\]
Appendix~\ref{app:regular-two-probability-proof} proves the local
probability inequality
\begin{equation}
\label{eq:regular-two-probability-certificate}
\Pr[X_{j^\star}=1]\geq3\Pr[X_{i^\star}=0].
\end{equation}
Consequently,
\[
\begin{aligned}
\Phi_G(i^\star,j^\star;\tilde f_2^\star)
&=\frac34+\frac14\bigl(\Pr[X_{j^\star}=1]
-3\Pr[X_{i^\star}=0]\bigr)\geq\frac34.
\end{aligned}
\]
Taking the infimum over regular graph-edge pairs gives the lower
bound. The reverse inequality follows from
Lemma~\ref{lem:regular-two-upper-bound}. Equivalently, its six-cycle
obstruction has score $\Psi_{1,0}(\tilde f_2^\star)=3/4$.
\end{proof}

Since $\tilde f_2^\star\in\mathcal F$, the two lemmas establish
$\tilde\Gamma(2)=3/4$ and attainment of the supremum.

\subsection{Uniqueness of the Optimal Profile}
\label{subsec:regular-two-uniqueness}

Let $g\in\mathcal F$ attain $\tilde\Gamma(2)=3/4$, and let
$\pi_g$ be its Stieltjes probability measure. The two obstructions
and~\eqref{eq:regular-two-kernel-identity} give
\[
\begin{aligned}
\frac34=\tilde\chi(g;2)
&\leq\frac14H_g(1)+\frac34\Psi_{1,0}(g)
=\frac34-\frac18\int_{[0,1]}x(1-x)(2-x)\,\pi_g(dx).
\end{aligned}
\]
The integrand is strictly positive on $(0,1)$, so
$\operatorname{supp}(\pi_g)\subseteq\{0,1\}$. Write
\[
\pi_g=\theta\delta_0+(1-\theta)\delta_1,
\qquad\theta\in[0,1].
\]
Then $g(y)=\theta$ for $0\leq y<1$, and $g(1)=1$. Since both
obstruction values are at least $\tilde\chi(g;2)=3/4$,
\[
\theta=H_g(1)\geq\frac34,
\qquad
1-\frac\theta3=\Psi_{1,0}(g)\geq\frac34.
\]
Thus $\theta=3/4$. The bijection between $\mathcal F$ and
$\mathcal P([0,1])$ implies $g=\tilde f_2^\star$ pointwise.
This completes the certifying-value and optimizer assertions of
Theorem~\ref{coro:Gamma-two-regular}.


\section{Proof of the Two-Regular Competitiveness Assertion in
Theorem~\ref{coro:Gamma-two-regular}}
\label{sec:proof-regular-two-competitiveness}

In this section, we prove that $\tilde\gamma(2)=119/144$ in
Theorem~\ref{coro:Gamma-two-regular}, and then derive the asserted
certifying gap. Subsection~\ref{subsec:reg-two-cycle-analysis}
decomposes each cycle according to its arrival order and characterizes
the safety events of the resulting path endpoints.
Subsection~\ref{subsec:regular-two-global-lower} bounds the total
expected loss by charging it to the sizes of these path blocks.
Subsection~\ref{subsec:regular-two-tight-instance} gives a matching
upper-bound instance and completes the gap calculation. The
auxiliary proofs are collected in
Appendix~\ref{app:regular-two-competitiveness-auxiliary}.

\subsection{Cycle Decomposition and Endpoint Survival}
\label{subsec:reg-two-cycle-analysis}

Every connected component of a finite simple 2-regular bipartite graph
is an even cycle and has a perfect matching. Arrivals in distinct
components do not affect one another, so the expected matching size
and the optimum both add across components. It therefore suffices to
analyze one cycle $C$ with $n$ offline and $n$ online nodes. The case
$n=2$ is $K_{2,2}$, on which both online nodes are matched for every
rank realization. Assume $n\geq3$ below.

Label the offline nodes cyclically as $i_1,\ldots,i_n$, with
$N_C(j_r)=\{i_r,i_{r+1}\}$ and indices interpreted modulo $n$.
Here $r$ specifies position on the cycle, not arrival order. Let
$t_r$ be the position of $j_r$ in the prescribed arrival order.
We call $j_r$ a \emph{peak} if
$t_r>t_{r-1}$ and $t_r>t_{r+1}$.

\begin{lemma}[Cycle decomposition and endpoint survival]
\label{lem:reg-cycle-blocks}
Every nonpeak online node is matched. For any fixed distinct offline
ranks, postponing all peak arrivals to the end leaves the complete
matching unchanged. Removing the peaks partitions the cycle into
vertex-disjoint alternating path blocks. Each block has a unique
minimum-time online node, with arrival times increasing along both
arms away from that node.

For an endpoint reached after $r-2$ arm extensions from the minimum
arrival, call $r\geq2$ its \emph{chain length}. Immediately before
the postponed peaks, its safety event satisfies
\begin{equation}
\label{eq:reg-chain-conditional}
\Pr[\mathcal E_i\mid Y_i=y]=\frac{y^{r-1}}{(r-1)!},
\qquad
\Pr[\mathcal E_i]=\frac1{r!}.
\end{equation}
The conditional version fixes $Y_i=y$ and samples the other ranks
independently. The two endpoint events of the same block are mutually
exclusive; endpoint events from distinct blocks are independent. A
block whose endpoint chain lengths are $r,s$ has $r+s-2$ offline
nodes.
\end{lemma}

The proof is given in Appendix~\ref{app:regular-two-block-proof}.
The chain length records both an endpoint's survival probability and
the number of offline nodes needed to produce it. Keeping both
quantities is essential to the global lower bound.

\subsection{Global Loss Accounting}
\label{subsec:regular-two-global-lower}

For each peak $j$, let $r_j,s_j\geq2$ be the chain lengths of its
two offline neighbors. Distinct peaks have disjoint offline
neighborhoods, so earlier postponed peaks do not change these safety
events. If $C$ has at least two peaks, the neighbors of each peak
belong to different blocks, and
\begin{equation}
\label{eq:reg-peak-failure-exact}
\Pr[X_j=0]=\left(1-\frac1{r_j!}\right)
\left(1-\frac1{s_j!}\right).
\end{equation}
If $C$ has only one peak, its two neighbors are endpoints of the same
block, and the failure probability is
$1-1/r_j!-1/s_j!$, which is smaller than the right-hand side
of~\eqref{eq:reg-peak-failure-exact}.

Every block endpoint is incident to exactly one removed peak. Summing
its chain length minus one first over blocks and then over peaks gives
\begin{equation}
\label{eq:reg-chain-accounting}
\sum_{j\text{ peak}}(r_j+s_j-2)=n.
\end{equation}
Appendix~\ref{app:regular-two-peak-accounting} gives the details of
these failure probabilities and the accounting identity.

For every integer $r\geq2$,
\begin{equation}
\label{eq:regular-two-one-endpoint-charge}
\left(1-\frac1{r!}\right)^2\leq\frac{25}{72}(r-1),
\end{equation}
with equality exactly when $r=3$. This is immediate for $r=2,3$;
for $r\geq4$, the right-hand side is at least $25/24>1$.
Consequently, for $r,s\geq2$,
\[
\begin{aligned}
\left(1-\frac1{r!}\right)\left(1-\frac1{s!}\right)
&\leq\frac12\left[\left(1-\frac1{r!}\right)^2
+\left(1-\frac1{s!}\right)^2\right]
\leq\frac{25}{144}(r+s-2).
\end{aligned}
\]
Only peaks can be unmatched. Therefore,
\[
\begin{aligned}
n-\mathbb E[\rank(C)]
&=\sum_{j\text{ peak}}\Pr[X_j=0]
\leq\frac{25}{144}\sum_{j\text{ peak}}(r_j+s_j-2)
=\frac{25n}{144}.
\end{aligned}
\]
Since $\operatorname{OPT}(C)=n$, this proves
$\mathbb E[\rank(C)]/\operatorname{OPT}(C)\geq119/144$.
The same inequality holds for $K_{2,2}$ and extends to every
$G\in\mathcal B_2$ by summing over connected components. Thus
\begin{equation}
\label{eq:regular-two-competitiveness-lower}
\tilde\gamma(2)\geq\frac{119}{144}.
\end{equation}

\subsection{A Tight Cycle and the Certifying Gap}
\label{subsec:regular-two-tight-instance}

The endpoint charge is tight when both chain lengths equal three.
This suggests two path blocks, each with four offline nodes, joined
by two final peaks. The resulting instance is the degree-two
obstruction of~\cite[Lemma~3.7]{feng2025degree}; we give its
construction and evaluation explicitly.

Introduce offline nodes $i_1,\ldots,i_4$ and
$\tilde i_1,\ldots,\tilde i_4$, and online nodes arriving in the
order $j_1\prec\cdots\prec j_8$ with neighborhoods
\begin{equation}
\label{eq:reg-eight-cycle}
\begin{array}{c|c@{\qquad}c|c}
j_1&\{i_2,i_3\}&j_2&\{\tilde i_2,\tilde i_3\}\\
j_3&\{i_1,i_2\}&j_4&\{i_3,i_4\}\\
j_5&\{\tilde i_1,\tilde i_2\}&j_6&\{\tilde i_3,\tilde i_4\}\\
j_7&\{i_4,\tilde i_1\}&j_8&\{\tilde i_4,i_1\}.
\end{array}
\end{equation}
This is a cycle with eight offline and eight online nodes.
The first six arrivals are always matched. Each final peak joins
independent endpoints of chain length three and is matched with
probability $1-(5/6)^2=11/36$. The matching events of the two peaks
need not be independent; linearity of expectation suffices. Thus
\begin{equation}
\label{eq:regular-two-tight-cycle-value}
\frac{\mathbb E[\rank(C)]}{\operatorname{OPT}(C)}
=\frac{6+2(11/36)}8=\frac{119}{144}.
\end{equation}
Appendix~\ref{app:regular-two-tight-cycle-proof} verifies the
endpoint events and an explicit perfect matching. This gives the
reverse inequality to~\eqref{eq:regular-two-competitiveness-lower},
and hence $\tilde\gamma(2)=119/144$.

\xhdr{The certifying gap.} Combining this value with
$\tilde\Gamma(2)=3/4$, established in
Section~\ref{sec:proof-Gamma-two-regular}, gives
\[
\tilde\Del(2)=\tilde\gamma(2)-\tilde\Gamma(2)
=\frac{119}{144}-\frac34=\frac{11}{144}
=0.0763\ldots>0.
\]
This proves the remaining assertion of
Theorem~\ref{coro:Gamma-two-regular}.

\section{Proof of the Certifying Assertions for Large Regular Degree in
Theorem~\ref{thm:Gamma-regular-large-degree}}
\label{app:Gamma-regular-large-degree}

In this section, we prove the asymptotic expression
\eqref{eq:Gamma-regular-asymptotic-value} for $\tilde\Gamma(d)$ and
the optimizer-convergence assertion
\eqref{eq:regular-optimal-profiles-converge} in
Theorem~\ref{thm:Gamma-regular-large-degree}.

Subsection~\ref{subsec:regular-large-degree-lower-certificate} uses
all incident arrivals of an offline node to obtain a
regularity-sensitive lower certificate.
Subsection~\ref{subsec:regularizing-rooted-instance} then completes
bounded-degree graph-edge obstructions to regular instances, controlling
the increase in their scores. This regularization and $L^1$ compactness
yield the limiting value and exponential rigidity in
Subsection~\ref{subsec:regular-limiting-value}. Finally,
Subsection~\ref{subsec:regular-quantitative-core-mixture} records the
reduced-threshold information needed from the common-pool mixture, and
Subsection~\ref{subsec:regular-sharp-upper-constant} uses it to prove the
matching leading constant. Auxiliary proofs are collected in
Appendix~\ref{app:regular-large-degree-certifying-auxiliary}.

\subsection{A Regularity-Sensitive Lower Certificate}
\label{subsec:regular-large-degree-lower-certificate}

Regularity gives every offline node $d$ opportunities to be matched.
We use its matching threshold over the full execution: truncating after
a distinguished arrival would discard some of these opportunities and
need not preserve membership in $\mathcal B_d$. For $d\geq2$, define
\begin{equation}
\label{eq:regular-survival-envelope}
S_d(y):=\left(1-(1-y)^{d-1}\right)^d,
\qquad y\in[0,1].
\end{equation}

\begin{lemma}[Full-history threshold certificate]
\label{lem:regular-full-history-threshold}
Fix $G\in\mathcal B_d$ and an offline node $i$. Define
\begin{equation}
\label{eq:regular-full-history-threshold}
\beta_i:=\max_{j\in N_G(i)}R_{G-i}(j),
\end{equation}
with partner rank one for an unmatched arrival. Then
\begin{equation}
\label{eq:regular-threshold-cdf-bound}
\Pr[\beta_i\leq y]\leq S_d(y)
\qquad\text{for every }y\in[0,1).
\end{equation}
For every $(i,j)\in E(G)$ and $f\in\mathcal F$, moreover,
\begin{equation}
\label{eq:regular-full-history-score-bound}
\Phi_G(i,j;f)\geq\mathbb E[H_f(\beta_i)].
\end{equation}
\end{lemma}

The proof in Appendix~\ref{app:regular-full-history-proof} exposes
selected ranks sequentially. Conditional on that transcript, the ranks
of safe alternatives remain independent uniforms truncated from below.
Thus, while $i$ is safe with fixed rank $y$, each incident arrival selects
it with conditional probability at least $(1-y)^{d-1}$.

The envelope makes very small thresholds unlikely. A flat initial
segment followed by the exponential profile trades a loss at these
thresholds for an improvement everywhere else.

\begin{lemma}[An explicit regularity-sensitive lower bound]
\label{lem:regular-explicit-lower-certificate}
For every integer $d\geq2$ and every $a\in[0,1]$,
\begin{equation}
\label{eq:regular-explicit-lower-certificate}
\tilde\Gamma(d)\geq
1-e^{a-1}\left(1-a+\int_0^a S_d(t)\,dt\right).
\end{equation}
Consequently,
\begin{equation}
\label{eq:regular-asymptotic-liminf}
\liminf_{d\to\infty}\frac{d^2}{(\ln d)^2}
\left[\tilde\Gamma(d)-\left(1-\frac1e\right)\right]
\geq\frac1{2e}.
\end{equation}
\end{lemma}

\begin{proof}
For $a\in[0,1]$, let $f_a(y):=e^{\max\{a,y\}-1}$. This profile
belongs to $\mathcal F$, and direct integration gives
\begin{equation}
\label{eq:regular-flat-exponential-H}
H_{f_a}(t)=1-(1-a)e^{a-1}-e^{a-1}(a-t)_+.
\end{equation}
For every regular graph-edge pair,
Lemma~\ref{lem:regular-full-history-threshold} and
\[
\mathbb E[(a-\beta_i)_+]
=\int_0^a\Pr[\beta_i\leq t]\,dt
\leq\int_0^a S_d(t)\,dt
\]
therefore imply the lower bound in
\eqref{eq:regular-explicit-lower-certificate}.

For sufficiently large $d$, choose
\begin{equation}
\label{eq:regular-lower-threshold-choice}
a_d:=\frac{\ln d-2\ln\ln d}{d-1}.
\end{equation}
Then $a_d\in(0,1)$, $a_d\sim(\ln d)/d$, and
$d(1-a_d)^{d-1}=(1+o(1))(\ln d)^2$. Hence
\[
S_d(a_d)\leq\exp\bigl(-d(1-a_d)^{d-1}\bigr)
=\exp\bigl(-(1+o(1))(\ln d)^2\bigr),
\qquad
\int_0^{a_d}S_d(t)\,dt\leq a_dS_d(a_d)=o(a_d^2).
\]
Using
\[
1-(1-a)e^{a-1}=1-\frac1e+\frac{a^2}{2e}+O(a^3)
\qquad(a\downarrow0)
\]
in~\eqref{eq:regular-explicit-lower-certificate} proves
\eqref{eq:regular-asymptotic-liminf}.
\end{proof}

\subsection{Regularizing a Graph-Edge Instance}
\label{subsec:regularizing-rooted-instance}

For the upper bound, we complete bounded-degree obstructions to regular
instances. Both sides must have degree at most $d$: a vertex whose
original degree exceeds $d$ cannot be completed to degree $d$. Define
\begin{equation}
\label{eq:regularization-hit-bound}
b_d(y):=\min\{1,d(1-y)^{d-1}\},\qquad y\in[0,1].
\end{equation}

\begin{lemma}[Regularization of an edge-score obstruction]
\label{lem:regularization-edge-score}
Let $d\geq2$, and let $G$ be a finite bipartite instance whose offline
and online degrees are at most $d$. Fix $(i^\star,j^\star)\in E(G)$
and put $\tau_G:=R_{G-i^\star}(j^\star)$. For every $\varepsilon>0$,
there is a finite $H\in\mathcal B_d$ containing this distinguished
edge such that, simultaneously for every $f\in\mathcal F$,
\begin{equation}
\label{eq:regularization-edge-score}
\Phi_H(i^\star,j^\star;f)
\leq\Phi_G(i^\star,j^\star;f)
+\int_0^1 f(y)b_d(y)\Pr[\tau_G\leq y]\,dy+\varepsilon.
\end{equation}
\end{lemma}

The construction, proved in Appendix~\ref{app:regularization-proof},
first saturates the degree of $i^\star$, then fills deficient online
neighborhoods with nodes that are already matched with high probability,
and finally completes the remaining offline degrees after the original
execution. The last stage adds no edge incident to $i^\star$, so it
cannot change either contribution of the distinguished edge.

We will use the bounds
\begin{align}
\int_0^1 b_d(y)\,dy
&\leq\frac{\ln d+1}{d-1},
\label{eq:regularization-unweighted-integral}\\
\int_0^1 yb_d(y)\,dy
&\leq\frac{\frac12(\ln d)^2+\ln d+1}{(d-1)^2}.
\label{eq:regularization-weighted-integral}
\end{align}
Both follow from $b_d(y)\leq\min\{1,de^{-(d-1)y}\}$ by extending
the integrals to $[0,\infty)$ and splitting at $y=(\ln d)/(d-1)$.

\subsection{The Limiting Value and Stability of Optimizers}
\label{subsec:regular-limiting-value}

Regularization first gives the qualitative limit. Compactness then
forces near-optimal profiles to approach the unique unrestricted
optimizer; this rigidity will determine the sharp upper constant.

\begin{lemma}[Compactness and continuity]
\label{lem:regular-profile-compactness}
The space $\mathcal F$ is compact under the $L^1$ metric. For every
fixed finite graph-edge pair and all $f,g\in\mathcal F$,
\begin{equation}
\label{eq:regular-fixed-edge-L1}
|\Phi_G(i,j;f)-\Phi_G(i,j;g)|
\leq\bigl(1+|N_G(j)|\bigr)\|f-g\|_1.
\end{equation}
Consequently, for each fixed finite $d$,
\[
|\tilde\chi(f;d)-\tilde\chi(g;d)|\leq(d+1)\|f-g\|_1,
\]
and the supremum defining $\tilde\Gamma(d)$ is attained.
\end{lemma}

The proof appears in Appendix~\ref{app:regular-compactness-proof}.
Right-continuity and the common terminal normalization make $L^1$
equality pointwise equality within $\mathcal F$.

For every fixed finite graph-edge pair and every $d$ at least its
maximum offline and online degrees,
Lemma~\ref{lem:regularization-edge-score} gives, after letting the
padding error tend to zero,
\begin{equation}
\label{eq:regular-fixed-core-upper}
\tilde\chi(f;d)\leq\Phi_G(i,j;f)+\frac{\ln d+1}{d-1}.
\end{equation}
Since $\chi(f;\infty)\leq\tilde\chi(f;d)$, taking the upper limit
in~\eqref{eq:regular-fixed-core-upper} and then the infimum over fixed
finite graph-edge pairs proves
\begin{equation}
\label{eq:regular-pointwise-inner-limit}
\lim_{d\to\infty}\tilde\chi(f;d)=\chi(f;\infty)
\qquad\text{for every fixed }f\in\mathcal F.
\end{equation}
We do not interchange this limit with the outer supremum. Instead,
compactness and the fixed-edge estimate give the following result.

\begin{lemma}[Limiting value and exponential rigidity]
\label{lem:regular-exponential-rigidity}
One has
\begin{equation}
\label{eq:regular-limiting-value}
\lim_{d\to\infty}\tilde\Gamma(d)=1-\frac1e.
\end{equation}
More generally, every sequence $f_d\in\mathcal F$ satisfying
\begin{equation}
\label{eq:regular-near-benchmark-condition}
\liminf_{d\to\infty}\tilde\chi(f_d;d)\geq1-\frac1e
\end{equation}
also satisfies
\begin{equation}
\label{eq:regular-profile-rigidity}
\|f_d-f_\infty^\star\|_1\longrightarrow0.
\end{equation}
In particular, $f_d(t)\to e^{t-1}$ at every fixed $t\in(0,1)$.
\end{lemma}

\begin{proof}
Take any subsequence along which $f_d$ converges in $L^1$ to some
$f\in\mathcal F$. For any fixed finite graph-edge pair $(G,i,j)$,
\eqref{eq:regular-fixed-core-upper},
\eqref{eq:regular-fixed-edge-L1}, and
\eqref{eq:regular-near-benchmark-condition} imply
$\Phi_G(i,j;f)\geq1-1/e$. Taking the infimum over all such pairs
gives $\chi(f;\infty)\geq1-1/e$. Theorem~\ref{thm:Gamma-infinity}
therefore forces $f=f_\infty^\star$.

Every convergent subsequence has this limit, so compactness gives
\eqref{eq:regular-profile-rigidity} for the full sequence. Monotonicity
bounds $f_d(t)$ between its averages on short intervals to the left and
right of $t$. Their $L^1$ convergence, followed by continuity of
$f_\infty^\star$, gives pointwise convergence at every interior point.

Finally, choose an optimizer $f_d$ of $\tilde\Gamma(d)$ for each $d$.
The exponential certificate gives $\tilde\Gamma(d)\geq1-1/e$, so
these optimizers satisfy~\eqref{eq:regular-near-benchmark-condition}
and converge to $f_\infty^\star$ in $L^1$. Applying
\eqref{eq:regular-fixed-core-upper} to each fixed graph-edge pair yields
\[
\limsup_{d\to\infty}\tilde\Gamma(d)
\leq\inf_{\substack{G\in\mathcal B_\infty^{\mathrm{on}}\\(i,j)\in E(G)}}
\Phi_G(i,j;f_\infty^\star)=1-\frac1e.
\]
Together with the lower bound, this proves
\eqref{eq:regular-limiting-value}.
\end{proof}

\subsection{The Quantitative Mixture Used for Regularization}
\label{subsec:regular-quantitative-core-mixture}

To obtain the leading upper constant, we need more than the scalar
bound on $\Gamma(d)$. The common-pool mixture also controls the
reduced-threshold laws, and its component graphs have bounded degrees
on both sides.

\begin{lemma}[Properties of the common-pool mixture]
\label{lem:regular-core-mixture}
There is an absolute constant $C>0$ such that, for every sufficiently
large $d$, there are finitely many graph-edge pairs
$(G_r,i_r^\star,j_r^\star)$, $r\in[m_d]$, and nonnegative weights
$\lambda_r$ of total mass one with the following properties.
Every offline and online degree of every $G_r$ is at most $d$, and
\begin{equation}
\label{eq:regular-core-mixture-score}
\sum_{r=1}^{m_d}\lambda_r\Phi_{G_r}(i_r^\star,j_r^\star;f)
\leq1-\frac1e+C\frac{\ln d}{d^2}
\qquad\text{for every }f\in\mathcal F.
\end{equation}
For $\tau_r:=R_{G_r-i_r^\star}(j_r^\star)$, moreover,
\begin{equation}
\label{eq:regular-core-mixture-cdf}
\sum_{r=1}^{m_d}\lambda_r\Pr[\tau_r\leq y]
\leq\left(1+\frac{C}{d^2}\right)y,
\qquad y\in[0,1).
\end{equation}
\end{lemma}

Appendix~\ref{app:regular-core-mixture-proof} verifies both properties
from the positive mixture of
Lemma~\ref{lem:large-degree-positive-mixture}, with $D=d-1$ and its
integer parameter $N=\Theta(\ln D)$. The CDF estimate uses the density
of the target measure, not merely its threshold-response approximation.

\begin{lemma}[Regularized mixed-score bound]
\label{lem:regularized-mixed-score}
For sufficiently large $d$ and every $f\in\mathcal F$,
\begin{equation}
\label{eq:regularized-mixed-score}
\tilde\chi(f;d)
\leq1-\frac1e+C\frac{\ln d}{d^2}
+\left(1+\frac{C}{d^2}\right)\int_0^1 yf(y)b_d(y)\,dy.
\end{equation}
\end{lemma}

\begin{proof}
Regularize every component from Lemma~\ref{lem:regular-core-mixture}
using Lemma~\ref{lem:regularization-edge-score}, with padding error at
most $\varepsilon$. The resulting instances $H_r\in\mathcal B_d$
satisfy
\[
\begin{aligned}
\tilde\chi(f;d)
&\leq\sum_{r=1}^{m_d}\lambda_r\Phi_{H_r}(i_r^\star,j_r^\star;f)\\
&\leq\sum_{r=1}^{m_d}\lambda_r\Phi_{G_r}(i_r^\star,j_r^\star;f)
+\int_0^1 f(y)b_d(y)
\left(\sum_{r=1}^{m_d}\lambda_r\Pr[\tau_r\leq y]\right)dy
+\varepsilon.
\end{aligned}
\]
Apply~\eqref{eq:regular-core-mixture-score} and
\eqref{eq:regular-core-mixture-cdf}, then let $\varepsilon\downarrow0$.
\end{proof}

The reduced-threshold CDF supplies the extra factor $y$ in
\eqref{eq:regularized-mixed-score}. It lowers the regularization cost
from the scale $\ln d/d$ in
\eqref{eq:regularization-unweighted-integral} to the scale
$(\ln d)^2/d^2$ in~\eqref{eq:regularization-weighted-integral}.

\subsection{The Sharp Upper Constant}
\label{subsec:regular-sharp-upper-constant}

\begin{lemma}[Asymptotic upper bound]
\label{lem:regular-asymptotic-upper-bound}
One has
\begin{equation}
\label{eq:regular-asymptotic-limsup}
\limsup_{d\to\infty}\frac{d^2}{(\ln d)^2}
\left[\tilde\Gamma(d)-\left(1-\frac1e\right)\right]
\leq\frac1{2e}.
\end{equation}
\end{lemma}

\begin{proof}
Choose optimizers $f_d\in\mathcal F$. By
Lemma~\ref{lem:regular-exponential-rigidity}, $f_d(t)\to e^{t-1}$
for every fixed $t\in(0,1)$. For such a $t$, monotonicity gives
\[
\begin{aligned}
\int_0^1 yf_d(y)b_d(y)\,dy
&\leq f_d(t)\int_0^1 yb_d(y)\,dy
+d\int_t^1(1-y)^{d-1}\,dy\\
&=f_d(t)\int_0^1 yb_d(y)\,dy+(1-t)^d.
\end{aligned}
\]
Using~\eqref{eq:regularization-weighted-integral} and then taking
$d\to\infty$ yields
\[
\limsup_{d\to\infty}\frac{d^2}{(\ln d)^2}
\int_0^1 yf_d(y)b_d(y)\,dy\leq\frac12e^{t-1}.
\]
Letting $t\downarrow0$ gives
\begin{equation}
\label{eq:regular-optimal-regularization-cost}
\limsup_{d\to\infty}\frac{d^2}{(\ln d)^2}
\int_0^1 yf_d(y)b_d(y)\,dy\leq\frac1{2e}.
\end{equation}
Now apply Lemma~\ref{lem:regularized-mixed-score} to $f_d$.
Its $O(\ln d/d^2)$ term is negligible at the scale $(\ln d)^2/d^2$,
and its multiplicative factor $1+C/d^2$ tends to one. Since
$\tilde\chi(f_d;d)=\tilde\Gamma(d)$, the desired bound follows.
\end{proof}

\subsection{Completion of the Certifying Proof}
\label{subsec:regular-large-degree-proof-completion}

Combining~\eqref{eq:regular-asymptotic-liminf} and
\eqref{eq:regular-asymptotic-limsup} proves
\[
\tilde\Gamma(d)=1-\frac1e+
\left(\frac1{2e}+o(1)\right)\frac{(\ln d)^2}{d^2},
\]
which is~\eqref{eq:Gamma-regular-asymptotic-value}.
If $f_d\in\mathcal F$ satisfies
$\tilde\Gamma(d)-\tilde\chi(f_d;d)\to0$, then
$\tilde\Gamma(d)\geq1-1/e$ implies
\eqref{eq:regular-near-benchmark-condition}.
Lemma~\ref{lem:regular-exponential-rigidity} therefore gives
$\|f_d-f_\infty^\star\|_1\to0$, proving
\eqref{eq:regular-optimal-profiles-converge} and completing the
certifying assertions of Theorem~\ref{thm:Gamma-regular-large-degree}.


\section{Proof of the Competitiveness Assertions for Large Regular Degree in
Theorem~\ref{thm:Gamma-regular-large-degree}}
\label{sec:proof-regular-large-degree-competitiveness}

In this section, we prove the finite-degree lower bound
\eqref{eq:gamma-regular-finite-degree-lower}, the limiting value
\eqref{eq:gamma-regular-limit}, and the gap assertions
\eqref{eq:Delta-regular-asymptotic-bounds} in
Theorem~\ref{thm:Gamma-regular-large-degree}.

Subsection~\ref{subsec:regular-global-threshold-cut} uses exact degree
balance across a rank-threshold cut to prove the lower bound.
Subsection~\ref{subsec:regular-global-limiting-upper} constructs exactly
regular graphs that approximate a complete triangular batch process,
and establishes the upper limit for every sufficiently large degree.
Subsection~\ref{subsec:regular-large-degree-gap} then combines these
competitiveness bounds with the certifying asymptotics. The global
upper construction is proved in
Appendix~\ref{app:regular-large-degree-competitiveness-auxiliary};
it is separate from the edge-score regularization of
Section~\ref{app:Gamma-regular-large-degree}.

\subsection{A Global Threshold-Cut Lower Bound}
\label{subsec:regular-global-threshold-cut}

Every $G\in\mathcal B_d$ has equally sized sides and a perfect
matching. Indeed, degree counting gives equality of the side sizes,
and $d|S|\leq d|N_G(S)|$ verifies Hall's condition for every
$S\subseteq I$. Write $n:=|I|=|J|=\operatorname{OPT}(G)$.
Using the full-history thresholds from
\eqref{eq:regular-full-history-threshold}, set
\[
q_G(t):=\frac1n\sum_{i\in I}\Pr[\beta_i>t],
\qquad
A_G(t):=\frac1n\mathbb E\!\left[
\sum_{i\in I}X_i\mathbf1\{Y_i\leq t\}\right].
\]
The matching-threshold characterization and independence of $Y_i$
and $\beta_i$ give
\begin{equation}
\label{eq:regular-global-A-integral}
A_G(t)=\int_0^t q_G(y)\,dy,
\qquad
A_G(1)=\frac{\mathbb E[\rank(G)]}{n}.
\end{equation}

\begin{lemma}[Threshold-cut lower bound]
\label{lem:regular-global-cut-lower}
For every integer $d\geq2$ and every $G\in\mathcal B_d$,
\begin{equation}
\label{eq:regular-global-cut-lower}
\frac{\mathbb E[\rank(G)]}{\operatorname{OPT}(G)}
\geq1-\left(1-\frac1d\right)^d.
\end{equation}
\end{lemma}

\begin{proof}
For fixed $t<1$ and fixed distinct ranks, define
\[
I_t:=\{i\in I:\beta_i>t\},
\qquad
J_t^{\mathrm{high}}:=\{j\in J:R_G(j)>t\},
\qquad
J_t^{\mathrm{low}}:=J\setminus J_t^{\mathrm{high}}.
\]
Insertion monotonicity gives $R_G(j)\leq R_{G-i}(j)\leq\beta_i$
for every edge $(i,j)$. Thus every neighbor of
$J_t^{\mathrm{high}}$ belongs to $I_t$, and regularity implies
\begin{equation}
\label{eq:regular-global-cut-count}
|E(I_t,J_t^{\mathrm{low}})|=d|I_t|-d|J_t^{\mathrm{high}}|.
\end{equation}
Every offline node with $\beta_i>t$ and $Y_i\leq t$ is matched,
and its matching edge belongs to this cut. Taking expectations yields
\[
dnq_G(t)-dn(1-A_G(t))
\geq\sum_{i\in I}\Pr[\beta_i>t,\ Y_i\leq t]
=tnq_G(t).
\]
Here the low online nodes correspond exactly to matched offline nodes
with ranks at most $t$. By~\eqref{eq:regular-global-A-integral},
\begin{equation}
\label{eq:regular-global-cut-differential}
\left(1-\frac td\right)\frac{dA_G(t)}{dt}+A_G(t)\geq1
\quad\text{for Lebesgue-almost every }t\in(0,1).
\end{equation}
The function $A_G$ is absolutely continuous and $A_G(0)=0$.
The last inequality makes
$(1-A_G(t))(1-t/d)^{-d}$ nonincreasing. Hence
$1-A_G(t)\leq(1-t/d)^d$. Continuity at $t=1$ proves the lemma.
\end{proof}

Taking the infimum gives~\eqref{eq:gamma-regular-finite-degree-lower}.
In particular, the exact lower bound has expansion
\begin{equation}
\label{eq:regular-global-lower-asymptotic}
1-\left(1-\frac1d\right)^d
=1-\frac1e+\frac1{2ed}+\frac5{24ed^2}+O(d^{-3}).
\end{equation}
The threshold-cut argument uses degree balance across the entire
instance, rather than a uniform lower bound on separate edge scores.

\subsection{An Exactly Regular Limiting Upper Construction}
\label{subsec:regular-global-limiting-upper}

Regularizing a distinguished edge does not control an instance's total
matching size. We instead approximate a complete triangular batch
process by simple regular graphs. The comparison is deterministic for
fixed ranks and uses block pairs with no large empty bipartite
rectangle.

\begin{lemma}[Algorithmic limiting upper bound]
\label{lem:regular-algorithmic-limiting-upper}
For every $\zeta>0$ and every sufficiently large integer $d$, there
is a finite $G\in\mathcal B_d$ such that
\begin{equation}
\label{eq:regular-algorithmic-limiting-upper}
\frac{\mathbb E[\rank(G)]}{\operatorname{OPT}(G)}
\leq1-\frac1e+\zeta.
\end{equation}
\end{lemma}

Here is the structure of the construction. Fix a number of batches
$m\geq2$ and a prime $p$. Use offline blocks $I_1,\ldots,I_m$ and
online blocks $J_1,\ldots,J_m$ with sizes
\begin{equation}
\label{eq:regular-triangular-block-sizes}
|I_i|=Lp^{i-1}(p-1)^{m-i},
\qquad
|J_j|=Lp^{m-j}(p-1)^{j-1},
\end{equation}
where $L$ is a positive integer scale. Process the online blocks in
order. Connect $J_j$ completely to $I_j$, to no $I_i$ with $i<j$,
and by a biregular graph of density $1/p$ to each $I_i$ with $i>j$.
Geometric sums give the same degree at every vertex:
\begin{equation}
\label{eq:regular-triangular-degree}
|I_j|+\frac1p\sum_{i>j}|I_i|
=|J_i|+\frac1p\sum_{j<i}|J_j|
=Lp^{m-1}=:d_0.
\end{equation}

The appendix constructs the biregular blocks by finite-field incidence
graphs and uniform vertex blow-ups. Their absence of large empty
rectangles controls the matching-size error relative to completing all
allowed block pairs. With $m$ fixed, taking $p$ large makes the
normalized block sizes nearly equal; then taking $m$ large makes the
complete-batch performance approach $1-1/e$.

The final degree-completion step is essential. The initial construction
produces selected multiples of a fixed degree, whereas the classes
$\mathcal B_d$ are not nested. Appendix~\ref{app:regular-degree-rounding}
completes these instances to every sufficiently large target degree
with a controlled increase in total matching size. The parameters are
chosen in the order: target accuracy, number of batches, prime, incidence
dimension, and finally the target degree. This proves convergence, but
asserts no quantitative upper rate in $d$.

Lemma~\ref{lem:regular-algorithmic-limiting-upper} and the lower bound
above imply $\tilde\gamma(d)\to1-1/e$, proving
\eqref{eq:gamma-regular-limit}.

\subsection{The Certifying Gap}
\label{subsec:regular-large-degree-gap}

Both $\tilde\gamma(d)$ and $\tilde\Gamma(d)$ converge to $1-1/e$,
so $\tilde\Del(d)\to0$. On the other hand,
\eqref{eq:regular-global-lower-asymptotic} and
\eqref{eq:Gamma-regular-asymptotic-value} give
\[
d\,\tilde\Del(d)
\geq\frac1{2e}+O(d^{-1})
-\left(\frac1{2e}+o(1)\right)\frac{(\ln d)^2}{d}.
\]
Consequently,
\[
\tilde\Del(d)=o(1),
\qquad
\liminf_{d\to\infty}d\,\tilde\Del(d)\geq\frac1{2e},
\]
which proves~\eqref{eq:Delta-regular-asymptotic-bounds} and completes
Theorem~\ref{thm:Gamma-regular-large-degree}. These bounds do not give
a matching upper convergence rate for $\tilde\Del(d)$.

\section{Conclusion and Future Directions}
\label{sec:conclusion}

This paper studies the certifying capacity of the rank-dependent
primal--dual protocol of~\cite{DJK13} by optimizing its gain-sharing
function while keeping \rank{} and the protocol fixed. The resulting
perspective distinguishes three choices: the instance class, the
admissible function class, and the underlying certification protocol.
Optimization removes losses caused by a suboptimal function choice
within the prescribed class, but need not eliminate the certifying
gap. Our degree-two results demonstrate that a strict gap can persist
even over all measurable nonnegative rank-only splits. At large
degree, the gaps over $\cF$ vanish, although the regular-graph results
show that agreement of limiting values can conceal a substantial
separation at the scale of the finite-degree improvement. These
observations suggest several directions.

\xhdr{Full-class gaps at every finite degree.}
Our degree-two results motivate the conjecture that the rank-dependent
protocol is not tight at any finite degree $d\geq2$, even over the
full admissible class:
\[
\Gamma_{\cA}(d)<\gamma(d),\qquad \tilde\Gamma_{\cA}(d)<\tilde\gamma(d)
\qquad\text{for every integer }d\text{ with }2\leq d<\infty.
\]
The corresponding gaps already vanish as $d\to\infty$. Indeed,
class inclusion and soundness give
\[
0\leq\gamma(d)-\Gamma_{\cA}(d)\leq\Del(d),\qquad
0\leq\tilde\gamma(d)-\tilde\Gamma_{\cA}(d)\leq\tilde\Del(d),
\]
and both right-hand sides vanish by
Theorems~\ref{thm:Gamma-large-degree}
and~\ref{thm:Gamma-regular-large-degree}.
The remaining questions concern strict positivity at each finite
degree and the sharp convergence rates. Does removing monotonicity
change the order of either gap, or only its leading constant?
In particular, does the $\Omega(1/d)$ lower bound for the regular
gap over $\cF$ persist over $\cA$? For online-degree-bounded
instances, obtaining a positive asymptotic lower bound remains a
challenge even over $\cF$.

\xhdr{Bounding certifying gaps directly.}
A more systematic approach would compare certifying capacity and
algorithmic performance without first determining both exactly.
For lower bounds on a gap, one can already combine a lower certificate
for the algorithm with an upper certificate for the protocol; exact
values are unnecessary. Developing structural comparisons between
rank-only, state-dependent, and global certificates may make such
separations easier to establish.

Direct upper bounds on the gap would be particularly useful when
combined with certified functional optimization. If
$\Del_{\mathfrak P(\cS)}(\cG)\leq\varepsilon$ and a function
$f\in\cS$ satisfies
$\Gamma_{\mathfrak P(\cS)}(\cG)-\chi(f,\cG)\leq\eta$, then
\[
\chi(f,\cG)\leq\gamma(\cG)\leq\chi(f,\cG)+\eta+\varepsilon.
\]
Thus a near-optimal function and a gap bound give a two-sided estimate
of the actual competitiveness. This conclusion requires a certified
optimality error, or an upper bound on the certifying capacity, in
addition to a lower certificate.

\xhdr{Structural conditions for tightness.}
For fixed $\mathfrak P$, $\cS$, and \rank{}, can tightness be
characterized through structural properties of the instance class
$\cG$? A useful starting point is the distinction between a worst
edge score and the aggregate score over a maximum matching.
For any $f\in\cS$, let $Z_v$ denote the resulting endpoint charges.
If $M^\star$ is a maximum matching and $V(M^\star)$ is its set of
covered vertices, gain conservation gives
\[
\mathbb E[\rank(G)]=\sum_{(i,j)\in M^\star}\Phi_G(i,j;f)
+\mathbb E\!\left[\sum_{v\notin V(M^\star)}Z_v\right].
\]
The edgewise certificate replaces the matching-edge scores by a
uniform lower bound and discards the remaining nonnegative charges.
This suggests seeking structural conditions under which, for
gain-sharing functions approaching certifying capacity, nearly
extremal edge-score obstructions can be realized across a maximum
matching in instances belonging to $\cG$, with vanishing average
surplus above $\chi(f,\cG)$ and vanishing normalized expected charge
on uncovered vertices. Such realizations would imply tightness.
The substantive question is which graph structures guarantee them.
Closure under disjoint unions alone cannot suffice: the two-regular
class has this property and still exhibits a strict certifying gap.

A structural tightness theorem would reduce the remaining performance
analysis to functional optimization over $\cS$. Conversely,
identifying the structural obstruction to tightness could indicate
which additional information or accounting mechanism a stronger
protocol must retain.

\xhdr{Beyond rank-only gain sharing.}
The state-dependent certificate for online degree two demonstrates
that more informative dual updates can yield a tight analysis without
changing \rank{}. A natural direction is to develop protocols whose
updates depend jointly on rank and additional information, such as
the number of earlier incident arrivals, currently available
neighbors, or prefix-dependent safety distributions. An analytical
certificate may use the arrival index or revealed history under
either random-order or adversarial arrivals. The adversarial model,
however, does not provide the distributional symmetry available under
random arrivals. The challenge is to identify summaries of this
information that support stronger dual updates while remaining
amenable to uniform analysis.

\xhdr{Algorithm-defining frameworks.}
The extension in Appendix~\ref{sec:other-related-work} raises analogous
questions when gain-sharing parameters also determine the allocation
algorithm. In that setting, the best certified guarantee must be
distinguished from both the best actual competitiveness within the
induced algorithm family and the optimum over all online algorithms
allowed by the model. A problem-relative gap can therefore reflect
parameter restrictions, limitations of the certificate, or limitations
of the induced family.

Developing bounds that separate these effects would help determine
whether a desired guarantee can be reached by further functional
optimization or instead requires a different protocol. Unlike in the
fixed-algorithm setting, varying the function can also change the
allocation distribution, so the affine structure used here need not
survive. Extending certifying-capacity methods to this setting is a
substantive direction. The broader goal is to use an understanding
of certifying barriers to guide the design of more effective
randomized primal--dual protocols, ideally tight for their target
algorithm or problem.



\newpage
\bibliographystyle{alpha} 
\bibliography{EC_21}
 
\clearpage
\appendix

\section{Related Work and Extensions to Algorithm-Defining Frameworks}
\label{sec:other-related-work}

This appendix reviews related work on degree-restricted online
matching and places our certifying-capacity perspective in a broader
primal--dual context. We distinguish frameworks that analyze a fixed
algorithm from those in which gain-sharing parameters also determine
the allocation rule. For the latter, we extend the notions of
certifying capacity and gap, separating the effects of parameter
restrictions, limitations of the induced algorithm family, and the
gap between its best competitiveness and its best certified guarantee. For broader background,
see the surveys on online matching and allocation~\cite{huang2024online,Mehta2013} and on online primal--dual methods by
Buchbinder and Naor~\cite{buchbinder2009design}.

\subsection{Degree-Restricted Online Matching}
\label{subsec:related-degree-restrictions}

Degree restrictions take several distinct forms: an upper bound on
online degree, an upper or lower bound on offline degree, or exact
regularity. Their algorithmic consequences also depend on the arrival
model and on whether allocations are integral or fractional. These
assumptions must therefore be distinguished when comparing guarantees.

\xhdr{Bounded-degree allocation and matching.} Buchbinder et
al.~\cite{buchbinder2007online} studied a bounded-degree ad-allocation
setting in which each arriving product interests at most $d$ buyers.
This is an upper bound on online degree, not offline degree.
Azar, Cohen, and Roytman~\cite{azar2017online} developed a duality-based
method for proving online impossibility results, including for
bounded-degree ad allocation and matching. Their construction implies
that no randomized online matching algorithm on
$\mathcal B_{\leq d}^{\mathrm{on}}$ has competitiveness exceeding
$1-(1-1/d)^d$. In particular, it gives the algorithm-independent bound
\[
\gamma(d)\leq1-\left(1-\frac1d\right)^d=1-\frac1e+\frac1{2ed}+O(d^{-2}).
\]
Thus the first-order algorithmic upper bound in
Theorem~\ref{thm:Gamma-large-degree} is also implied by this earlier
construction; it is distinct from our upper bound on the certifying
capacity of the fixed rank-dependent protocol.

Naor and Wajc~\cite{naor2018near} introduced the $(k,d)$-bounded
model, in which every offline vertex has degree at least $k$ and
every online vertex has degree at most $d$. For $k\geq d\geq2$,
they obtained the deterministic guarantee $1-(1-1/d)^k$, with
matching bounds for the vertex-weighted problem. Unlike an upper
bound on online degree alone, this model also guarantees sufficiently
many opportunities for each offline resource. Albers and
Schubert~\cite{albers2022tight} extended this line to unweighted and
vertex-weighted matching with offline capacities, obtaining tight
deterministic bounds whose competitiveness approaches one as
capacities grow when $k\geq d$. Their related work on
AdWords~\cite{albers2022online} gives optimal deterministic guarantees
on $(k,d)$-graphs, approaching one in the small-bid regime when
$k\geq d$.

\xhdr{Regular graphs and the degree-two regime.} Cohen and
Wajc~\cite{cohen2018randomized} obtained competitiveness
$1-O(\sqrt{\ln d/d})$ on $d$-regular graphs under adversarial
arrivals, while showing that \rank{} remains bounded away from one.
Feng et al.~\cite{feng2025degree} subsequently compared \rank{}
with online correlated selection (OCS) on $(d,d)$-bounded graphs,
which require offline degree at least $d$ and online degree at most
$d$ and include all $d$-regular graphs. They gave stronger upper
bounds for \rank{} and OCS-based guarantees exceeding its
competitiveness for every fixed $d\geq2$.

For online degree at most two, Bhangale, Chakraborty, and
Harsha~\cite{BCH25} proved the optimality of \textsc{Half-Half} for
randomized integral matching and separated its competitiveness from
the optimal fractional value $3/4$. Our degree-two results address a
different comparison: the actual competitiveness of \rank{} versus
the strongest guarantee obtained by its rank-dependent gain-sharing
certificate. Identifying an optimal online algorithm does not by
itself determine whether a particular primal--dual protocol analyzes
that algorithm tightly.

\xhdr{Offline-degree bounds and degree information.} Under random
arrival order, Xu~\cite{xu2026degree} studies edge-weighted online
matching parameterized by maximum offline degree and a sampling
fraction. That work gives a tight analysis of a greedy sampling
algorithm and a transfer theorem that converts offline approximation
guarantees into online guarantees for a sampling--matching framework.
Its degree restriction, arrival model, and algorithmic family differ
from the adversarial-arrival \rank{} setting considered here.
Aamand et al.~\cite{aamand2022optimal} instead study access to
predicted offline degrees and prove optimality of
\textsc{MinPredictedDegree} in a specified stochastic graph model.
Degree predictions provide additional information to the algorithm;
they should not be conflated with a worst-case degree restriction on
the instance class.

\subsection{Algorithm-Analysis and Algorithm-Defining Frameworks}
\label{subsec:related-algorithm-defining-frameworks}

The distinction between a gain-sharing protocol and its functional
parameters applies beyond the fixed-algorithm setting of this paper.
We call a protocol an \emph{algorithm-analysis framework} (AAF) when
the gain-sharing function affects only the certificate, as in the
analysis of \rank{} by~\cite{DJK13}. In an
\emph{algorithm-defining framework} (ADF), the function also enters
the allocation rule. An admissible choice then specifies both an
online algorithm and its accompanying certificate. This distinction
concerns dependence on the function, not whether the algorithm was
historically introduced before its analysis. Different functions may
also induce the same allocation algorithm.

\xhdr{Load-dependent and time-dependent allocation rules.} Huang
and Zhang~\cite{huang2020online} use randomized primal--dual analysis
with a configuration LP for online matching with stochastic rewards.
In their vertex-weighted extension of Stochastic Balance, the
function determines both dual gains and the offers used for
allocation. For example, in the common-success-probability setting,
an unsuccessful offline vertex $i$ of weight $w_i$ and accumulated
load $\ell_i$ offers a quantity proportional to
$w_i(1-f(\ell_i))$. Changing $f$ can therefore change the selected
vertex. Their use of a configuration LP rather than the standard
matching LP also illustrates a change to the underlying protocol,
not merely to its gain-sharing function.

Huang et al.~\cite{huang2019online} use a two-dimensional
gain-sharing function for vertex-weighted matching under random
arrivals. The function depends on the offline rank and online arrival
time, and determines the weighted offer compared by the algorithm as
well as the endpoint dual charges. Optimizing this function changes
the algorithm's allocation priorities, unlike optimizing the
analytical rank-dependent split for unweighted \rank{}.

\xhdr{Gain sharing and online rounding.} Fahrbach et
al.~\cite{fahrbach2022edge} combine OCS with a primal--dual algorithm
for edge-weighted matching under free disposal. Their gain-sharing
parameters determine both the dual updates and the offers used to
choose offline candidates. These parameters are sequences indexed by
algorithmic state, rather than a single scalar-rank function, and are
optimized through a factor-revealing LP. The OCS subroutine is another
part of the construction: optimizing the gain-sharing parameters
while fixing this subroutine is different from optimizing the entire
algorithmic framework.

Chen et al.~\cite{chen2024stochastic} develop stochastic online
correlated selection as a family of online rounding procedures for
stochastic allocation with independent, possibly heterogeneous arrivals.
Their framework applies to stochastic matching, AdWords, and display
ads. This is related algorithmic machinery, rather than an instance
of the single rank-only splitting rule studied here. It emphasizes
that an algorithm-defining protocol may include a fractional
allocation procedure and a rounding subroutine in addition to gain
sharing.

The distinction also applies to deterministic primal--dual design.
For small-bid AdWords, Buchbinder et al.~\cite{buchbinder2007online}
use primal--dual prices to obtain the optimal $1-1/e$ guarantee,
matching that of Mehta et al.~\cite{mehta2007adwords}. Thus the
general issue is not randomization alone, but which parts of an
algorithm--analysis construction are fixed and which are optimized.

\subsection{Certifying Gaps for Algorithm-Defining Frameworks}
\label{subsec:related-design-certifying-gaps}

The notions of capacity and gap extend to ADFs, but the comparison
target must be specified. There are two different targets: the best
algorithm obtainable within the framework, and the best admissible
online algorithm for the problem. A gap to the latter can reflect
limitations of both the induced algorithms and their certificates.

\xhdr{The induced algorithm family.} Fix an online maximization
problem, an instance class $\cG$, its benchmark
$\operatorname{OPT}(G)$, and the allowed information and allocation
model. Let $\mathfrak Q$ be an algorithm-defining protocol; we
continue to reserve $\mathfrak P$ for the DJK protocol studied in
this paper. Write $\cA_{\mathfrak Q}$ for its full admissible
parameter class and $\cS\subseteq\cA_{\mathfrak Q}$ for a
nonempty subclass. Here a parameter $f$ may be a function, a
multivariate function, or a fixed collection of gain-sharing
sequences; its domain and admissibility conditions belong to the
specification of $\mathfrak Q$ and need not coincide with those
of $\cA$ in this paper.

For each $f$, let $\mathsf{ALG}_f$ be the induced online algorithm,
and let $\chi_{\mathfrak Q}(f,\cG)$ be the strongest uniform
guarantee supplied by the protocol's prescribed expected dual
criterion. Generalizing the competitiveness notation to an arbitrary
online algorithm, write
\[
\gamma(\mathsf{ALG},\cG):=\inf_{\substack{G\in\cG\\\operatorname{OPT}(G)>0}}\frac{\mathbb E[\mathsf{ALG}(G)]}{\operatorname{OPT}(G)}.
\]
The benchmark and the algorithmic model are held fixed throughout;
for example, integral and fractional algorithms, or stochastic
benchmarks with different information, cannot be interchanged in
these comparisons. Soundness gives
$\chi_{\mathfrak Q}(f,\cG)\leq\gamma(\mathsf{ALG}_f,\cG)$.
Define
\begin{equation}
\label{eq:related-adf-capacity-and-performance}
\Gamma_{\mathfrak Q(\cS)}(\cG):=\sup_{f\in\cS}\chi_{\mathfrak Q}(f,\cG),\qquad
\gamma_{\mathfrak Q(\cS)}(\cG):=\sup_{f\in\cS}\gamma(\mathsf{ALG}_f,\cG).
\end{equation}
The first is the best certified guarantee; the second is the best
actual competitiveness within the induced algorithm family. They
need not coincide, and the functions optimizing them need not be
the same. In particular, optimizing the certificate does not by
itself identify the best-performing algorithm in the family.

\xhdr{A problem-relative certifying gap.} Let
\[
\gamma_{\mathrm{opt}}(\cG):=\sup_{\mathsf{ALG}}\gamma(\mathsf{ALG},\cG),
\]
where the supremum ranges over all online algorithms permitted by the
fixed model, not only those induced by $\mathfrak Q$. Then
\begin{equation}
\label{eq:related-adf-soundness-chain}
\Gamma_{\mathfrak Q(\cS)}(\cG)\leq\gamma_{\mathfrak Q(\cS)}(\cG)\leq\gamma_{\mathrm{opt}}(\cG).
\end{equation}
The \emph{problem-relative certifying gap} is
\begin{equation}
\label{eq:related-adf-problem-gap}
\Del^{\mathrm{opt}}_{\mathfrak Q(\cS)}(\cG):=\gamma_{\mathrm{opt}}(\cG)-\Gamma_{\mathfrak Q(\cS)}(\cG)\geq0.
\end{equation}
This is the natural extension when the goal is to design an optimal
algorithm for the problem, rather than tightly analyze a specified
algorithm. It includes more than a possible weakness in the analysis:
the framework may fail to induce algorithms approaching the problem
optimum in the first place.

The gap separates into three nonnegative components:
\begin{equation}
\label{eq:related-adf-gap-decomposition}
\begin{aligned}
\Del^{\mathrm{opt}}_{\mathfrak Q(\cS)}(\cG)
={}&\underbrace{\Gamma_{\mathfrak Q(\cA_{\mathfrak Q})}(\cG)-\Gamma_{\mathfrak Q(\cS)}(\cG)}_{\text{loss from restricting the parameter class}}\\
&+\underbrace{\gamma_{\mathfrak Q(\cA_{\mathfrak Q})}(\cG)-\Gamma_{\mathfrak Q(\cA_{\mathfrak Q})}(\cG)}_{\text{gap between optimized performance and certification}}+\underbrace{\gamma_{\mathrm{opt}}(\cG)-\gamma_{\mathfrak Q(\cA_{\mathfrak Q})}(\cG)}_{\text{limitation of the induced algorithm family}}.
\end{aligned}
\end{equation}
The identity follows by adding and subtracting the two full-class
optima; nonnegativity follows from class inclusion and soundness.
The first term is the loss in certified performance due to restricting
$f$. The remaining terms persist after allowing every admissible
parameter: one concerns certification within the induced family,
and the other concerns the expressive power of that family. For an
AAF, all induced algorithms coincide with the fixed target, so the
middle term becomes the full-class certifying gap studied here.
The last term is then the target algorithm's own suboptimality, which
our definition of a certifying gap deliberately excludes.

\xhdr{Assessing a protocol before optimizing its parameters.} This
formulation separates two complementary tasks: determining what a
protocol can certify, and finding functions that approach that
capacity. If
$\Del^{\mathrm{opt}}_{\mathfrak Q(\cS)}(\cG)\leq\varepsilon$
and $f\in\cS$ has a certificate within $\delta$ of the capacity,
then soundness implies
\[
\gamma(\mathsf{ALG}_f,\cG)\geq\gamma_{\mathrm{opt}}(\cG)-\varepsilon-\delta.
\]
Thus a bound on the framework's gap quantifies how much can be gained
by solving its remaining parameter-optimization problem. Conversely,
an upper bound on capacity below a desired guarantee rules out
achieving that guarantee through parameter tuning alone. Attainment
is unnecessary for these statements: functions approaching the
supremum suffice.

Existing work already uses differential inequalities and
factor-revealing LPs to choose gain-sharing functions or
parameters~\cite{huang2020online,fahrbach2022edge}.
Xu~\cite{xu2026gainsharing} studies gain-sharing optimization for the
Stochastic Balance template, using Volterra comparison and validated
residuals alongside auxiliary LPs with refinement and coarsening error
bounds. These tools certify functional optima within a specified
template; they do not establish that the template captures the
optimal online competitiveness. Similarly, Feng et
al.~\cite{feng2025degree} explicitly distinguish optimality of the
candidate function in their OCS analysis from optimality of the
resulting online algorithm.

The extension is therefore conceptual, not an automatic transfer of
our technical results. In an ADF, varying $f$ changes the distribution
of allocations, so the certificate need not retain the affine
structure available for a fixed algorithm. Moreover, optimizing a
sufficient relaxation of the dual constraints does not establish the
full certifying capacity unless that relaxation is proved exact.
Functional optimization remains substantive, but capacity and gap
bounds determine whether further optimization within the same
protocol can reach the target or whether the protocol itself must
change. Developing techniques for this distinction, in both AAFs and
ADFs, is a broader motivation for the present work.

\section{Auxiliary Proof for the Certifying-Capacity Formulation}
\label{app:intro-gain-sharing-soundness}

\begin{proof}[Proof of Proposition~\ref{prop:intro-gain-sharing-soundness}]
Fix $f\in\cS$ and an instance $G\in\cG$ with
$\operatorname{OPT}(G)>0$. Let $M^\star$ be a maximum matching
in $G$.

By construction, each successful match contributes exactly one
unit to the total dual gain. Thus, pathwise,
\[
\sum_{i\in I}Z_i+\sum_{j\in J}Z_j=\rank(G).
\]
Moreover, all dual variables are nonnegative. Since the edges of
$M^\star$ have disjoint endpoints,
\[
\begin{aligned}
\operatorname{OPT}(G)\,\chi(f,\cG)
&\leq
\sum_{(i,j)\in M^\star}\Phi_G(i,j;f)=
\sum_{(i,j)\in M^\star}
\mathbb E[Z_i+Z_j]\\
&\leq
\mathbb E\left[
\sum_{i\in I}Z_i+\sum_{j\in J}Z_j
\right]
=
\mathbb E[\rank(G)].
\end{aligned}
\]
Dividing by $\operatorname{OPT}(G)$ and taking the infimum over
$G\in\cG$ gives
$\chi(f,\cG)\leq\gamma(\cG)$.
Finally, taking the supremum over $f\in\cS$ proves
$\Gamma_{\mathfrak P(\cS)}(\cG)\leq\gamma(\cG)$.
\end{proof}

\section{Auxiliary Proof for the Unrestricted-Degree Characterization
in Theorem~\ref{thm:Gamma-infinity}}
\label{app:unrestricted-degree-characterization}

\begin{proof}[\textbf{Proof of
Lemma~\ref{thm:chi-infinity-characterization}}]
We prove the two directions separately.

\xhdr{Lower bound.}
Fix $G\in\mathcal B_\infty^{\mathrm{on}}$ and an edge
$(i^\star,j^\star)\in E(G)$. Truncating the instance immediately
after $j^\star$ leaves its online contribution unchanged and can only
decrease the offline contribution of $i^\star$. Since the truncated
instance remains in $\mathcal B_\infty^{\mathrm{on}}$, it suffices to
assume that $j^\star$ is the final arrival.

Condition on all offline ranks other than $Y_{i^\star}$, and run
\rank{} on $G^{-i^\star}:=G-i^\star$. Let
\[
\tau:=R_{G^{-i^\star}}(j^\star).
\]
The value $\tau$ is fixed under this conditioning.

Insert $i^\star$ with rank $Y_{i^\star}=y$. By
Lemma~\ref{lem:offline-insertion-boundary} and monotonicity of $f$,
the online contribution of $j^\star$ is at least $1-f(\tau)$.
When $\tau=1$, this lower bound is zero.

Moreover, if $y<\tau$, then $i^\star$ is matched by the end. Indeed,
either it is selected before $j^\star$, or it remains safe until
$j^\star$. In the latter case, all earlier decisions agree with the
reduced execution, and $j^\star$ selects $i^\star$ because
$y<\tau$. Rank ties occur with probability zero.

Averaging first over $Y_{i^\star}$ and then over the remaining ranks
therefore gives
\[
\Phi_G(i^\star,j^\star;f)
\geq
\mathbb E\!\left[
\int_0^\tau f(y)\,dy+1-f(\tau)
\right]
=
\mathbb E[H_f(\tau)]
\geq
\min_{t\in[0,1]}H_f(t).
\]
Taking the infimum over graph-edge pairs proves
\[
\chi(f;\infty)\geq\min_{t\in[0,1]}H_f(t).
\]

\xhdr{Upper bound at continuity points.}
Fix a continuity point $t\in[0,1)$ of $f$. Choose
$q_m\in[m]$ such that $q_m/m\to t$; when $t=0$, take $q_m=1$.

Consider $K_{m,q_m}$ with $m$ offline nodes and $q_m$ online nodes,
and let $j^\star$ be the final online arrival. Distinguish any edge
$(i^\star,j^\star)$. Write
\[
U_{(1:m)}<\cdots<U_{(m:m)}
\]
for the order statistics of the $m$ offline ranks. The successive
arrivals are matched to the first $q_m$ order statistics, and symmetry
gives
\begin{equation}
\label{eq:complete-quantile-edge-score}
\Phi_{K_{m,q_m}}(i^\star,j^\star;f)
=
\frac1m\sum_{r=1}^{q_m}\mathbb E[f(U_{(r:m)})]
+
1-\mathbb E[f(U_{(q_m:m)})].
\end{equation}

Since $U_{(q_m:m)}\to t$ in probability and $f$ is bounded and
continuous at $t$,
\[
\mathbb E[f(U_{(q_m:m)})]\longrightarrow f(t).
\]
If $U_1$ is one of the original ranks and
$\operatorname{rk}_m(U_1)$ denotes its rank among the $m$ samples,
then
\[
\frac1m\sum_{r=1}^{q_m}\mathbb E[f(U_{(r:m)})]
=
\mathbb E\!\left[
f(U_1)\mathbf1\{\operatorname{rk}_m(U_1)\leq q_m\}
\right].
\]
Conditional on $U_1=y$,
\[
\operatorname{rk}_m(U_1)-1\sim\operatorname{Bin}(m-1,y).
\]
Thus
\[
\begin{aligned}
\frac1m\sum_{r=1}^{q_m}\mathbb E[f(U_{(r:m)})]
&=
\int_0^1
f(y)\Pr[\operatorname{Bin}(m-1,y)\leq q_m-1]\,dy\\
&\longrightarrow
\int_0^t f(y)\,dy
\end{aligned}
\]
by dominated convergence. Hence
\[
\lim_{m\to\infty}
\Phi_{K_{m,q_m}}(i^\star,j^\star;f)
=
H_f(t),
\]
and therefore $\chi(f;\infty)\leq H_f(t)$ at every continuity point
$t<1$.

\xhdr{Discontinuities and the terminal endpoint.}
For arbitrary $t<1$, choose continuity points
$t_\ell\downarrow t$ with $t_\ell>t$. Since $F_f$ is continuous and
$f$ is right-continuous,
\[
\chi(f;\infty)
\leq
\lim_{\ell\to\infty}H_f(t_\ell)
=
H_f(t).
\]

For $t=1$, take one offline node $i^\star$ and two online nodes, both
adjacent only to $i^\star$, with the distinguished node $j^\star$
arriving second. The first arrival matches $i^\star$ and
$j^\star$ is unmatched, so
\[
\Phi_G(i^\star,j^\star;f)
=
\int_0^1f(y)\,dy
=
H_f(1).
\]
Thus $\chi(f;\infty)\leq H_f(t)$ for every $t\in[0,1]$. Taking the
minimum completes the proof.
\end{proof}

\section{Proof of the Online-Degree-One Assertions in
Theorem~\ref{thm:Gamma-degree-one}}
\label{sec:degree-one}

When every online node has at most one offline neighbor, the inner
certifying value has a simple exact expression.

\begin{lemma}[Exact degree-one characterization]
\label{thm:chi-degree-one}
For every $f\in\mathcal F$,
\begin{equation}
\label{eq:chi-degree-one}
\chi(f;1)
=
H_f(1)
=
\int_0^1f(y)\,dy.
\end{equation}
\end{lemma}

\begin{proof}
Fix $G\in\mathcal B_{\leq1}^{\mathrm{on}}$ and an edge
$(i,j)\in E(G)$. The unique offline neighbor of $j$ is $i$.
If $i$ is safe when $j$ arrives, then $j$ selects it; otherwise,
$i$ has already been matched. Thus $X_i=1$ for every rank
realization. Since the online contribution is nonnegative,
\[
\Phi_G(i,j;f)
\geq
\mathbb E[f(Y_i)]
=
\int_0^1f(y)\,dy.
\]
Taking the infimum over all graph-edge pairs gives
$\chi(f;1)\geq\int_0^1f(y)\,dy$.

For the reverse inequality, take one offline node $i$ and two online
nodes $j_1\prec j_2$, both adjacent only to $i$, and distinguish
$(i,j_2)$. The first arrival matches to $i$, while $j_2$ is
unmatched. Hence
\[
\Phi_G(i,j_2;f)
=
\mathbb E[f(Y_i)]
=
\int_0^1f(y)\,dy.
\]
This instance belongs to $\mathcal B_{\leq1}^{\mathrm{on}}$, so
$\chi(f;1)\leq\int_0^1f(y)\,dy$. Finally, $f(1)=1$ gives
$H_f(1)=\int_0^1f(y)\,dy$, completing the proof.
\end{proof}

\begin{proof}[\textbf{Proof of Theorem~\ref{thm:Gamma-degree-one}}]
\xhdr{Certifying value and uniqueness.} By
Lemma~\ref{thm:chi-degree-one},
\[
\Gamma(1)
=
\sup_{f\in\mathcal F}\int_0^1f(y)\,dy
=
1,
\]
because $0\leq f\leq1$ and $f_1^\star\equiv1$ belongs to
$\mathcal F$.

Suppose $f\in\mathcal F$ also attains this value. Then
$f=1$ Lebesgue-almost everywhere. Right-continuity implies
$f(t)=1$ for every $t\in[0,1)$: otherwise, $f(t)<1$ at some such
point, and right-continuity would make $f$ bounded strictly below
one on an interval of positive length immediately to its right,
a contradiction. The normalization $f(1)=1$ supplies the remaining
endpoint. Therefore $f_1^\star\equiv1$ is the unique optimizer.

\xhdr{Competitiveness and gap.} Soundness gives
$\gamma(1)\geq\Gamma(1)=1$. Conversely, every matching produced
by \rank{} has cardinality at most $\operatorname{OPT}(G)$, so
$\gamma(1)\leq1$. Consequently,
\[
\gamma(1)=\Gamma(1)=1,
\qquad
\Del(1)=\gamma(1)-\Gamma(1)=0.
\]
\end{proof}

\section{Auxiliary Proofs for the Online-Degree-Two Certifying Assertions in
Theorem~\ref{thm:Gamma-two}}
\label{app:degree-two-certifying-auxiliary}

\subsection{Degree-Two Order-State Characterization}
\label{app:degree-two-state-characterization}

This subsection proves
Lemmas~\ref{lem:rooted-order-state-representation} and
\ref{lem:independent-or-exclusive-safety-events}. Fix an online-degree-two instance 
\[
G=(I,[n],E).
\]
 For $r\in\{0,1,\ldots,n\}$, let the \emph{prefix at time $r$} consist
of arrivals $1,\ldots,r$ and the matching decisions made by \rank on
those arrivals. For an offline node $i$, define
\begin{equation}
\label{eq:E-i-r-appendix}
\mathcal E_i^{(r)}
:=
\left\{
i\text{ is safe after \rank processes arrivals }1,\ldots,r
\right\}.
\end{equation}
We call $i$ \emph{positive at time $r$} if
\[
\Pr[\mathcal E_i^{(r)}]>0;
\]
otherwise, $i$ is \emph{dead}. A dead offline node remains dead at all
later times.

The main body fixes one particular prefix, namely the prefix
immediately preceding the distinguished arrival $j^\star$. Its event
$\mathcal E_i$ is the corresponding event
$\mathcal E_i^{(r)}$ in the notation of this appendix.

Since the offline ranks are independent and continuously distributed,
the event
\begin{equation}
\label{eq:tie-free-event-appendix}
\Omega_{\mathsf{tf}}
:=
\left\{
Y_i\neq Y_{\widehat i}
\text{ for all distinct }i,\widehat i\in I
\right\}
\end{equation}
has probability one. All event identities below are understood after
intersection with $\Omega_{\mathsf{tf}}$. This allows us to use strict
rank comparisons throughout.

\subsubsection{Rooted Order States}
\label{subsec:rooted-order-states-appendix}

Let $\mathcal W\subseteq I$ be finite, and let $\prec$ be a strict
partial order on $\mathcal W$. Recall that its order event is
\begin{equation}
\label{eq:order-event-appendix}
\mathcal O(\mathcal W,\prec)
:=
\bigcap_{\substack{
i_1,i_2\in\mathcal W\\
i_1\prec i_2
}}
\{Y_{i_1}<Y_{i_2}\}.
\end{equation}
A rooted order state for an offline node $i$ is a triple
\[
(\mathcal W,\prec;i)
\]
such that $i\in\mathcal W$ and $i$ is the unique maximal element of
the finite poset $(\mathcal W,\prec)$. We call $\mathcal W$ its
witness set.

We prove the two main-body lemmas through the following stronger
inductive invariant.

\begin{lemma}[Inductive order-state invariant]
\label{lem:inductive-order-state-invariant}
After every arrival prefix $r\in\{0,\ldots,n\}$, the following
properties hold.

\begin{enumerate}
    \item Every positive offline node $i$ admits a rooted order state
    \[
    (\mathcal W_i^{(r)},\prec_i^{(r)};i)
    \]
    satisfying
    \begin{equation}
    \label{eq:order-state-rep-inductive}
    \mathcal E_i^{(r)}
    =
    \mathcal O(\mathcal W_i^{(r)},\prec_i^{(r)}).
    \end{equation}

    \item For any two distinct positive offline nodes $i_1,i_2$,
    either their witness sets are disjoint,
    \begin{equation}
    \label{eq:pairwise-disjoint-witness-appendix}
    \mathcal W_{i_1}^{(r)}
    \cap
    \mathcal W_{i_2}^{(r)}
    =
    \varnothing,
    \end{equation}
    or their safety events are mutually exclusive,
    \begin{equation}
    \label{eq:pairwise-exclusive-events-appendix}
    \Pr\!\left[
    \mathcal E_{i_1}^{(r)}
    \cap
    \mathcal E_{i_2}^{(r)}
    \right]
    =
    0.
    \end{equation}
\end{enumerate}
\end{lemma}

\begin{proof}
We proceed by induction over $r$.

\xhdr{Base case.}
Before any online arrival, every offline node is safe. For each
$i\in I$, set
\[
\mathcal W_i^{(0)}:=\{i\},
\]
and let $\prec_i^{(0)}$ be the empty relation. Then
\[
\mathcal O(\mathcal W_i^{(0)},\prec_i^{(0)})
=
\Omega_{\mathsf{tf}},
\]
so it represents $\mathcal E_i^{(0)}$. The singleton witness sets are
pairwise disjoint.

\xhdr{Inductive step.}
Assume the invariant holds after the first $r$ arrivals, and consider
arrival $r+1$.

An offline node outside the neighborhood of arrival $r+1$ has the same
safety event and retains its current rooted order state. We therefore
focus on the at most two offline neighbors of the new arrival.

If the arrival has no offline neighbor, nothing changes.

Suppose it has exactly one offline neighbor $i$. Whenever $i$ is safe,
the new arrival matches to $i$; otherwise, $i$ has already been
matched. Hence $i$ is dead after the arrival. All other positive
states remain unchanged, so the invariant is preserved.

Now suppose the new arrival has two offline neighbors $i_1,i_2$.

If at least one of them is dead before the arrival, then the other,
whenever safe, is the unique safe neighbor and is therefore matched.
Thus both nodes are dead after the arrival.

Assume next that $i_1$ and $i_2$ are both positive. Write
\[
\mathcal E_k:=\mathcal E_{i_k}^{(r)},
\qquad
\mathcal W_k:=\mathcal W_{i_k}^{(r)},
\qquad
\prec_k:=\prec_{i_k}^{(r)},
\qquad
k\in\{1,2\}.
\]

If
\[
\mathcal W_1\cap\mathcal W_2\neq\varnothing,
\]
the inductive invariant implies
\[
\Pr[\mathcal E_1\cap\mathcal E_2]=0.
\]
Thus, at most one of $i_1,i_2$ can be safe. Whenever one of them is
safe, the new arrival matches to that node. Consequently, both nodes
are dead after the arrival.

It remains to consider the case
\[
\mathcal W_1\cap\mathcal W_2=\varnothing.
\]
The post-arrival safety events are then
\begin{align}
\mathcal E_{i_1}^{(r+1)}
&=
\mathcal E_1\cap\mathcal E_2
\cap\{Y_{i_2}<Y_{i_1}\},
\label{eq:post-event-i1-order-appendix}\\
\mathcal E_{i_2}^{(r+1)}
&=
\mathcal E_1\cap\mathcal E_2
\cap\{Y_{i_1}<Y_{i_2}\}.
\label{eq:post-event-i2-order-appendix}
\end{align}
Indeed, if both nodes are safe, \rank matches the one with smaller
rank. If exactly one is safe, that node is matched and therefore does
not remain safe.

Set
\[
\mathcal W^+:=\mathcal W_1\cup\mathcal W_2.
\]
Let $\prec_1^+$ be the transitive closure of
\[
\prec_1\cup\prec_2\cup\{(i_2,i_1)\},
\]
and let $\prec_2^+$ be the transitive closure of
\[
\prec_1\cup\prec_2\cup\{(i_1,i_2)\}.
\]

Because $\mathcal W_1$ and $\mathcal W_2$ are disjoint, neither added
comparison creates a directed cycle. Hence $\prec_1^+$ and
$\prec_2^+$ are strict partial orders. Furthermore, $i_1$ is the
unique maximal element of $(\mathcal W^+,\prec_1^+)$, while $i_2$ is
the unique maximal element of
$(\mathcal W^+,\prec_2^+)$.

By
\eqref{eq:post-event-i1-order-appendix} and
\eqref{eq:post-event-i2-order-appendix},
\begin{align}
\mathcal E_{i_1}^{(r+1)}
&=
\mathcal O(\mathcal W^+,\prec_1^+),
\label{eq:new-order-state-i1-appendix}\\
\mathcal E_{i_2}^{(r+1)}
&=
\mathcal O(\mathcal W^+,\prec_2^+).
\label{eq:new-order-state-i2-appendix}
\end{align}
Thus both output nodes admit rooted order states.

The two output events are mutually exclusive because they impose the
opposite strict comparisons
\[
Y_{i_2}<Y_{i_1}
\qquad\text{and}\qquad
Y_{i_1}<Y_{i_2}.
\]

It remains to compare the two new states with an unaffected positive
offline node $i$. If
\[
\mathcal W_i^{(r)}
\cap
(\mathcal W_1\cup\mathcal W_2)
=
\varnothing,
\]
then its witness set is disjoint from the new witness set
$\mathcal W^+$.

Otherwise, $\mathcal W_i^{(r)}$ intersects at least one of
$\mathcal W_1,\mathcal W_2$. Suppose, without loss of generality, that
it intersects $\mathcal W_1$. The inductive invariant gives
\[
\Pr[\mathcal E_i^{(r)}\cap\mathcal E_1]=0.
\]
Since each new event in
\eqref{eq:post-event-i1-order-appendix} and
\eqref{eq:post-event-i2-order-appendix} is a subset of
$\mathcal E_1$, we have
\[
\Pr\!\left[
\mathcal E_i^{(r)}
\cap
\mathcal E_{i_k}^{(r+1)}
\right]
=
0,
\qquad
k\in\{1,2\}.
\]
Therefore, every overlapping pair of witness sets corresponds to
mutually exclusive safety events.

All remaining pairs are unaffected by the arrival. This completes the
induction.
\end{proof}

\subsubsection{Proof of Lemma~\ref{lem:rooted-order-state-representation}}
\label{subsec:proof-rooted-order-state-representation}

\begin{proof}[Proof of the Rooted Order-State Representation]
Fix the arrival prefix appearing in the statement of the lemma, and
let $r$ denote its length.

 If
$
\Pr[\mathcal E_i^{(r)}]>0$,
then $i$ is positive at time $r$. Applying
Lemma~\ref{lem:inductive-order-state-invariant} gives a rooted order
state
$
(\mathcal W_i^{(r)},\prec_i^{(r)};i)
$
such that
$
\mathcal E_i^{(r)}
=
\mathcal O(\mathcal W_i^{(r)},\prec_i^{(r)})
$
on the tie-free event $\Omega_{\mathsf{tf}}$.

Since
\[
\Pr[\Omega_{\mathsf{tf}}]=1,
\]
the equality holds up to a probability-zero set, as claimed.
\end{proof}

\subsubsection{Proof of Lemma~\ref{lem:independent-or-exclusive-safety-events}}
\label{subsec:proof-independent-or-exclusive}

\begin{proof}[Proof of the Pairwise Dichotomy]
Fix the arrival prefix in the statement and let $r$ denote its length.
Apply
Lemma~\ref{lem:inductive-order-state-invariant} to the two positive
offline nodes $i_1,i_2$.

If
\[
\mathcal W_1\cap\mathcal W_2=\varnothing,
\]
then $\mathcal E_{i_1}^{(r)}$ is measurable with respect to the ranks
\[
\{Y_i:i\in\mathcal W_1\},
\]
while $\mathcal E_{i_2}^{(r)}$ is measurable with respect to
\[
\{Y_i:i\in\mathcal W_2\}.
\]
These two rank collections are independent, so the two safety events
are independent.

Moreover, after conditioning on
\[
Y_{i_1}=x,
\qquad
Y_{i_2}=y,
\]
the remaining ranks in the two witness sets are still independent.
Using
Corollary~\ref{cor:monomial-safety-profile-Gamma}, we obtain, for  every $x,y\in[0,1]$,
\[
\begin{aligned}
&
\Pr\!\left[
\mathcal E_{i_1}^{(r)}
\cap
\mathcal E_{i_2}^{(r)}
\mathrel{}\middle|\mathrel{}
Y_{i_1}=x,\,
Y_{i_2}=y
\right]=
\Pr[
\mathcal E_{i_1}^{(r)}
\mid
Y_{i_1}=x
]
\Pr[
\mathcal E_{i_2}^{(r)}
\mid
Y_{i_2}=y
]=
a_1a_2x^{p_1}y^{p_2}.
\end{aligned}
\]

If the two witness sets are not disjoint,
Lemma~\ref{lem:inductive-order-state-invariant} gives
\[
\Pr[
\mathcal E_{i_1}^{(r)}
\cap
\mathcal E_{i_2}^{(r)}
]
=
0.
\]

Finally, the two alternatives cannot hold simultaneously. Indeed, if
the witness sets are disjoint, then the positive safety events are
independent and have positive marginal probabilities, so
\[
\Pr[
\mathcal E_{i_1}^{(r)}
\cap
\mathcal E_{i_2}^{(r)}
]
=
\Pr[\mathcal E_{i_1}^{(r)}]
\Pr[\mathcal E_{i_2}^{(r)}]
>
0.
\]
Thus exactly one of the two structural cases applies.
\end{proof}

\subsubsection{Proof of Lemma~\ref{lem:power-filter-gadget-score}}
\label{app:power-filter-gadget-score}

\begin{proof}[Proof of Score of the Power-Filter Gadget]
Fix $p,q\in\mathbb Z_{\geq0}$ and $f\in\cF$. We compute
the score of the distinguished edge $(i^\star,j^\star)$ in the gadget
$G_{p,q}$.

Let
\[
X_{i^\star},\qquad X_{j^\star}
\]
be the matched indicators of $i^\star$ and $j^\star$, respectively.
Recall that
\[
\Phi_{G_{p,q}}(i^\star,j^\star;f)
=
\mathbb E[f(Y_{i^\star})X_{i^\star}]
+
\mathbb E[(1-f(Y_{j^\star}))X_{j^\star}].
\]

\xhdr{Step 1: survival probabilities of the two roots.}
Conditional on $Y_{i^\star}=x$, the root $i^\star$ remains safe after
the $p$ private filters if and only if every private co-neighbor
$u_1,\ldots,u_p$ has rank smaller than $x$. Since these ranks are
independent and uniform on $[0,1]$,
\begin{equation}
\label{eq:root-survival-left}
\Pr[i^\star\text{ is safe before }j^\star \mid Y_{i^\star}=x]
=
x^p.
\end{equation}
Similarly,
\begin{equation}
\label{eq:root-survival-right}
\Pr[\tilde i^\star\text{ is safe before }j^\star \mid
Y_{\tilde i^\star}=y]
=
y^q.
\end{equation}
These two survival events are independent.

\xhdr{Step 2: the offline contribution of $i^\star$.}
The node $i^\star$ fails to be matched at the end only in the following
case: both $i^\star$ and $\tilde i^\star$ are safe immediately before
$j^\star$, and
\[
Y_{\tilde i^\star}<Y_{i^\star},
\]
so that \rank matches $j^\star$ to $\tilde i^\star$.

Therefore, conditional on $Y_{i^\star}=x$,
\[
\Pr[X_{i^\star}=0\mid Y_{i^\star}=x]
=
x^p\int_0^x y^q\,dy
=
\frac{x^{p+q+1}}{q+1}.
\]
Hence
\[
\Pr[X_{i^\star}=1\mid Y_{i^\star}=x]
=
1-\frac{x^{p+q+1}}{q+1},
\]
and thus
\begin{equation}
\label{eq:offline-contribution-power-filter}
\begin{aligned}
\mathbb E[f(Y_{i^\star})X_{i^\star}]
&=
\int_0^1
\left(
1-\frac{x^{p+q+1}}{q+1}
\right)
f(x)\,dx=
\int_0^1 f(x)\,dx
-
\frac{1}{q+1}\int_0^1x^{p+q+1}f(x)\,dx.
\end{aligned}
\end{equation}

\xhdr{Step 3: the online contribution of $j^\star$.}
The node $j^\star$ is matched if and only if at least one of
$i^\star$ and $\tilde i^\star$ is safe immediately before its
arrival. Since the two survival events are independent and have
probabilities $1/(p+1)$ and $1/(q+1)$, respectively,
\begin{equation}
\label{eq:jstar-matched-probability}
\Pr[X_{j^\star}=1]
=
\frac1{p+1}+\frac1{q+1}
-\frac1{(p+1)(q+1)}
=
\frac{p+q+1}{(p+1)(q+1)}.
\end{equation}

Next, we compute the expected gain-sharing value associated with the
offline partner selected by $j^\star$. If $j^\star$ is matched to
$i^\star$, then $i^\star$ must be safe and either
$\tilde i^\star$ is not safe or
$Y_{i^\star}<Y_{\tilde i^\star}$. The symmetric statement holds when
$j^\star$ is matched to $\tilde i^\star$. Therefore,
\begin{equation}
\label{eq:partner-rank-contribution}
\begin{aligned}
\mathbb E[f(Y_{j^\star})X_{j^\star}]
={}&
\int_0^1 x^p f(x)\,dx
+
\int_0^1 y^q f(y)\,dy\\
&-
\left(
\frac1{p+1}+\frac1{q+1}
\right)
\int_0^1 z^{p+q+1}f(z)\,dz.
\end{aligned}
\end{equation}
Indeed, the final integral subtracts the double-counted contribution
when both roots are safe: on the event
$Y_{\tilde i^\star}<Y_{i^\star}$ the selected rank is
$Y_{\tilde i^\star}$, while on the complementary strict ordering it
is $Y_{i^\star}$.

Consequently,
\begin{equation}
\label{eq:online-contribution-power-filter}
\begin{aligned}
\mathbb E[(1-f(Y_{j^\star}))X_{j^\star}]
={}&
\frac{p+q+1}{(p+1)(q+1)}
-
\int_0^1 y^p f(y)\,dy
-
\int_0^1 y^q f(y)\,dy\\
&+
\left(
\frac1{p+1}+\frac1{q+1}
\right)
\int_0^1 y^{p+q+1}f(y)\,dy.
\end{aligned}
\end{equation}

\xhdr{Step 4: combine the two parts.}
Adding
\eqref{eq:offline-contribution-power-filter} and
\eqref{eq:online-contribution-power-filter}, the coefficient of
$\int_0^1 y^{p+q+1}f(y)\,dy$ simplifies as
\[
-\frac1{q+1}
+
\left(
\frac1{p+1}+\frac1{q+1}
\right)
=
\frac1{p+1}.
\]
Hence
\[
\Phi_{G_{p,q}}(i^\star,j^\star;f)
=
\frac{p+q+1}{(p+1)(q+1)}
+
\int_0^1
\left(
1-y^p-y^q+\frac{y^{p+q+1}}{p+1}
\right)f(y)\,dy.
\]
This is exactly $\Psi_{p,q}(f)$.
\end{proof}

\subsection{Finite Verification of the Filter Inequalities}
\label{app:finite-filter-verification}

This subsection completes the finite verification used in the proof of
Lemma~\ref{lem:all-filtered-root-constraints}. Recall from
\eqref{eq:Psi-p-q-minus-gamma-Q} that
\[
\Psi_{p,q}(f_2^\star)-r^\star
=
\frac{w^\star}{
(p+1)(q+1)(p+q+2)
}
Q_{p,q}(s^\star),
\]
where $w^\star>0$ and
$Q_{p,q}(s^\star)$ is defined in
\eqref{eq:Q-p-q-definition}. The main-body proof already establishes
$Q_{p,q}(s^\star)>0$ whenever $p+q+2\geq10$. It remains to verify
\[
Q_{p,q}(s^\star)\geq0
\qquad
\text{for }p+q+2\leq9.
\]

For the algebraic verification below, extend
\eqref{eq:Q-p-q-definition} to the polynomial
\begin{equation}
\label{eq:Q-p-q-polynomial-appendix}
\begin{aligned}
Q_{p,q}(z)
:={}&
(q+1)(p+q+2)z^{p+1}
+(p+1)(p+q+2)z^{q+1}\\
&-(q+1)z^{p+q+2}
-(p+1)+
\frac{z^2-4z+1}{2}
(p+q+2)(p+q+1).
\end{aligned}
\end{equation}

\subsubsection{Localization of \texorpdfstring{$s^\star$}{s*}}
\label{subsec:root-localization-filter-appendix}

To certify the signs below using exact rational arithmetic, we first
localize $s^\star$ in a short rational interval. Define
\begin{equation}
\label{eq:root-interval-filter-appendix}
\ell:=\frac{101381}{500000},
\qquad
u:=\frac{202763}{1000000}.
\end{equation}
Direct rational evaluation of the quartic in
\eqref{eq:s-two-star-polynomial} gives
\[
2\ell^4+2\ell^3-8\ell^2-33\ell+7>0
\]
and
\[
2u^4+2u^3-8u^2-33u+7<0.
\]
Since $s^\star$ is the unique root in $(0,1)$ of
\eqref{eq:s-two-star-polynomial},
\begin{equation}
\label{eq:s-star-tight-rational-interval}
\ell<s^\star<u.
\end{equation}

\subsubsection{Reduction to Cubic Polynomials}
\label{subsec:cubic-reduction-filter-appendix}

For each ordered pair
$p,q\in\mathbb Z_{\geq0}$, let $R_{p,q}(z)$ denote the remainder of
$Q_{p,q}(z)$ upon division by the quartic in
\eqref{eq:s-two-star-polynomial}. Then
\[
\deg R_{p,q}\leq3,
\]
and, since $s^\star$ satisfies
\eqref{eq:s-two-star-polynomial},
\begin{equation}
\label{eq:Q-equals-remainder-at-root}
Q_{p,q}(s^\star)=R_{p,q}(s^\star).
\end{equation}
The reduction can be performed recursively using
\begin{equation}
\label{eq:quartic-power-reduction}
z^4
\equiv
-z^3+4z^2+\frac{33}{2}z-\frac72
\pmod{2z^4+2z^3-8z^2-33z+7}.
\end{equation}

To certify the sign of $R_{p,q}$ on $[\ell,u]$, write its degree-three
Bernstein representation as
\begin{equation}
\label{eq:Bernstein-representation-filter}
R_{p,q}\bigl(\ell+(u-\ell)x\bigr)
=
\sum_{k=0}^3
\beta_{p,q,k}
\binom{3}{k}
x^k(1-x)^{3-k}.
\end{equation}
All coefficients $\beta_{p,q,k}$ are rational. Since the Bernstein
basis functions are nonnegative and sum to one on $[0,1]$,
\begin{equation}
\label{eq:Bernstein-lower-bound-filter}
R_{p,q}(z)
\geq
\min_{0\leq k\leq3}\beta_{p,q,k}
\qquad
\text{for every }z\in[\ell,u].
\end{equation}

For completeness, if
\[
R_{p,q}\bigl(\ell+(u-\ell)x\bigr)
=
r_0+r_1x+r_2x^2+r_3x^3,
\]
then
\begin{align}
\beta_{p,q,0}
&=
r_0,
\label{eq:Bernstein-beta-zero}\\
\beta_{p,q,1}
&=
r_0+\frac{r_1}{3},
\label{eq:Bernstein-beta-one}\\
\beta_{p,q,2}
&=
r_0+\frac{2r_1}{3}+\frac{r_2}{3},
\label{eq:Bernstein-beta-two}\\
\beta_{p,q,3}
&=
r_0+r_1+r_2+r_3.
\label{eq:Bernstein-beta-three}
\end{align}
Thus the remaining verification uses only exact rational arithmetic.

\subsubsection{The Finite Verification}
\label{subsec:finite-filter-table}

There are two equality cases. First,
\begin{equation}
\label{eq:Q-zero-zero}
Q_{0,0}(z)\equiv0.
\end{equation}
Second,
\begin{equation}
\label{eq:Q-two-one-factorization}
Q_{2,1}(z)
=
(1-z)
\left(
2z^4+2z^3-8z^2-33z+7
\right),
\end{equation}
so
\[
Q_{2,1}(s^\star)=0.
\]

For every other ordered pair satisfying $p+q+2\leq9$, reducing
$Q_{p,q}$ modulo the quartic in
\eqref{eq:s-two-star-polynomial} and applying
\eqref{eq:Bernstein-lower-bound-filter} gives the exact rational
bounds in Table~\ref{tab:finite-filter-verification}. For each value
of $p+q+2$, the third column records a non-equality pair attaining the
smallest Bernstein coefficient.

\begin{table}[t]
\centering
\caption{Exact rational verification of the finite filtered-root
constraints.}
\label{tab:finite-filter-verification}
\renewcommand{\arraystretch}{1.2}
\begin{tabular}{c c c c}
\toprule
$p+q+2$
&
equality pair
&
minimizing non-equality pair
&
certified lower bound
\\
\midrule
$2$
&
$(0,0)$
&
---
&
---
\\
$3$
&
---
&
$(1,0)$
&
$\displaystyle \frac1{50}$
\\
$4$
&
---
&
$(1,1)$
&
$\displaystyle \frac1{50}$
\\
$5$
&
$(2,1)$
&
$(1,2)$
&
$\displaystyle \frac9{10}$
\\
$6$
&
---
&
$(3,1)$
&
$\displaystyle \frac25$
\\
$7$
&
---
&
$(3,2)$
&
$1$
\\
$8$
&
---
&
$(4,2)$
&
$\displaystyle \frac74$
\\
$9$
&
---
&
$(5,2)$
&
$\displaystyle \frac{27}{10}$
\\
\bottomrule
\end{tabular}
\end{table}

More precisely, for every ordered pair
$p,q\in\mathbb Z_{\geq0}$ with $3\leq p+q+2\leq9$, excluding
$(p,q)=(2,1)$ when $p+q+2=5$, each coefficient
$\beta_{p,q,k}$, $k\in\{0,1,2,3\}$, is at least the certified lower
bound in the corresponding row of
Table~\ref{tab:finite-filter-verification}.

As a representative calculation, for $(p,q)=(1,0)$,
\[
R_{1,0}(z)
=
1-6z+6z^2-z^3.
\]
Its smallest Bernstein coefficient over $[\ell,u]$ is
\[
\frac{21762843419391053}{1000000000000000000}
>
\frac1{50}.
\]
All remaining entries are obtained from
\eqref{eq:quartic-power-reduction} and
\eqref{eq:Bernstein-beta-zero}--
\eqref{eq:Bernstein-beta-three} in the same way.

Since $s^\star\in[\ell,u]$, it follows from
\eqref{eq:Q-equals-remainder-at-root} and
\eqref{eq:Bernstein-lower-bound-filter} that
\begin{equation}
\label{eq:finite-Q-nonnegative}
Q_{p,q}(s^\star)\geq0
\qquad
\text{for all }p,q\in\mathbb Z_{\geq0}
\text{ with }p+q+2\leq9,
\end{equation}
with equality only for
\[
(p,q)=(0,0)
\qquad\text{and}\qquad
(p,q)=(2,1).
\]
Together with the case $p+q+2\geq10$ proved in
Lemma~\ref{lem:all-filtered-root-constraints}, this establishes
\[
Q_{p,q}(s^\star)\geq0
\qquad
\text{for all }p,q\in\mathbb Z_{\geq0},
\]
with the same two equality cases. By
\eqref{eq:Psi-p-q-minus-gamma-Q},
\[
\Psi_{p,q}(f_2^\star)\geq r^\star
\qquad
\text{for all }p,q\in\mathbb Z_{\geq0},
\]
with equality only for
\[
(p,q)=(0,0)
\qquad\text{and}\qquad
(p,q)=(2,1).
\]
This completes the finite verification required for
Lemma~\ref{lem:all-filtered-root-constraints}.

\section{Auxiliary Proofs for the Online-Degree-Two Competitiveness
Assertion in Theorem~\ref{thm:Gamma-two}}
\label{app:degree-two-competitiveness-auxiliary}

This section supplies the two technical verifications used in
Section~\ref{sec:proof-degree-two-competitiveness}: the general
dual-budget criterion and feasibility of the explicit state-dependent
budget.

\subsection{Proof of the Dual-Budget Criterion}
\label{app:degree-two-dual-budget-criterion-proof}

\begin{proof}[\textbf{Proof of
Lemma~\ref{lem:degree-two-dual-budget-criterion}}]
Fix an instance $G$. Maintain deterministic offline analysis variables
$z_i\in[0,c]$, initially zero, satisfying
$z_i\geq c-b(\mathsf s_i)$ after each prefix. At an arrival $j$
with nonempty neighborhood, update
\[
z_i^+:=\max\{z_i,c-b(\mathsf s_i^+)\}
\quad(i\in N_G(j)),
\qquad
z_j:=c-\min_{i\in N_G(j)}z_i^+.
\]
Unaffected offline variables are unchanged. All variables are
nonnegative, offline variables never decrease, and every edge
incident to $j$ satisfies $z_i^++z_j\geq c$.

We show that the increase in the total analysis mass is at most the
matching probability of $j$. For two independent positive inputs,
set $u_k:=c-z_{i_k}$. Then
$0\leq u_k\leq b(\mathsf s_k)$ and
$u_k^+=\min\{u_k,b(\mathsf s_k^+)\}$. Therefore,
\[
\begin{aligned}
\sum_{k=1}^2(z_{i_k}^+-z_{i_k})+z_j
&=u_1+u_2-\min\{u_1^+,u_2^+\}\\
&=\max\left\{
 u_1+(u_2-b(\mathsf s_2^+))_+,\,
 u_2+(u_1-b(\mathsf s_1^+))_+
\right\}\\
&\leq\max\left\{
 b(\mathsf s_1),\,b(\mathsf s_2),\,
 b(\mathsf s_1)+b(\mathsf s_2)-b(\mathsf s_1^+),\,
 b(\mathsf s_1)+b(\mathsf s_2)-b(\mathsf s_2^+)
\right\}\\
&\leq P(\mathsf s_1,\mathsf s_2).
\end{aligned}
\]
The last inequality uses
\eqref{eq:degree-two-dual-budget-inequalities} together with
$b(\mathsf s_k)\leq q_k\leq P(\mathsf s_1,\mathsf s_2)$.

For mutually exclusive positive inputs, both output states are dead,
so the increase is at most
$b(\mathsf s_1)+b(\mathsf s_2)\leq q_1+q_2$, which equals the
matching probability. If an arrival has only one positive neighbor,
the same calculation reduces to the singleton case and the increase
is at most $b(\mathsf s)\leq q(\mathsf s)$. An isolated arrival
receives $z_j=0$.

Summing the per-arrival inequalities gives
\[
\sum_{i\in I}z_i+\sum_{j\in J}z_j
\leq
\sum_{j\in J}\Pr[X_j=1]
=
\mathbb E[\rank(G)].
\]
Every edge becomes $c$-feasible when its online endpoint arrives and
remains so because the offline variables never decrease. Summing the
final edge inequalities over a maximum matching yields
\[
c\operatorname{OPT}(G)
\leq
\sum_{i\in I}z_i+\sum_{j\in J}z_j
\leq
\mathbb E[\rank(G)].
\]
Taking the infimum over
$G\in\mathcal B_{\leq2}^{\mathrm{on}}$ proves $\gamma(2)\geq c$.
\end{proof}

\subsection{Proof of Feasibility of the State-Dependent Budget}
\label{app:degree-two-state-budget-feasibility-proof}

\begin{proof}[\textbf{Proof of
Lemma~\ref{lem:degree-two-state-budget-feasible}}]
The defining series for $c$ gives
\begin{equation}
\label{eq:degree-two-competitiveness-rational-bracket}
\frac{71777}{100000}<c<\frac{359}{500}.
\end{equation}
For example, retain the first four terms and bound the remaining
positive tail by twice its first term. Substituting these bounds in
the seven non-dyadic entries of
Table~\ref{tab:degree-two-state-budgets} gives
\[
0<b(m,q)\leq q-\frac{q}{512}.
\]
The dyadic bounds follow from
\eqref{eq:degree-two-symmetric-budget-identities}. Every nonfresh
positive state has $q\leq1/2<c$, so all prescribed budgets belong
to $[0,\min\{q,c\}]$.

If either input budget is zero, each left side of
\eqref{eq:degree-two-dual-budget-inequalities} is at most the other
input budget and hence at most the matching probability. It remains
to check pairs with positive budgets.

\xhdr{Equal dyadic inputs.}
Both outputs are the next dyadic state, so the second identity in
\eqref{eq:degree-two-symmetric-budget-identities} proves equality
for both output choices.

\xhdr{Four further equality pairs.}
The additional tight transitions are
\[
\begin{array}{c|c}
\text{inputs}&\text{outputs}\\ \hline
(1,1),(2,1/2)&(3,1/6),(3,1/3)\\
(2,1/2),(3,1/6)&(5,1/30),(5,1/20)\\
(3,1/6),(3,1/6)&(6,1/72),(6,1/72)\\
(3,1/6),(4,1/8)&(7,1/112),(7,1/84).
\end{array}
\]
Substituting the table entries proves
\eqref{eq:degree-two-dual-budget-inequalities} with equality in
each row.

\xhdr{The remaining small pairs.}
There are $66$ unordered pairs, with repetition, among the eleven
states in Table~\ref{tab:degree-two-state-budgets}. Eight are the
four equal-dyadic pairs and the four further pairs above. For every
other pair, exact rational substitution using $c<359/500$ gives
\begin{equation}
\label{eq:degree-two-small-pair-slack}
P(\mathsf s_1,\mathsf s_2)-b(\mathsf s_1)-b(\mathsf s_2)
\geq\frac{27}{128000}>0.
\end{equation}
For completeness, the minimum rational lower bounds, grouped by the
first table index and taking the second index at least the first, are
\[
\begin{array}{c|c@{\qquad}c|c}
\text{first index}&\text{minimum}&\text{first index}&\text{minimum}\\
\hline
1&16/125&6&787/32000\\
2&9/2000&7&7897/192000\\
3&917/96000&8&1219/128000\\
4&211/1500&9&2021/1075200\\
5&27/128000&10&433/89600.
\end{array}
\]
The pair with both indices eleven is an equality pair. Since output
budgets are nonnegative,~\eqref{eq:degree-two-small-pair-slack} is
stronger than the required inequalities.

\xhdr{Pairs involving the dyadic tail.}
Every state in Table~\ref{tab:degree-two-state-budgets} satisfies
$q-b(m,q)\geq q/512$. For the four displayed dyadic states, this
follows from $q-b_{\mathrm{sym}}(q)\geq q^2/4$ and
$q\geq1/128$; the non-dyadic entries were checked above. Every
dyadic state of witness size at least sixteen has
$q\leq1/32768$. Thus, if the first input is one of the displayed
small states and the second lies in the dyadic tail,
\[
P-b(\mathsf s_1)-b(\mathsf s_2)
\geq q_1\left(\frac1{512}-\frac1{32768}\right)>0.
\]

For distinct tail inputs, assume $q_1>q_2$. Their dyadic form implies
$q_2\leq q_1^2/2$. Using
$b_{\mathrm{sym}}(q)\leq q-q^2/4$ for both inputs,
\[
P-b(\mathsf s_1)-b(\mathsf s_2)
\geq\frac{q_1^2}{4}-q_1q_2
\geq q_1^2\left(\frac14-\frac{q_1}{2}\right)\geq0.
\]
Equal tail inputs were handled by the exact dyadic identity. These
cases exhaust all pairs with positive budgets and complete the proof.
\end{proof}

\section{Auxiliary Proofs for the Certifying Bounds for Large Online Degree in
Theorem~\ref{thm:Gamma-large-degree}}
\label{app:large-degree-auxiliary-estimates}

We prove the conditional moment estimates used in
Lemma~\ref{lem:large-degree-conditional-slack} and the positive
mixed-kernel construction in
Lemma~\ref{lem:large-degree-positive-mixture}. Throughout, $D:=d-1$.
The moment estimates hold for every $D\geq1$; the kernel construction
uses sufficiently large $D$. In the latter construction, $N$ is the
integer parameter of the common-pool graphs, and the mixed kernel is
denoted by $\breve K_{D,N}$.

\subsection{A Joint Shell Inequality}
\label{subsec:large-degree-shell-inequality}

Recall the critical ranks $\alpha,\tau$ from
\eqref{eq:large-degree-alpha-tau}, defined in the reduced execution on
$G-i^\star$. The final arrival $j^\star$ has at most $D$ neighbors
in this reduced graph. Raising the rank selected by $j^\star$
preserves every preceding decision, and hence preserves $\alpha$.

\begin{lemma}[Joint shell inequality]
\label{lem:large-degree-shell-inequality}
For every Borel set $B\subseteq[0,1]$ and $0\leq s<t<1$,
\begin{equation}
\label{eq:large-degree-shell-inequality}
\Pr[\alpha\in B,\ s\leq\tau<t]
\leq
\frac{D(t-s)}{1-t}\Pr[\alpha\in B,\ \tau>s].
\end{equation}
\end{lemma}

\begin{proof}
Fix a reduced offline neighbor $\hat i$ of $j^\star$, and condition
on all reduced ranks other than $Y_{\hat i}$. Suppose that, at a
value $Y_{\hat i}=y$ with no rank ties, $j^\star$ selects $\hat i$.
Then $\hat i$ is safe at every earlier arrival. Every earlier arrival
adjacent to $\hat i$ therefore selects another node with rank smaller
than $y$. Increasing $Y_{\hat i}$ leaves all these decisions unchanged
and preserves $\alpha$. At $j^\star$, the selected rank becomes
strictly larger than $y$.

As $Y_{\hat i}$ varies, the section of
\[
\{\alpha\in B,\ s\leq\tau<t,\ j^\star\text{ selects }\hat i\}
\]
has Lebesgue measure at most $t-s$. If this section has positive
measure, choose a tie-free value in it. Raising that value to any
$y'\in(t,1)$, except for finitely many tie values, preserves
$\alpha\in B$ and gives $\tau>s$. Thus the corresponding section of
$\{\alpha\in B,\tau>s\}$ has measure at least $1-t$.
The claimed comparison follows by integrating these section bounds
and summing over the at most $D$ possible selected neighbors.
\end{proof}

For every fixed $t<1$, the variable $\tau$ has no atom at $t$:
it is either the rank of one of finitely many offline nodes or the
unmatched value one. Apply~\eqref{eq:large-degree-shell-inequality}
to the successive intervals of a sufficiently fine partition
$s=t_0<\cdots<t_m=t$. Each interval gives
\[
\Pr[\alpha\in B,\tau>t_r]
\geq
\left(1-\frac{D(t_r-t_{r-1})}{1-t_r}\right)
\Pr[\alpha\in B,\tau>t_{r-1}].
\]
Multiplying and letting the mesh tend to zero yields
\begin{equation}
\label{eq:large-degree-residual-survival}
\Pr[\alpha\in B,\tau>t]
\geq
\left(\frac{1-t}{1-s}\right)^D
\Pr[\alpha\in B,\tau>s].
\end{equation}

Fix a regular conditional law of $\tau$ given $\alpha$. Since
\eqref{eq:large-degree-residual-survival} holds for every Borel set
$B$, it implies the corresponding conditional inequality for
$\mathcal L(\alpha)$-almost every conditioning value. Applying this
first to rational endpoints gives a single exceptional set of
$\mathcal L(\alpha)$-measure zero. Right-continuity of the
conditional survival function then extends the inequality to all
endpoints. Thus, outside that exceptional set,
\begin{equation}
\label{eq:large-degree-conditional-survival}
\Pr[\tau>t\mid\alpha=a]
\geq
\left(\frac{1-t}{1-s}\right)^D
\Pr[\tau>s\mid\alpha=a],
\qquad 0\leq s<t<1.
\end{equation}

There is also no mass on the diagonal below one:
\begin{equation}
\label{eq:large-degree-diagonal-exclusion}
\Pr[\alpha=\tau<1]=0.
\end{equation}
Indeed, a value of $\alpha$ in $(0,1)$ is the rank of a node selected
before $j^\star$, whereas a value of $\tau$ in $(0,1)$ is selected
at $j^\star$. These nodes are distinct and have different ranks,
almost surely. If $\alpha=0$, then $\tau>0$ almost surely. Enlarge
the exceptional set of conditioning values so that, for every
remaining $a<1$, both
$\Pr[\tau=a\mid\alpha=a]=0$ and
$\Pr[\tau=0\mid\alpha=a]=0$ hold.

Fix such an $a\in[0,1)$, and let
$p_a:=\Pr[\tau<a\mid\alpha=a]$. Then
$\Pr[\tau>a\mid\alpha=a]=1-p_a$, so
\[
\begin{aligned}
\mathbb E[(\tau-a)_+^2\mid\alpha=a]
&=2\int_a^1(t-a)\Pr[\tau>t\mid\alpha=a]\,dt\\
&\geq
2(1-p_a)\int_a^1(t-a)
\left(\frac{1-t}{1-a}\right)^Ddt=\frac{2(1-a)^2}{(D+1)(D+2)}(1-p_a).
\end{aligned}
\]
This proves~\eqref{eq:large-degree-upper-gap-moment} for
$\mathcal L(\alpha)$-almost every $a\in[0,1)$.

For $0\leq s<t\leq a<1$, equation
\eqref{eq:large-degree-conditional-survival} and Bernoulli's
inequality give
\[
\begin{aligned}
\Pr[s<\tau\leq t\mid\alpha=a]
&\leq
\left[1-\left(1-\frac{t-s}{1-s}\right)^D\right]
\Pr[\tau>s\mid\alpha=a]\leq\frac{D(t-s)}{1-a}.
\end{aligned}
\]
Together with the absence of an atom at zero, this shows that the
conditional law on $[0,a)$ has density at most $D/(1-a)$. Its mass
is $p_a$, and hence $p_a(1-a)/D\leq a$. For
$0\leq u\leq p_a(1-a)/D$,
\[
\Pr[\tau<a-u\mid\alpha=a]
\geq p_a-\frac{Du}{1-a}.
\]
Integrating this bound yields
\[
\begin{aligned}
\mathbb E[(a-\tau)_+\mid\alpha=a]
&\geq
\int_0^{p_a(1-a)/D}
\left(p_a-\frac{Du}{1-a}\right)du=\frac{(1-a)p_a^2}{2D}.
\end{aligned}
\]
For $a=0$, both sides are zero. This proves
\eqref{eq:large-degree-lower-gap-moment} with the same almost-everywhere
qualification.

\subsection{A Positive Order-Statistic Approximation}
\label{subsec:large-degree-order-statistic-approximation}

For $D\geq2$, let
\[
W_{(1:D)}<\cdots<W_{(D:D)}
\]
be the order statistics of $D$ independent uniform ranks. We use
$W_{(0:D)}:=0$ and $W_{(D+1:D)}:=1$ as endpoint conventions.
For $k\in[D]$, define
\begin{equation}
\label{eq:large-degree-target-coefficients}
\lambda_{D,k}
:=\frac1D e^{-(k-1)/(D-1)}
\left(1-\frac{(k-1)(D-k)}{2(D-1)^3}\right),
\end{equation}
and let
\[
\lambda_{D,\infty}:=1-\sum_{k=1}^D\lambda_{D,k}.
\]
Since $(k-1)(D-k)\leq(D-1)^2/4$, the correction factor is positive,
and $0<\lambda_{D,k}\leq1/D$. Thus all coefficients are nonnegative,
and
\begin{equation}
\label{eq:large-degree-target-measure}
\nu_D
:=\sum_{k=1}^D\lambda_{D,k}\mathcal L(W_{(k:D)})
+\lambda_{D,\infty}\delta_1
\end{equation}
is a probability measure.

\begin{lemma}[Corrected Bernstein approximation]
\label{lem:large-degree-Bernstein-approximation}
Uniformly over $s\in[0,1]$,
\begin{equation}
\label{eq:large-degree-Bernstein-approximation}
\int_{[0,1]}H_{h_s}(t)\,\nu_D(dt)
=1-\frac1e+O(D^{-2}).
\end{equation}
Moreover, $\lambda_{D,\infty}=e^{-1}+O(D^{-2})$.
\end{lemma}

\begin{proof}
The density of the continuous part of $\nu_D$ is
\[
\rho_D(t)
=\sum_{k=1}^D D\lambda_{D,k}
\binom{D-1}{k-1}t^{k-1}(1-t)^{D-k}.
\]
For a function $u$ on $[0,1]$, write
\[
B_{D-1}u(t)
:=\sum_{r=0}^{D-1}u\!\left(\frac r{D-1}\right)
\binom{D-1}{r}t^r(1-t)^{D-1-r}.
\]
For each fixed $u\in C^4([0,1])$, Taylor expansion and the binomial
central moments give, uniformly in $t$,
\[
B_{D-1}u(t)
=u(t)+\frac{t(1-t)}{2(D-1)}\frac{d^2u(t)}{dt^2}
+O(D^{-2}).
\]
In particular, the third central moment and the fourth-moment
remainder both contribute $O(D^{-2})$.

Take $u(t):=e^{-t}$ and $v(t):=t(1-t)e^{-t}$. The coefficients
in~\eqref{eq:large-degree-target-coefficients} give exactly
\[
\rho_D(t)=B_{D-1}u(t)-\frac1{2(D-1)}B_{D-1}v(t).
\]
Because $d^2u(t)/dt^2=u(t)$ and $B_{D-1}v=v+O(D^{-1})$,
the terms of order $D^{-1}$ cancel. Hence
\[
\sup_{t\in[0,1]}|\rho_D(t)-e^{-t}|=O(D^{-2}).
\]
Integrating proves
$\lambda_{D,\infty}=1-\int_0^1\rho_D(t)\,dt=e^{-1}+O(D^{-2})$.

Recall the equalizing identity
\eqref{eq:large-degree-exponential-mixture}. Since $\nu_D$ and
$e^{-t}\,dt+e^{-1}\delta_1$ both have mass one,

\[
\begin{aligned}
&\int_{[0,1]}H_{h_s}(t)\,\nu_D(dt)-\left(1-\frac1e\right)=
\int_0^1(\rho_D(t)-e^{-t})
\bigl(H_{h_s}(t)-H_{h_s}(1)\bigr)\,dt.
\end{aligned}
\]
The second factor has absolute value at most one for all $s,t\in[0,1]$.
This proves the uniform estimate, including $s=0$ and $s=1$.
\end{proof}

\subsection{Positive Covering by Common-Pool Laws}
\label{subsec:large-degree-positive-covering}

Let $M:=|\mathcal U|=\lfloor D/2\rfloor$ be the shared-pool size,
so the private pool has size $D-M$. We assume $D$ sufficiently large
and $2\leq N\leq D^{1/3}$; in particular, $N\leq M$.
The private/shared labels in increasing rank order form a uniform word
with $M$ shared labels and $D-M$ private labels. This word is
independent of the ordered rank values.

Let $R_N$ be the position of the $N$th shared label in a uniform word
of this type. For each $0\leq q\leq D-N$, exchangeability of the label
positions gives
\begin{align}
\Pr[T_q=W_{(k:D)}]
&=\Pr[R_N=k-q],
\qquad k\in[D],
\label{eq:large-degree-shift-law}\\
\Pr[T_q=1]
&=\Pr[R_N>D-q].
\label{eq:large-degree-shift-overflow}
\end{align}
These are distributional identities: the suffix after position $q$
has the same label law as an initial segment of the same length.
They do not assert a simultaneous samplewise identity for all $q$.

Before the $N$th shared label appears, the conditional probability
that the next label is shared is at least
$(M-N+1)/D\geq1/3$. Thus $R_N$ is stochastically dominated by the
sum of $N$ independent geometric random variables with success
probability $1/3$, each supported on the positive integers. It follows
that
\begin{equation}
\label{eq:large-degree-shift-moments}
\mathbb E[R_N]=O(N),
\qquad
\mathbb E[e^{cR_N/D}]=1+O(N/D)
\end{equation}
for every fixed $c>0$, with the second implicit constant depending
only on $c$. There are also absolute constants $c_0,C_0>0$ such that
\[
\Pr[R_N>C_0N+u]\leq C_0e^{-c_0u},
\qquad u\geq0.
\]

\begin{lemma}[Positive coefficient covering]
\label{lem:large-degree-coefficient-covering}
There are absolute constants $A,L,C>0$ such that, for sufficiently
large $D$ and integers $A\ln D\leq N\leq D^{1/3}$,
\begin{equation}
\label{eq:large-degree-positive-decomposition}
\nu_D
=\sum_{q=0}^{D-N}\omega_q\mathcal L(T_q)
+\sum_{k=1}^D\delta_k\mathcal L(W_{(k:D)})
+\delta_\infty\delta_1
\end{equation}
for nonnegative coefficients of total mass one satisfying
\begin{equation}
\label{eq:large-degree-covering-bounds}
\omega_q\leq\frac1D,
\qquad
\delta_k\leq\frac1D,
\qquad
\delta_k\leq C\frac{N}{D^2}
\quad\text{for }k\geq LN.
\end{equation}
\end{lemma}

\begin{proof}
The function
\[
x\longmapsto e^{-x}
\left(1-\frac{x(1-x)}{2(D-1)}\right)
\]
is positive and decreasing on $[0,1]$, with uniformly bounded
logarithmic derivative. Hence the coefficients $\lambda_{D,k}$
decrease with $k$, and for an absolute constant $c>0$,
\[
\lambda_{D,k-r+1}\leq e^{cr/D}\lambda_{D,k},
\qquad 1\leq r\leq k\leq D.
\]

Set $\overline\omega_q:=\lambda_{D,q+1}$ for $0\leq q\leq D-N$.
By~\eqref{eq:large-degree-shift-law}, their induced coefficient at
order-statistic index $k$ is
\[
\widehat\lambda_{D,k}
:=\sum_{r=N}^{k}\lambda_{D,k-r+1}\Pr[R_N=r],
\]
where an empty sum is zero. There is no additional upper truncation: on
$R_N\leq k\leq D$, the index $q=k-R_N$ satisfies
$0\leq q\leq D-N$ because $R_N\geq N$.
Equation~\eqref{eq:large-degree-shift-moments} implies
\[
\widehat\lambda_{D,k}
\leq(1+C_0N/D)\lambda_{D,k}
\]
for an absolute constant $C_0$. Choose $L$ sufficiently large in the
exponential tail bound, and then choose $A$ sufficiently large.
For $N\geq A\ln D$ and $k\geq LN$, this gives
$\Pr[R_N>k]\leq D^{-3}$. Monotonicity of the target coefficients
therefore yields
\[
\widehat\lambda_{D,k}
\geq(1-D^{-3})\lambda_{D,k}
\qquad(k\geq LN).
\]

Choose an absolute constant $C_1>C_0$ and set
\[
\omega_q:=\left(1-C_1\frac ND\right)\overline\omega_q.
\]
For sufficiently large $D$, these weights are nonnegative, and their
induced finite coefficients do not exceed $\lambda_{D,k}$.
Define
\[
\delta_k
:=\lambda_{D,k}
-\left(1-C_1\frac ND\right)\widehat\lambda_{D,k}.
\]
Then $0\leq\delta_k\leq\lambda_{D,k}\leq1/D$, and for $k\geq LN$,
\[
\delta_k
\leq\lambda_{D,k}\left(C_1\frac ND+D^{-3}\right)
\leq C\frac{N}{D^2}.
\]

It remains to check the atom at one. Since $N\leq R_N\leq D$,
the total overflow mass of the common-pool terms is at most
\[
\begin{aligned}
\sum_{q=0}^{D-N}\omega_q\Pr[T_q=1]
&\leq\frac1D\sum_{q=0}^{D-N}\Pr[R_N>D-q]=\frac1D\mathbb E[(R_N-N)_+]
=O(N/D).
\end{aligned}
\]
This is smaller than $\lambda_{D,\infty}$ for sufficiently large $D$,
because $N/D\leq D^{-2/3}$ and
$\lambda_{D,\infty}\to1/e$. Define $\delta_\infty$ as their
difference. It is nonnegative.

Independence of the label word and the ordered ranks, together with
\eqref{eq:large-degree-shift-law}--\eqref{eq:large-degree-shift-overflow},
now gives~\eqref{eq:large-degree-positive-decomposition}.
Each law has mass one, so equality of total masses implies
\[
\sum_{q=0}^{D-N}\omega_q+\sum_{k=1}^D\delta_k+\delta_\infty=1.
\]
The remaining bound $\omega_q\leq1/D$ is immediate from its definition.
\end{proof}

\subsection{Exact-Score Excess of the Common-Pool Graphs}
\label{subsec:large-degree-common-pool-excess}

We couple the common-pool graphs for different $q$ using the same
pool ranks and label word. Fix the pool ranks, and write
$B_q=\beta$, $S_q=a$, and $T_q=t$, with $a\leq t$.
The remaining expectation is over the independent rank of $i^\star$.

If $Y_{i^\star}<\beta$, the distinguished node is selected in the
private phase. If $\beta\leq Y_{i^\star}<a$, it is selected at the
penultimate shared-pool arrival. Otherwise, when it remains safe,
the final arrival compares its rank with $t$. The private singleton
phase ensures that no private-node availability discrepancy persists
into the shared phase. Splitting the root rank at $\beta,a,t$ gives
\[
\begin{aligned}
&\mathbb E\!\left[
f(Y_{i^\star})X_{i^\star}
+(1-f(Y_{j^\star}))X_{j^\star}
\,\middle|\,\text{pool ranks}\right]=
\begin{cases}
\mathscr R_f(a,t)+(a-\beta)(f(t)-f(a)),&\beta<a,\\[1mm]
\mathscr R_f(\beta,t),&\beta\geq a.
\end{cases}
\end{aligned}
\]
The formula remains valid with nonexistent pool ranks represented by
one, using $f(1)=1$; equalities between actual sampled ranks are null
events.

For $f=h_s$, the conditional excess over $H_{h_s}(t)$ is
\[
\begin{cases}
(s-\beta)\mathbf1\{a<s\leq t\},&\beta<a,\\[1mm]
(s-\beta)\mathbf1\{\beta<s\leq t\},&a\leq\beta\leq t,\\[1mm]
(\beta-s)_+-(t-s)_+,&\beta>t.
\end{cases}
\]
Each expression is nonnegative. In the last case it is at most
$\beta-t$; in the other cases it is bounded by the localized term
below. Taking expectations gives
\begin{equation}
\label{eq:large-degree-common-pool-excess}
\begin{aligned}
0\leq{}&\Xi_{D,q,N}(s)-\mathbb E[H_{h_s}(T_q)]
\mathbb E\!\left[
(T_q-B_q)_+\mathbf1\{S_q<s\leq T_q\}
\right]
+\mathbb E[(B_q-T_q)_+].
\end{aligned}
\end{equation}

\xhdr{Exceptional ordering.} The event $B_q>T_q$ requires the first
$N$ labels after position $q$ to be shared. By exchangeability of
these label positions,
\[
\Pr[B_q>T_q]
\leq\frac{(M)_N}{(D)_N}
\leq\left(\frac MD\right)^N
\leq2^{-N},
\]
where $(x)_N:=x(x-1)\cdots(x-N+1)$. Since the common-pool weights
have total mass at most one,
\begin{equation}
\label{eq:large-degree-wrong-order-bound}
\sum_{q=0}^{D-N}\omega_q\mathbb E[(B_q-T_q)_+]\leq2^{-N}.
\end{equation}

\xhdr{Localized excess.} We prove
\begin{equation}
\label{eq:large-degree-localized-excess}
\sup_{s\in[0,1]}
\sum_{q=0}^{D-N}
\mathbb E\!\left[
(T_q-B_q)_+\mathbf1\{S_q<s\leq T_q\}
\right]
=O(N/D).
\end{equation}
Since $\omega_q\leq1/D$, this gives a weighted bound $O(N/D^2)$.

Let $Z_s$ count the pool ranks strictly smaller than $s$, so
$Z_s\sim\operatorname{Bin}(D,s)$. Condition on $Z_s=m$ and on the
label word. The event $S_q<s\leq T_q$ is equivalent, up to sampled
rank ties with $s$, to positions $q+1,\ldots,m$ containing exactly
$N-1$ shared labels. Define
\[
\mathcal Q_m
:=\left\{
\begin{array}{@{}l@{}}
q\in\mathbb Z_{\geq0}:\ q\leq\min\{m,D-N\},\ \text{and positions}\\
q+1,\ldots,m\text{ contain exactly }N-1\text{ shared labels}
\end{array}
\right\}.
\]
Let $b_m$ be the first shared position after $m$, with $b_m:=D+1$
if none exists. For $q\in\mathcal Q_m$, one has
$T_q=W_{(b_m:D)}$ and $B_q\geq W_{(q:D)}$, so
\[
(T_q-B_q)_+\leq W_{(b_m:D)}-W_{(q:D)}.
\]
Conditional on $Z_s=m$, the first $m$ order statistics have the law
of the order statistics of $m$ independent uniform samples on $[0,s]$.
The remaining order statistics are similarly generated on $[s,1]$.
Both groups are independent of the label word. Their means give
\[
\begin{aligned}
&\mathbb E[W_{(b_m:D)}-W_{(q:D)}
\mid Z_s=m,\text{ labels}]\\
&\qquad=
\frac{s(m+1-q)}{m+1}
+\frac{(1-s)(b_m-m)}{D-m+1}.
\end{aligned}
\]
The endpoint conventions also give this identity when $q=0$ or
$b_m=D+1$, and when $s\in\{0,1\}$ at its possible conditioning value.

We next establish the label-gap estimates
\begin{equation}
\label{eq:large-degree-label-gap-estimates}
\mathbb E\!\left[
\sum_{q\in\mathcal Q_m}(m+1-q)
\right]=O(N),
\qquad
\mathbb E[|\mathcal Q_m|(b_m-m)]=O(1),
\end{equation}
uniformly in $m\in\{0,\ldots,D\}$, where the expectations are over
the label word.

Encode a uniform word by the numbers $\ell_0,\ldots,\ell_M$ of
private labels before, between, and after its shared labels. These
form a uniform weak composition of $D-M$ into $M+1$ parts. For each
fixed gap index $h$ and $0\leq r\leq D-M$,
\[
\Pr[\ell_h\geq r]
=\frac{(D-M)_r}{(D)_r}\leq(2/3)^r.
\]
For larger $r$ the probability is zero. Hence
$\mathbb E[(1+\ell_h)^2]$ is bounded by an absolute constant,
uniformly in $h,D$.

Reverse the first $m$ positions, leaving the others in place. This
fixed permutation again produces a uniform label word. In this word,
the lengths $m-q$, $q\in\mathcal Q_m$, lie between its $(N-1)$st
shared position and one less than its $N$th shared position.
Using the gap counts of the permuted word, we therefore have
\[
|\mathcal Q_m|\leq1+\ell_{N-1},
\qquad
m+1-q\leq\sum_{h=0}^{N-1}(1+\ell_h)
\quad(q\in\mathcal Q_m).
\]
Cauchy--Schwarz and the uniform second-moment bounds give the first
estimate in~\eqref{eq:large-degree-label-gap-estimates}, as well as
$\mathbb E[|\mathcal Q_m|^2]=O(1)$.

Likewise, $b_m-m$ is at most the first shared-position waiting time
in the cyclic ordering $m+1,\ldots,D,1,\ldots,m$. This is another
fixed permutation of a uniform word, so the waiting time has uniformly
bounded second moment. Cauchy--Schwarz gives the second estimate in
\eqref{eq:large-degree-label-gap-estimates}. No independence between
these waiting times and $\mathcal Q_m$ is required.

Averaging the conditional order-statistic bound over the labels gives
\[
\begin{aligned}
&\mathbb E\!\left[
\sum_{q=0}^{D-N}(T_q-B_q)_+
\mathbf1\{S_q<s\leq T_q\}
\,\middle|\,Z_s=m\right]\leq
CN\frac{s}{m+1}+C\frac{1-s}{D-m+1}.
\end{aligned}
\]
Finally, the binomial identities
\[
s\,\mathbb E\!\left[\frac1{Z_s+1}\right]
=\frac{1-(1-s)^{D+1}}{D+1},
\qquad
(1-s)\,\mathbb E\!\left[\frac1{D-Z_s+1}\right]
=\frac{1-s^{D+1}}{D+1}
\]
prove~\eqref{eq:large-degree-localized-excess}, uniformly including
the endpoints.

Combining the localized and exceptional-ordering estimates with
\eqref{eq:large-degree-common-pool-excess} yields
\begin{equation}
\label{eq:large-degree-total-common-pool-excess}
\sup_{s\in[0,1]}
\sum_{q=0}^{D-N}\omega_q
\left(\Xi_{D,q,N}(s)-\mathbb E[H_{h_s}(T_q)]\right)
\leq C\frac{N}{D^2}+2^{-N}.
\end{equation}

\subsection{Residual Corrections and Completion of the Mixture}
\label{subsec:large-degree-residual-corrections}

Consider the complete bipartite graph $K_{D+1,k}$, $k\in[D]$, with
an edge incident to its final arrival distinguished. Deleting the
distinguished offline node leaves the final reduced partner rank
$W_{(k:D)}$. Conditional on all reduced ranks, integrating over the
inserted rank gives
\[
\begin{aligned}
&\mathbb E\!\left[
f(Y_{i^\star})X_{i^\star}
+(1-f(Y_{j^\star}))X_{j^\star}
\,\middle|\,\text{reduced ranks}\right]\\
&\quad=
1+F_f(W_{(k-1:D)})-W_{(k-1:D)}f(W_{(k-1:D)})
-(1-W_{(k:D)})f(W_{(k:D)}).
\end{aligned}
\]
For $f=h_s$, the excess over $H_{h_s}(W_{(k:D)})$ is therefore
\[
s\,\mathbf1\{W_{(k-1:D)}<s\leq W_{(k:D)}\}.
\]
Consequently,
\begin{equation}
\label{eq:large-degree-complete-kernel}
\Xi_{D,k}^{\mathsf{cmp}}(s)
=\mathbb E[H_{h_s}(W_{(k:D)})]
+s\Pr[\operatorname{Bin}(D,s)=k-1].
\end{equation}
The identity also holds at $s=0$ and $s=1$, where the correction is
zero.

We bound the aggregate correction under the residual coefficients of
Lemma~\ref{lem:large-degree-coefficient-covering}. Recalling
$Z_s\sim\operatorname{Bin}(D,s)$, the bounds
\eqref{eq:large-degree-covering-bounds} give
\[
s\sum_{k=1}^D\delta_k\Pr[Z_s=k-1]
\leq\frac{s}{D}\Pr[Z_s<\lceil LN\rceil]
+C\frac{N}{D^2}.
\]
For $r\in\{0,\ldots,D\}$,
\[
s\Pr[Z_s=r]
=\frac{r+1}{D+1}
\Pr[\operatorname{Bin}(D+1,s)=r+1].
\]
Summing shows that
\[
s\Pr[Z_s<\lceil LN\rceil]
\leq\frac{\lceil LN\rceil}{D+1},
\]
and hence the total residual correction is $O(N/D^2)$, uniformly
in $s$.

\begin{proof}[\textbf{Proof of Lemma~\ref{lem:large-degree-positive-mixture}}]
Take the coefficients from
Lemma~\ref{lem:large-degree-coefficient-covering} and form
\[
\breve K_{D,N}(s)
=\sum_{q=0}^{D-N}\omega_q\Xi_{D,q,N}(s)
+\sum_{k=1}^D\delta_k\Xi_{D,k}^{\mathsf{cmp}}(s)
+\delta_\infty H_{h_s}(1),
\]
as in~\eqref{eq:large-degree-mixed-kernel}. All coefficients are
nonnegative and sum to one. By
\eqref{eq:large-degree-positive-decomposition}, its baseline is
exactly \\ $\int_{[0,1]}H_{h_s}(t)\,\nu_D(dt)$.
Lemma~\ref{lem:large-degree-Bernstein-approximation},
\eqref{eq:large-degree-total-common-pool-excess}, and the residual
correction bound therefore imply
\[
\begin{aligned}
\sup_{s\in[0,1]}\breve K_{D,N}(s)
&\leq1-\frac1e+O(D^{-2})+O(N/D^2)+2^{-N}=1-\frac1e+O(N/D^2)+2^{-N}.
\end{aligned}
\]
This is the required bound with nonnegative weights of total mass one.
\end{proof}

\section{Auxiliary Proofs for the Competitiveness Bounds for Large Online Degree in
Theorem~\ref{thm:Gamma-large-degree}}
\label{app:large-degree-competitiveness-proofs}

We establish the forward reduction and the global threshold identities,
then prove the finite-degree surplus estimate by combining loss
accounting with binomial concentration. The final part constructs the
layered upper-bound instances and estimates their limiting
competitiveness. All instances and arrival orders are fixed before
the offline ranks are sampled.

\subsection{Forward Reduction and the Rank-Ordered Execution}
\label{app:large-degree-forward-proof}

\begin{proof}[\textbf{Proof of Lemma~\ref{lem:large-degree-forward-reduction}}]
We first prove the rank-order equivalence for arbitrary finite
instances with distinct offline ranks. Consider the lowest-ranked
remaining offline node. Before its earliest remaining online neighbor
arrives, it cannot have been selected; at that arrival, it is preferred
to every other remaining offline node. Thus both procedures select the
same pair. Remove the pair and continue inductively. An offline node
with no remaining neighbor can be removed without a match. This proves
that the rank-ordered and online-ordered executions produce the same
matching.

For fixed strict priorities, inserting one offline node cannot reduce
the number of matched online nodes, by
Lemma~\ref{lem:offline-insertion-boundary}. In the equivalent
rank-ordered execution, inserting one online node is an insertion on
the choice side of the transposed instance, with priorities given by
the online arrival order. The same insertion argument therefore shows
that it cannot reduce matching cardinality either. Thus deleting
vertices on either side cannot increase the cardinality returned by
\rank{} under this coupling.

Fix an instance with a maximum matching of size $n>0$, and retain only
the vertices covered by that matching. The induced instance has the
same optimum, a perfect matching covering both sides, and no larger
online degrees. Its expected matching cardinality is no larger than
that of the original instance. Relabel its online nodes in arrival
order as $j_1,\ldots,j_n$, and relabel their perfect-matching partners
as $i_1,\ldots,i_n$.

Delete every edge $(i_k,j_r)$ with $k<r$, and couple the two executions
using the same offline ranks. Immediately before $j_r$, the safe nodes
among $i_r,\ldots,i_n$ in the original execution contain the safe
nodes in the pruned execution. This holds initially. If the original
execution selects a node that is also a safe candidate in the pruned
suffix, it must be the pruned execution's lowest-ranked choice.
Otherwise, its selection lies outside that common safe set. In either
case, the inclusion persists in the next suffix. In particular, every
arrival matched in the pruned execution is also matched in the
original execution.

The pruning preserves every diagonal edge and does not increase
online degrees. It therefore preserves the optimum $n$ while not
increasing the expected matching cardinality. Since every pruned
instance is itself feasible for the original infimum, the forward
restriction leaves $\gamma(d)$ unchanged.
\end{proof}

\subsection{The Global Threshold Identities}
\label{app:large-degree-surplus-identities}

\begin{proof}[\textbf{Proof of Lemma~\ref{lem:large-degree-global-surplus}}]
Fix all ranks other than $Y_{i_r}$. Until $i_r$ is first selected,
the executions on $G$ and $G-i_r$ agree. Thus $i_r$ is matched
exactly when $Y_{i_r}<\beta_r$, apart from ties. Independence of
$\beta_r$ and $Y_{i_r}$ gives
\[
A_G(t)
=\frac1n\sum_{r=1}^n\int_0^t\Pr[\beta_r>y]\,dy
=\int_0^t q_G(y)\,dy.
\]
Each $\beta_r$ is positive almost surely, since $i_r$ has its diagonal
neighbor and every reduced partner rank belongs to $(0,1]$ almost
surely. Thus $q_G(0)=1$. For $t<1$, the number of online nodes whose
selected ranks are at most $t$ equals the number of matched offline nodes with ranks at most
$t$. Hence
\[
\frac1n\sum_{r=1}^n\Pr[R_r\leq t]=A_G(t).
\]
Since $R_r\leq\beta_r$, subtracting their survival probabilities
proves
\[
q_G(t)+A_G(t)-1
=\frac1n\sum_{r=1}^n\Pr[R_r\leq t<\beta_r]\geq0.
\]

We have $A_G(0)=0$ and
\[
\frac{dA_G(t)}{dt}+A_G(t)=1+\mathcal D_G(t)
\]
for almost every $t\in(0,1)$. An integrating factor yields
\begin{equation}
\label{eq:large-degree-A-integrating-factor}
A_G(t)
=1-e^{-t}+e^{-t}\int_0^t e^u\mathcal D_G(u)\,du.
\end{equation}
At $t=1$, this gives the first equality
in~\eqref{eq:large-degree-global-surplus-integral}. Integrating the
event representation and using Tonelli's theorem gives the second:
\[
\int_0^1e^{t-1}\mathcal D_G(t)\,dt
=\frac1n\sum_{r=1}^n
\mathbb E\!\left[\int_{R_r}^{\beta_r}e^{t-1}\,dt\right].
\]
Finally, the nonnegative integral in
\eqref{eq:large-degree-A-integrating-factor} is at most
$\int_0^1 e^u\mathcal D_G(u)\,du=e\delta_G$. Since $e^{-t}\leq1$,
this proves~\eqref{eq:large-degree-global-profile-control}.
\end{proof}

\subsection{Proof of the Finite-Degree Surplus Estimate}
\label{app:large-degree-finite-surplus-proof}

We use the rank-ordered execution from
Lemma~\ref{lem:large-degree-forward-reduction}. At rank time $t$, an
offline node is \emph{unprocessed} if its rank exceeds $t$, and is
\emph{active} if it is unprocessed and has an unmatched online
neighbor. These notions refer to the rank-ordered execution, not to
safety after a prefix of the original online arrival order.

\begin{proof}[\textbf{Proof of Lemma~\ref{lem:large-degree-finite-surplus-estimate}}]
\xhdr{Two nonnegative contributions.} Write
\begin{equation}
\label{eq:large-degree-birth-self-contributions}
\begin{aligned}
\mathcal B_G
&:=\frac1n\sum_{r=1}^n
\mathbb E[e^{\beta_r-1}-e^{\tau_r-1}],\\
\mathcal S_G
&:=\frac1n\sum_{r=1}^n
\mathbb E\!\left[
(e^{\tau_r-1}-e^{R_r-1})
\mathbf1\{j_r\text{ matches }i_r\}\right].
\end{aligned}
\end{equation}
Since $R_r\leq\tau_r\leq\beta_r$, the surplus identity gives
\begin{equation}
\label{eq:large-degree-birth-self-lower}
\delta_G\geq\mathcal B_G+\mathcal S_G.
\end{equation}
The omitted terms are the nonnegative insertion contributions of
nonself matches.

\xhdr{Displacement births and occupation.} An active node is
\emph{displaced} if its diagonal online partner has already been
matched. A \emph{displacement birth} occurs when an unprocessed node
loses its diagonal partner but retains another unmatched neighbor.
Each offline node has at most one such birth. Once it leaves the
displaced active population, it cannot return, because online nodes
only change from unmatched to matched.

Let $N_{\mathsf{disp}}$ be the total number of births, let $Z_G(t)$
be the size of the displaced active population at $t$, and define
\[
\zeta_G:=\frac{\mathbb E[N_{\mathsf{disp}}]}n.
\]
On the event $Y_{i_r}>t$, the rank-ordered executions on $G$ and
$G-i_r$ agree through time $t$. Thus $i_r$ is displaced and active
at $t$ exactly when $\tau_r\leq t<\beta_r$. These two thresholds
are independent of $Y_{i_r}$, so
\begin{equation}
\label{eq:large-degree-displaced-occupation}
\mathbb E[Z_G(t)]
=(1-t)\sum_{r=1}^n\Pr[\tau_r\leq t<\beta_r].
\end{equation}
It follows that
\begin{equation}
\label{eq:large-degree-birth-integral}
\mathcal B_G
=\frac1n\int_0^1\frac{e^{t-1}}{1-t}
\mathbb E[Z_G(t)]\,dt.
\end{equation}

Condition on the rank-exposure history immediately after a birth at
time $s<1$. Choose any unmatched neighbor of the newly displaced
node. At most $d$ of that online node's offline neighbors are still
unprocessed. Their ranks, conditional on the history, are independent
uniforms on $(s,1)$. If none is processed before $t>s$, the chosen
online node remains unmatched, so the displaced node remains active.
The conditional probability of this event is at least
$((1-t)/(1-s))^d$. Thus each birth contributes at least
\[
\int_s^1\frac{e^{t-1}(1-t)^{d-1}}{(1-s)^d}\,dt
\geq\frac{e^{s-1}}d\geq\frac1{ed}
\]
to the unnormalized occupation integral. Summing over births and using
Tonelli's theorem gives
\begin{equation}
\label{eq:large-degree-birth-surplus-bound}
\mathcal B_G\geq\frac{\zeta_G}{ed}.
\end{equation}

\xhdr{Loss rates.} A \emph{loss} occurs when an active offline node
becomes inactive because another node takes its last unmatched
neighbor. It is an \emph{ordinary loss} if that neighbor is the
node's diagonal partner, and a \emph{displaced loss} otherwise.
Every final unmatched offline node suffers exactly one loss.

Let $g_{\mathsf o}(t)$ and $g_{\mathsf b}(t)$ be the expected
ordinary- and displaced-loss rates, divided by $n$. These rates count
individual lost nodes, including simultaneous losses at one selection.
For a finite instance they are well defined for $t<1$; the identities
below may equivalently be read as identities of loss-density measures.

Let $a_G(t)$ count active offline nodes. The matching-threshold
characterization gives
\[
\mathbb E[a_G(t)]=n(1-t)q_G(t).
\]
Conditional on the rank-exposure history, each unprocessed node has
processing rate $1/(1-t)$. Thus active nodes are themselves processed
at expected rate $nq_G(t)$, in addition to the two loss rates.
Differentiating the preceding identity yields
\begin{equation}
\label{eq:large-degree-loss-density}
g_{\mathsf o}(t)+g_{\mathsf b}(t)
=-(1-t)\frac{dq_G(t)}{dt}.
\end{equation}
Every displaced loss follows a distinct displacement birth. Hence
\begin{equation}
\label{eq:large-degree-displaced-loss-count}
\int_0^1 g_{\mathsf b}(t)\,dt\leq\zeta_G.
\end{equation}

\xhdr{A self-match replacement estimate.} For fixed $t\in[0,1)$,
let $\mathcal A_r(t)$ be the event, over the ranks other than
$Y_{i_r}$, that inserting $i_r$ with rank $t$ makes $j_r$ select
$i_r$. Let $S_r^-$ be the set of safe neighbors of $j_r$ immediately
before its arrival in the reduced chronological execution on $G-i_r$,
and let $m_r^-:=|S_r^-|$.

The event $\mathcal A_r(t)$ requires all earlier incident reduced
partner ranks to be at most $t$ and every rank in $S_r^-$ to exceed
$t$. We claim that
\begin{equation}
\label{eq:large-degree-self-replacement-cost}
\mathcal S_G\geq
\frac1n\sum_{r=1}^n\int_0^1 e^{t-1}(1-t)
\mathbb E\!\left[
\frac{\mathbf1_{\mathcal A_r(t)}}{m_r^-+1}\right]dt.
\end{equation}

To prove the claim, condition on the full reduced chronological
transcript before $j_r$, including the identities and ranks of all
selected nodes. The identities of the remaining safe nodes are then
fixed. For each such node, the transcript imposes only lower bounds
on its rank: its rank must exceed the selected rank at each earlier
arrival where it was an available alternative. Consequently, the
remaining safe ranks are conditionally independent uniforms on
intervals truncated from below.

The restrictions on earlier incident partner ranks in $\mathcal A_r(t)$
are fixed by the transcript. Its remaining restrictions truncate the
ranks in $S_r^-$ further from below at $t$, preserving conditional
independence. Thus, conditional on the transcript and
$\mathcal A_r(t)$, these ranks stochastically dominate independent
uniforms on $[t,1]$. For $t\leq u<1$,
\[
\Pr[\tau_r>u\mid\text{transcript},\mathcal A_r(t)]
\geq\left(\frac{1-u}{1-t}\right)^{m_r^-}.
\]
The conditional expected replacement cost is therefore at least
\[
\int_t^1 e^{u-1}
\left(\frac{1-u}{1-t}\right)^{m_r^-}\,du
\geq\frac{e^{t-1}(1-t)}{m_r^-+1}.
\]
This also holds when $m_r^-=0$, since then $\tau_r=1$.
On a self-match $R_r=Y_{i_r}$, and $Y_{i_r}$ is independent of the
reduced execution. Integrating over $Y_{i_r}=t$ proves
\eqref{eq:large-degree-self-replacement-cost}. Conditional statements
here are made for almost every conditioning value under its law.

\xhdr{Relating ordinary losses to available competitors.} Define
\begin{equation}
\label{eq:large-degree-surviving-neighbor-count}
V_r(t):=
\sum_{i_k\in N_G(j_r)\setminus\{i_r\}}
\mathbf1\{Y_{i_k}>t\}.
\end{equation}
Marginally,
$V_r(t)\sim\operatorname{Bin}(\deg_G(j_r)-1,1-t)$.
On $\mathcal A_r(t)$, every member of $S_r^-$ has rank above $t$,
so $m_r^-\leq V_r(t)$.

Let $C_r(t)$ count the unprocessed nodes in the reduced rank-ordered
execution whose earliest unmatched neighbor is $j_r$. Let $g_r(t)$
be the ordinary-loss rate associated with $i_r$, without division by
$n$. Because $j_r$ is the last neighbor of $i_r$, the condition
$\mathcal A_r(t)$ is equivalent to saying that $j_r$ is its only
unmatched neighbor in the reduced rank-ordered execution at $t$.
On $Y_{i_r}>t$, that execution agrees with the original one through
$t$. Each of the $C_r(t)$ competitors has processing rate
$1/(1-t)$, while the independent event $Y_{i_r}>t$ has probability
$1-t$. These factors cancel, giving
\begin{equation}
\label{eq:large-degree-ordinary-loss-competitors}
g_r(t)=\mathbb E[C_r(t)\mathbf1_{\mathcal A_r(t)}]
\leq\mathbb E[V_r(t)\mathbf1_{\mathcal A_r(t)}],
\qquad
g_{\mathsf o}(t)=\frac1n\sum_{r=1}^n g_r(t).
\end{equation}
No independence between $V_r(t)$ and $\mathcal A_r(t)$ is used.

\xhdr{Concentration and integration.} Fix the parameters $T,\eta$
of the lemma, and set
\[
M_d(t):=(1+\eta)d(1-t),\qquad 0\leq t\leq T.
\]
The variable $V_r(t)$ is stochastically dominated by
$\operatorname{Bin}(d-1,1-t)$. The multiplicative binomial tail bound
therefore gives
\begin{equation}
\label{eq:large-degree-neighbor-tail}
\Pr[V_r(t)>M_d(t)]
\leq\exp\!\left(-\frac{\eta^2(1-t)(d-1)}{2+\eta}\right)
\leq\epsilon_{d,T,\eta}.
\end{equation}
Since $V_r(t)\leq d$, equation
\eqref{eq:large-degree-ordinary-loss-competitors} implies
\[
g_r(t)
\leq M_d(t)\Pr[\mathcal A_r(t),\ V_r(t)\leq M_d(t)]
+d\epsilon_{d,T,\eta}.
\]
Together with $m_r^-\leq V_r(t)$ on $\mathcal A_r(t)$, this yields
\begin{equation}
\label{eq:large-degree-self-competitor-bound}
\mathbb E\!\left[
\frac{\mathbf1_{\mathcal A_r(t)}}{m_r^-+1}\right]
\geq
\frac{g_r(t)-d\epsilon_{d,T,\eta}}
{M_d(t)(M_d(t)+1)}.
\end{equation}
Substitute into~\eqref{eq:large-degree-self-replacement-cost}, sum over
$r$, and use~\eqref{eq:large-degree-loss-density}. Because
$-dq_G(t)/dt\geq0$,
\begin{equation}
\label{eq:large-degree-self-estimate-explicit}
\begin{aligned}
\mathcal S_G\geq{}&
\frac{\displaystyle\int_0^T e^{t-1}
\left(-\frac{dq_G(t)}{dt}\right)\,dt}
{(1+\eta)d\bigl((1+\eta)d+(1-T)^{-1}\bigr)}-\frac{\zeta_G}{d^2(1-T)}
-\frac{\epsilon_{d,T,\eta}}d\ln\frac1{1-T}.
\end{aligned}
\end{equation}
For the first term, we used
\[
\frac{(1-t)^2}{M_d(t)(M_d(t)+1)}
\geq\frac1{(1+\eta)d((1+\eta)d+(1-T)^{-1})}.
\]
For the displaced-loss term, its coefficient is at most
$1/[d^2(1-T)]$, and
\eqref{eq:large-degree-displaced-loss-count} bounds its integral by
$\zeta_G$. For the binomial-tail error, the coefficient after
multiplication by $d$ is at most $1/[d(1-t)]$; its integral gives
the logarithm in~\eqref{eq:large-degree-self-estimate-explicit}.

Finally, combine~\eqref{eq:large-degree-birth-self-lower},
\eqref{eq:large-degree-birth-surplus-bound}, and
\eqref{eq:large-degree-self-estimate-explicit}. The coefficient of
$\zeta_G$ in the resulting lower bound is
\[
\frac1{ed}-\frac1{d^2(1-T)}\geq0
\]
by the assumed condition $d(1-T)\geq e$. Dropping this nonnegative
term proves~\eqref{eq:large-degree-finite-surplus-estimate}.
\end{proof}

\subsection{The Layered Upper-Bound Family}
\label{app:large-degree-layered-proof}

\begin{proof}[\textbf{Proof of Lemma~\ref{lem:large-degree-layered-upper}}]
\xhdr{A finite instance.} Fix a depth $L\geq1$, and introduce
$d^L$ offline nodes indexed by tuples
$(x_1,\ldots,x_L)\in[d]^L$. At layer $k\in[L]$, the roots still
designated for future use satisfy $x_1,\ldots,x_{k-1}\in[d-1]$.
Group these roots by fixing every coordinate except $x_k$, and send
one online arrival adjacent to each group of $d$ roots. The order of
groups within a layer is arbitrary and fixed in advance. Predetermine
that the root with $x_k=d$ receives no subsequent edges. After the
last layer, append a singleton arrival for each root still designated
for future use. These designations are independent of all ranks and
matching decisions.

Assign each layer arrival to its predetermined retiring root, and
assign each final singleton to its neighbor. This is a perfect
matching covering all $d^L$ offline nodes. Thus the instance has
optimum $d^L$ and online degree at most $d$.

\xhdr{Safety probabilities.} Before a layer-$k$ group is processed,
the histories of its $d$ roots are supported on disjoint rank sets:
for each root, earlier layers varied only the first $k-1$ coordinates,
while the group's different $x_k$ coordinates remained fixed.
The histories are isomorphic and hence identically distributed.
Other groups in the current layer do not change the safety of these
roots. Thus the $d$ input rank-and-safety collections are independent
and identically distributed.

If the common safety probability is $u_{k-1}$, the group arrival is
matched with probability $1-(1-u_{k-1})^d$. Symmetry assigns one
$d$th of that probability to each root. The common output safety
probability is therefore
\[
u_k=u_{k-1}-\frac{1-(1-u_{k-1})^d}{d},
\]
with $u_0=1$. This proves~\eqref{eq:large-degree-layered-recurrence}.
It uses independence of the complete input histories, not an assumption
that the ranks of safe nodes remain uniform.

At layer $k+1$, the fraction of original offline nodes that retire is
$d^{-1}(1-1/d)^k$. A retiring node remains unmatched exactly when it
is still safe after that layer's arrival. The final singletons ensure
that every nonretired root is eventually matched. Hence the expected
matching size, normalized by the optimum, is
\begin{equation}
\label{eq:large-degree-finite-layered-ratio}
1-\frac1d\sum_{k=0}^{L-1}
\left(1-\frac1d\right)^ku_{k+1}.
\end{equation}
The sequence $u_k$ is nonnegative and nonincreasing, because
$0\leq1-(1-u)^d\leq du$ for $u\in[0,1]$.
The distance between~\eqref{eq:large-degree-finite-layered-ratio}
and its infinite-depth limit is at most
$(1-1/d)^L u_L$, which tends to zero. Taking $L\to\infty$ proves
the inequality in~\eqref{eq:large-degree-layered-upper}.

\xhdr{Asymptotics of the recurrence.} Put
$v_k:=(1-k/d)_+$. Since each step decreases $u_k$ by at most $1/d$,
we have $u_k\geq v_k$. A finite geometric-sum calculation gives
\begin{equation}
\label{eq:large-degree-linear-profile-sum}
1-\frac1d\sum_{k=0}^{\infty}
\left(1-\frac1d\right)^kv_{k+1}
=1-\left(1-\frac1d\right)^d.
\end{equation}
For $1\leq k\leq d$, telescoping the recurrence gives
\[
0\leq u_k-v_k
=\frac1d\sum_{r=0}^{k-1}(1-u_r)^d
\leq\frac1d\sum_{r=0}^{k-1}(r/d)^d
<\frac{e^{k-d}}{d(e-1)}.
\]
The last inequality uses $(r/d)^d\leq e^{r-d}$ and a geometric
sum. Thus
\begin{equation}
\label{eq:large-degree-layered-initial-error}
\sum_{k=1}^d(u_k-v_k)<\frac{e}{d(e-1)^2},
\qquad
u_d<\frac1{d(e-1)}.
\end{equation}
The elementary lower bound
$1-(1-u)^d\geq du-\binom d2u^2$ implies
\[
u_{k+1}\leq\frac{d-1}{2}u_k^2<\frac{u_k}{3}
\qquad(k\geq d).
\]
Consequently,
$\sum_{k=d+1}^{\infty}u_k<u_d/2<1/[2d(e-1)]$.
By~\eqref{eq:large-degree-linear-profile-sum} and
\eqref{eq:large-degree-layered-initial-error}, the difference between
$1-(1-1/d)^d$ and the middle expression in
\eqref{eq:large-degree-layered-upper} is positive and smaller than
\[
\frac1{d^2}
\left(\frac{e}{(e-1)^2}+\frac1{2(e-1)}\right)
<\frac2{d^2}.
\]
Positivity follows, for example, from $u_2>v_2$ when $d\geq2$.
Finally,
\[
\left(1-\frac1d\right)^d
=e^{-1}\left(1-\frac1{2d}+O(d^{-2})\right),
\]
which proves the asymptotic expression and completes the lemma.
\end{proof}

\section{Auxiliary Proofs for the Two-Regular Certifying Assertions in
Theorem~\ref{coro:Gamma-two-regular}}
\label{app:regular-two-certifying-auxiliary}

We prove the regular-instance obstruction bounds used in
Subsection~\ref{subsec:regular-two-upper-bound} and the local
probability inequality underlying the constant certificate in
Subsection~\ref{subsec:regular-two-lower-bound}. All instances below
belong to $\mathcal B_2$, and their scores are evaluated after the
complete execution.

\subsection{The Regular-Instance Obstructions}
\label{app:regular-two-obstruction-proofs}

\begin{proof}[\textbf{Proof of the obstruction bounds in
Lemma~\ref{lem:regular-two-upper-bound}}]
\xhdr{The obstruction $H_f(1)$.} Fix an integer $m\geq1$ and
introduce distinct offline nodes
$i_0,\ldots,i_m,\tilde i_0,\ldots,\tilde i_m$. The online
neighborhoods are
\[
\begin{aligned}
N_G(j_0)&=\{i_0,\tilde i_0\},
&N_G(j^\star)&=\{i_m,\tilde i_m\},\\
N_G(j_r)&=\{i_{r-1},i_r\},
&N_G(\tilde j_r)&=\{\tilde i_{r-1},\tilde i_r\},
\qquad r\in[m].
\end{aligned}
\]
Use the fixed arrival order
\[
j_0\prec j_1\prec\cdots\prec j_m
\prec\tilde j_1\prec\cdots\prec\tilde j_m\prec j^\star,
\]
and distinguish $(i_m,j^\star)$. Every offline and online node has
degree two.

On the event of distinct ranks, the endpoint safety events before
$j^\star$ are
\begin{align}
\mathcal E_{i_m}
&=\{Y_{\tilde i_0}<Y_{i_0}<Y_{i_1}<\cdots<Y_{i_m}\},
\label{eq:regular-cycle-left-safety}\\
\mathcal E_{\tilde i_m}
&=\{Y_{i_0}<Y_{\tilde i_0}<Y_{\tilde i_1}
<\cdots<Y_{\tilde i_m}\}.
\label{eq:regular-cycle-right-safety}
\end{align}
Indeed, after each arm extension, its new endpoint remains safe
exactly when the preceding endpoint was safe and had smaller rank.
The two events impose opposite comparisons at $j_0$, so they are
mutually exclusive. Each specifies one strict ordering of $m+2$
independent ranks and has probability $1/(m+2)!$.

The offline node $i_m$ is matched by the end: either it was matched
earlier, or it is the only safe neighbor of $j^\star$. The final
arrival is matched with probability $2/(m+2)!$. Since $0\leq f\leq1$,
\[
\begin{aligned}
0\leq\Phi_G(i_m,j^\star;f)-H_f(1)
&=\mathbb E[(1-f(Y_{j^\star}))X_{j^\star}]
\leq\frac{2}{(m+2)!}.
\end{aligned}
\]
Letting $m\to\infty$ proves
\begin{equation}
\label{eq:regular-two-H-upper}
\tilde\chi(f;2)\leq H_f(1).
\end{equation}

\xhdr{The obstruction $\Psi_{1,0}(f)$.} Start with the power-filter
gadget $G_{1,0}$. Its offline nodes are
$i^\star,\tilde i^\star,i_1$, and the first two online arrivals
have neighborhoods $\{i^\star,i_1\}$ and
$\{i^\star,\tilde i^\star\}$, respectively. The second arrival
is $j^\star$, and $(i^\star,j^\star)$ is distinguished.
Append one arrival with neighborhood $\{i_1,\tilde i^\star\}$.
The complete instance is a six-cycle in $\mathcal B_2$.

The added arrival is not incident to $i^\star$, so it changes
neither the final matched status of $i^\star$ nor the contribution
of $j^\star$. Lemma~\ref{lem:power-filter-gadget-score} therefore
gives
\begin{equation}
\label{eq:regular-two-Psi-10}
\Phi_G(i^\star,j^\star;f)=\Psi_{1,0}(f)
=1-\int_0^1\left(y-\frac{y^2}{2}\right)f(y)\,dy.
\end{equation}
Hence $\tilde\chi(f;2)\leq\Psi_{1,0}(f)$. Combining this with
\eqref{eq:regular-two-H-upper} proves
\eqref{eq:regular-two-upper-obstructions}.
\end{proof}

\subsection{The Local Probability Certificate}
\label{app:regular-two-probability-proof}

\begin{proof}[\textbf{Proof of~\eqref{eq:regular-two-probability-certificate}}]
Fix $G\in\mathcal B_2$ and a distinguished edge
$(i^\star,j^\star)$. We distinguish whether $j^\star$ is the
earlier or later of the two online neighbors of $i^\star$.

\xhdr{$j^\star$ is the earlier neighbor.} The offline node
$i^\star$ is certainly safe when $j^\star$ arrives, so
$\Pr[X_{j^\star}=1]=1$. Let $\hat i$ and $\tilde i$ be the
other offline neighbors of the two arrivals incident to $i^\star$.
If $\hat i=\tilde i$, the same alternative cannot be selected by
both arrivals, so $i^\star$ is matched by the end.

Otherwise, for $i^\star$ to remain unmatched, both arrivals must
select their respective alternatives while $i^\star$ is safe.
Thus
\[
\begin{aligned}
\Pr[X_{i^\star}=0]
&\leq\Pr[Y_{\hat i}<Y_{i^\star},\ Y_{\tilde i}<Y_{i^\star}]
=\int_0^1y^2\,dy=\frac13.
\end{aligned}
\]
This proves the required inequality in this case.

\xhdr{$j^\star$ is the later neighbor.} Let $\tilde i^\star$ be
its other offline neighbor, and let $\mathcal E_{i^\star}$ and
$\mathcal E_{\tilde i^\star}$ be their safety events immediately
before $j^\star$. If these events have intersection probability
zero, $i^\star$ is matched by the end almost surely: whenever it
is still safe, it is the only safe neighbor of $j^\star$.

Otherwise, Lemma~\ref{lem:independent-or-exclusive-safety-events}
and Corollary~\ref{cor:monomial-safety-profile-Gamma} give disjoint
witness sets and monomial conditional safety probabilities
\[
\Pr[\mathcal E_{i^\star}\mid Y_{i^\star}=x]=ax^p,
\qquad
\Pr[\mathcal E_{\tilde i^\star}\mid Y_{\tilde i^\star}=y]=by^q,
\]
where $a,b\in(0,1]$ and $p,q\in\mathbb Z_{\geq0}$. In fact,
$p\geq1$. Surviving the earlier neighbor of $i^\star$ requires
that arrival to select its other offline neighbor at a smaller rank.
Consequently $ax^p\leq x$ for $0<x<1$, which excludes $p=0$
because $a>0$.

The node $i^\star$ has no neighbor after $j^\star$. It remains
unmatched exactly when both roots are safe and
$Y_{\tilde i^\star}<Y_{i^\star}$. Independence of the witness
sets therefore gives
\[
\begin{aligned}
\Pr[X_{i^\star}=0]
&=\int_0^1ax^p\left(\int_0^xby^q\,dy\right)dx
=\frac{ab}{(q+1)(p+q+2)}
\leq\frac{b}{3(q+1)}.
\end{aligned}
\]
The last inequality uses $a\leq1$ and $p+q+2\geq3$. Whenever
$\tilde i^\star$ is safe, $j^\star$ is matched, so
\[
\Pr[X_{j^\star}=1]
\geq\Pr[\mathcal E_{\tilde i^\star}]
=\frac{b}{q+1}\geq3\Pr[X_{i^\star}=0].
\]
This proves~\eqref{eq:regular-two-probability-certificate} without
truncating the regular instance.
\end{proof}


\section{Auxiliary Proofs for the Two-Regular Competitiveness Assertion in
Theorem~\ref{coro:Gamma-two-regular}}
\label{app:regular-two-competitiveness-auxiliary}

We establish the cycle decomposition and endpoint probabilities used
in Section~\ref{sec:proof-regular-two-competitiveness}, justify the
peak accounting, and evaluate the tight cycle. The final calculation
also records how the certifying gap appears as aggregate edge slack.

\subsection{Cycle Decomposition and Endpoint States}
\label{app:regular-two-block-proof}

\begin{proof}[\textbf{Proof of Lemma~\ref{lem:reg-cycle-blocks}}]
Use the cyclic labeling from Subsection~\ref{subsec:reg-two-cycle-analysis}.
A nonpeak $j_r$ has a neighboring online node that arrives later.
The offline node they share has not faced any earlier arrival, because
its only two neighbors are those two online nodes. It is therefore
safe when $j_r$ arrives, and $j_r$ is matched.

Each peak arrives after both neighboring online nodes and is the last
arrival incident to either of its offline neighbors. Its decision
therefore affects no later arrival. Distinct peaks cannot be adjacent
on the cycle and hence share no offline neighbor. We may postpone all
of them to the end, preserving the order of the remaining arrivals,
without changing any matching decision.

Remove the peak online nodes and their incident edges. The resulting
components are alternating paths whose endpoints are offline nodes.
Between successive peaks, the sequence of original online arrival
times first decreases and then increases: a change from increasing
to decreasing would create another peak. It follows that each block
has one minimum-time online node and that arrival times increase along
both arms away from it. This also applies when there is only one peak,
with the remaining path running around the rest of the cycle.

The minimum arrival of a block has two fresh offline neighbors. Call
them $i_0$ and $\hat i_0$. On one arm, let $i_1,\ldots,i_{r-2}$
be the successive fresh offline nodes exposed by its $r-2$ extensions.
Each extension joins the preceding endpoint to its new endpoint.
The final endpoint of this arm is safe exactly on the order event
\[
Y_{\hat i_0}<Y_{i_0}<Y_{i_1}<\cdots<Y_{i_{r-2}},
\]
with the event interpreted as $Y_{\hat i_0}<Y_{i_0}$ when $r=2$.
It has probability $1/r!$. Fixing the endpoint rank to $y$ gives
$y^{r-1}/(r-1)!$, proving~\eqref{eq:reg-chain-conditional};
these conditional probabilities use the order-event versions.

The opposite arm requires the reverse comparison between the two
initial roots, so the block's endpoint safety events are mutually
exclusive. Different blocks have disjoint offline vertex sets, and
their endpoint events depend only on those respective rank sets.
They are therefore independent across blocks.

If the other arm has $s-2$ extensions, the total number of offline
nodes is $2+(r-2)+(s-2)=r+s-2$. This proves every assertion.
\end{proof}

\subsection{Peak Failure Probabilities and Global Accounting}
\label{app:regular-two-peak-accounting}

Let $j$ be a postponed peak, with endpoint chain lengths $r_j,s_j$.
When there are at least two blocks, these endpoints belong to distinct
blocks, and their safety events are independent by
Lemma~\ref{lem:reg-cycle-blocks}. Their respective probabilities are
$1/r_j!$ and $1/s_j!$. The peak is unmatched exactly when neither
endpoint is safe, proving~\eqref{eq:reg-peak-failure-exact}.

When there is one block, its two endpoint events are mutually
exclusive. The unique peak therefore has failure probability
\[
\Pr[X_j=0]=1-\frac1{r_j!}-\frac1{s_j!}
<\left(1-\frac1{r_j!}\right)\left(1-\frac1{s_j!}\right).
\]
All these events may be evaluated just before the first postponed peak:
no peak is incident to an endpoint of another peak, so later peak
decisions do not change the available endpoints of $j$.

Assign an endpoint of chain length $r$ the weight $r-1$. The two
endpoint weights of each block sum to its number of offline nodes.
The blocks partition all $n$ offline nodes, so their endpoint weights
sum to $n$. Each endpoint is incident to exactly one removed peak;
regrouping the same weights by peaks proves
\[
\sum_{j\text{ peak}}(r_j+s_j-2)=n,
\]
which is~\eqref{eq:reg-chain-accounting}.

\subsection{The Tight Cycle and Its Edge Scores}
\label{app:regular-two-tight-cycle-proof}

Consider the instance in~\eqref{eq:reg-eight-cycle}. Every offline
and online node has degree two, and the following edges form a perfect
matching:
\[
\begin{aligned}
M^\star=\{&(i_2,j_1),(\tilde i_2,j_2),(i_1,j_3),(i_3,j_4),\\
&(\tilde i_1,j_5),(\tilde i_3,j_6),(i_4,j_7),(\tilde i_4,j_8)\}.
\end{aligned}
\]
Thus its optimum is eight. The first six arrivals are always matched:
$j_1,j_2$ each start with two fresh neighbors, and each of
$j_3,\ldots,j_6$ has a fresh endpoint.

Immediately before $j_7$, the endpoint events in the first block are
\[
\mathcal E_{i_1}=\{Y_{i_3}<Y_{i_2}<Y_{i_1}\},
\qquad
\mathcal E_{i_4}=\{Y_{i_2}<Y_{i_3}<Y_{i_4}\}.
\]
The two corresponding events in the other block have the same forms
with tilded indices. Each endpoint is safe with probability $1/6$;
the endpoint events within a block are mutually exclusive, while the
two blocks depend on disjoint rank sets.

The arrival $j_7$ joins $i_4$ and $\tilde i_1$, whose events are
independent, so
\[
\Pr[X_{j_7}=1]=1-\left(1-\frac16\right)^2=\frac{11}{36}.
\]
Its decision does not change the matched status of either neighbor
of $j_8$. Those two neighbors, $\tilde i_4$ and $i_1$, likewise
have independent safety events, so $\Pr[X_{j_8}=1]=11/36$.
Linearity of expectation yields
\[
\frac{\mathbb E[\rank(C)]}{\operatorname{OPT}(C)}
=\frac{6+2(11/36)}8=\frac{119}{144},
\]
proving~\eqref{eq:regular-two-tight-cycle-value}.

\xhdr{Scores on the perfect matching.} Each internal offline node
$i_2,i_3,\tilde i_2,\tilde i_3$ is unmatched exactly when its
rank exceeds both alternatives faced at its two incident arrivals.
Its matching probability is therefore $2/3$.
An endpoint of either block remains unmatched exactly when it and
the alternative endpoint at its final peak are both safe and the
alternative has smaller rank. Using their independent chain-length-three
profiles gives
\[
\Pr[X_i=0]
=\int_0^1\frac{y^2}{2}
\left(\int_0^y\frac{z^2}{2}\,dz\right)dy=\frac1{72}.
\]
Thus each endpoint is matched with probability $71/72$.
Under $\tilde f_2^\star$, the score of an edge $(i,j)$ is
$\frac34\Pr[X_i=1]+\frac14\Pr[X_j=1]$. The eight edges of
$M^\star$ consequently have the following scores:
\begin{equation}
\label{eq:regular-two-tight-cycle-score-types}
\begin{array}{c|c|c}
\text{offline/online type}&\text{number of edges}&\text{score}\\ \hline
\text{internal / always matched}&4&3/4\\
\text{endpoint / always matched}&2&95/96\\
\text{endpoint / peak}&2&235/288.
\end{array}
\end{equation}
Their minimum is $3/4$, whereas their average is $119/144$.
The positive excess in this average is exactly $11/144$.

\section{Auxiliary Proofs for the Certifying Assertions for Large Regular Degree in
Theorem~\ref{thm:Gamma-regular-large-degree}}
\label{app:regular-large-degree-certifying-auxiliary}

This appendix proves the full-history threshold bound, the
regularization lemma, compactness and continuity, and the additional
properties of the common-pool mixture used in
Section~\ref{app:Gamma-regular-large-degree}.

\subsection{The Full-History Threshold Certificate}
\label{app:regular-full-history-proof}

\begin{proof}[\textbf{Proof of Lemma~\ref{lem:regular-full-history-threshold}}]
\xhdr{The matching threshold.} Fix all ranks other than $Y_i$.
Until $i$ is first selected, the executions on $G$ and $G-i$ agree.
Consequently, inserting $i$ with rank $y<1$ leaves it unmatched exactly
when every reduced partner rank at an incident arrival is at most $y$,
apart from ties. For each fixed $y<1$, ties with the other random ranks
have probability zero. Thus, under the conditional version obtained by
fixing $Y_i=y$ and sampling all other ranks independently,
\begin{equation}
\label{eq:regular-threshold-survival-identity}
\Pr[\beta_i\leq y]=\Pr[X_i=0\mid Y_i=y],
\qquad y\in[0,1).
\end{equation}

\xhdr{Survival through the incident arrivals.} Fix $Y_i=y$. Reveal the
identity and rank of the selected offline node after each matched
arrival. Conditional on the transcript, the ranks of the currently safe
nodes other than $i$ are independent uniforms on intervals
$[\ell_{\hat i},1]$. Indeed, whenever an arrival selects rank $x$,
each adjacent node that remains safe acquires only the separate
constraint that its own rank exceeds $x$. Conditioning on selected
identities and ranks therefore preserves this product form.

When a neighbor of $i$ arrives while $i$ is safe, there are at most
$d-1$ safe alternatives. Each satisfies
\[
\Pr[Y_{\hat i}\geq y\mid\text{transcript},Y_i=y]\geq1-y.
\]
Conditional independence implies that all alternatives have rank at
least $y$ with probability at least $(1-y)^{d-1}$; then the arrival
selects $i$. At each of the $d$ incident arrivals, conditional on
survival up to that point, the probability of surviving again is at
most $1-(1-y)^{d-1}$. Iteration and
\eqref{eq:regular-threshold-survival-identity} give
\[
\Pr[\beta_i\leq y]=\Pr[X_i=0\mid Y_i=y]
\leq\left(1-(1-y)^{d-1}\right)^d=S_d(y).
\]

\xhdr{The edge score.} Condition again on all ranks other than $Y_i$.
The matching threshold is independent of $Y_i$, so the conditional
expected offline charge is $F_f(\beta_i)$. For every incident arrival
$j$, insertion monotonicity gives
\[
R_G(j)\leq R_{G-i}(j)\leq\beta_i.
\]
Monotonicity of $f$ and the normalization $f(1)=1$ imply that the
online charge is at least $1-f(\beta_i)$, including the unmatched
case. The conditional expected edge charge is therefore at least
$F_f(\beta_i)+1-f(\beta_i)=H_f(\beta_i)$. Averaging proves
\eqref{eq:regular-full-history-score-bound}.
\end{proof}

\subsection{Regularization of an Edge-Score Obstruction}
\label{app:regularization-proof}

\begin{proof}[\textbf{Proof of Lemma~\ref{lem:regularization-edge-score}}]
We construct $H$ in three stages, keeping the order of all original
online nodes unchanged. The construction is independent of $f$.

\xhdr{Saturating the distinguished offline degree.} Prepend
$d-\deg_G(i^\star)$ arrivals, each adjacent to $i^\star$ and
$d-1$ fresh private offline nodes. The private sets are disjoint.
Conditional on $Y_{i^\star}=y$, the probability that these arrivals
match $i^\star$ is
\[
1-\left(1-(1-y)^{d-1}\right)^{d-\deg_G(i^\star)}
\leq\min\{1,d(1-y)^{d-1}\}=b_d(y).
\]
The degree of $i^\star$ is now exactly $d$.

Couple the original and modified executions with the same original
ranks. If the preliminary arrivals do not match $i^\star$, the
execution on the original graph is unchanged. Otherwise, the original
arrival sequence runs on the remaining original nodes exactly as on
$G-i^\star$. Insertion monotonicity and monotonicity of $f$ imply
that this deletion cannot increase the online charge of $j^\star$.
The offline charge can increase only when $i^\star$ was unmatched
at the end of the original execution, and the increase is then
$f(Y_{i^\star})$.

An originally unmatched $i^\star$ necessarily satisfies
$Y_{i^\star}\geq\tau_G$, apart from ties. Since the preliminary
private ranks are independent of the original ranks, the expected
score increase is at most
\begin{equation}
\label{eq:regularization-preliminary-cost}
\int_0^1 f(y)b_d(y)\Pr[\tau_G\leq y]\,dy.
\end{equation}

\xhdr{Saturating the online degrees.} Fill every missing neighbor slot
of an original online node with the root of a separate padding tree.
For a depth $L$, this tree is a rooted $(d-1)$-ary tree of offline
nodes, with one online node adjacent to each internal node and its
$d-1$ children. Process its online nodes from the leaves upward.
Every tree arrival has degree $d$, every offline degree is at most two,
and the root has degree one before attachment.

If the root remains safe after these arrivals, its incident arrival
must have selected a safe child of smaller rank. Repeating this argument
down the tree gives a strictly decreasing rank path from root to leaf.
There are $(d-1)^L$ such paths, each involving $L+1$ independent
uniform ranks. Hence
\begin{equation}
\label{eq:regularization-padding-failure}
\Pr[\text{padding root remains safe}]
\leq\frac{(d-1)^L}{(L+1)!}\longrightarrow0
\qquad(L\to\infty).
\end{equation}

Use disjoint copies for the finitely many missing neighbor slots, and
process all padding-tree arrivals before the preliminary arrivals of
the first stage. The attached root has degree two, at most $d$.
Choose the depths so that the probability that any attached root remains
safe is at most $\varepsilon/2$. On the complementary event, the
original arrivals see exactly the same safe original neighbors as after
the first stage. A realized edge charge lies in $[0,2]$, so this stage
changes its expectation by at most $\varepsilon$, simultaneously for
all $f\in\mathcal F$.

At this point every online node has degree $d$, every offline node has
degree at most $d$, and $i^\star$ has degree exactly $d$.

\xhdr{Completing the remaining offline degrees.} Let $I^+,J^+$ be
the current sides. Degree counting gives $|I^+|\geq|J^+|$, with total
offline deficit $d(|I^+|-|J^+|)$. Add $d$ new offline nodes and
$|I^+|-|J^+|+d$ new online nodes. The total offline deficit is now
$d(|I^+|-|J^+|+d)$.

List all deficit slots consecutively, keeping the slots of each offline
node together, and assign the list cyclically to the new online nodes.
Each offline deficit is at most $d$, and there are at least $d$ new
online nodes. Thus an offline node is never assigned twice to the same
new online node, and every new online node receives exactly $d$ edges.
The resulting simple bipartite graph is $d$-regular.

Process all completion arrivals after the original sequence. They do
not change the charge of $j^\star$, and none is adjacent to the already
saturated node $i^\star$. They therefore do not change its charge either.
Combining~\eqref{eq:regularization-preliminary-cost} with the padding
error proves~\eqref{eq:regularization-edge-score}.
\end{proof}

\subsection{Compactness and Continuity of the Edge Scores}
\label{app:regular-compactness-proof}

\begin{proof}[\textbf{Proof of Lemma~\ref{lem:regular-profile-compactness}}]
The $L^1$ distance is a metric on $\mathcal F$: if two profiles differ
at some $t<1$, right-continuity preserves a nonzero difference on a
short interval to its right. They cannot differ only at $1$, because
both terminal values are fixed to one.

Given a sequence in $\mathcal F$, diagonal selection at the rational
points gives a subsequence converging at every continuity point of a
nondecreasing $[0,1]$-valued limit. Choose its right-continuous
representative on $[0,1)$ and set its terminal value to one. The
resulting profile belongs to $\mathcal F$. Its discontinuities are
at most countable, so dominated convergence gives $L^1$ convergence.
This proves sequential compactness and hence compactness in the metric.

For the edge score, the offline difference is at most
$\mathbb E[|f(Y_i)-g(Y_i)|X_i]\leq\|f-g\|_1$. When $j$ is matched,
its partner is one of its offline neighbors, so
\[
\mathbb E[|f(Y_j)-g(Y_j)|X_j]
\leq\sum_{\hat i\in N_G(j)}
\mathbb E[|f(Y_{\hat i})-g(Y_{\hat i})|]
=|N_G(j)|\,\|f-g\|_1.
\]
This proves~\eqref{eq:regular-fixed-edge-L1}. Taking the infimum over
regular graph-edge pairs gives the $(d+1)$-Lipschitz bound on
$\tilde\chi(\cdot;d)$. Its continuity and compactness of $\mathcal F$
imply attainment.
\end{proof}

\subsection{The Common-Pool Mixture and Its Reduced Thresholds}
\label{app:regular-core-mixture-proof}

\begin{proof}[\textbf{Proof of Lemma~\ref{lem:regular-core-mixture}}]
Apply Lemma~\ref{lem:large-degree-positive-mixture} with $D=d-1$
and $N$ chosen as in the proof of
Lemma~\ref{lem:large-degree-upper-bound}. Thus $N=\Theta(\ln D)$,
$N\geq2$, and $2^{-N}\leq D^{-3}$ for sufficiently large $D$.
Its components are the common-pool graphs, residual complete bipartite
graphs $K_{D+1,k}$, and the two-arrival singleton graph. The uniform
bound on their mixed kernel $\breve K_{D,N}$, integrated against
$\pi_f$, proves~\eqref{eq:regular-core-mixture-score}.

We also need the offline-degree bound. In a common-pool graph with
$q$ initial pool-wide arrivals, each shared node has degree $q+N$,
each private node has degree $q+2$, and the distinguished node has
degree three. Since $0\leq q\leq D-N$ and $N\geq2$, these degrees
are at most $d$ for sufficiently large $d$. The complete graph
$K_{D+1,k}$ has offline degree $k\leq D$, and the singleton graph
has offline degree two. All online degrees are at most $d$ by the
construction.

After deleting the distinguished node, the common-pool component has
final reduced threshold $T_q$, the complete component $K_{D+1,k}$
has threshold $W_{(k:D)}$, and the singleton component has threshold
one. Hence the positive decomposition
\eqref{eq:large-degree-positive-decomposition} gives
\[
\sum_{r=1}^{m_d}\lambda_r\mathcal L(\tau_r)=\nu_D,
\]
where $\nu_D$ is the target measure in
\eqref{eq:large-degree-target-measure}. The density estimate in the
proof of Lemma~\ref{lem:large-degree-Bernstein-approximation} states
\[
\nu_D(dt)=\rho_D(t)\,dt+\lambda_{D,\infty}\delta_1(dt),
\qquad
\sup_{t\in[0,1]}|\rho_D(t)-e^{-t}|=O(D^{-2}).
\]
For $y<1$, its atom at one contributes nothing, so
\[
\sum_{r=1}^{m_d}\lambda_r\Pr[\tau_r\leq y]
=\int_0^y\rho_D(t)\,dt
\leq1-e^{-y}+O(D^{-2})y
\leq\bigl(1+O(D^{-2})\bigr)y.
\]
Since $D=d-1$, choosing one sufficiently large absolute constant $C$
proves both~\eqref{eq:regular-core-mixture-score} and
\eqref{eq:regular-core-mixture-cdf}.
\end{proof}


\section{Auxiliary Proofs for the Competitiveness Assertions for Large Regular Degree  in
Theorem~\ref{thm:Gamma-regular-large-degree}}
\label{app:regular-large-degree-competitiveness-auxiliary}

We prove Lemma~\ref{lem:regular-algorithmic-limiting-upper} by comparing
sparse batches with complete batches, constructing exactly biregular
block pairs, and completing the resulting graphs to arbitrary sufficiently
large degrees. The comparison holds for every fixed rank realization;
randomness enters only in evaluating the complete-batch limit.

\subsection{Comparing Sparse and Complete Batches}
\label{app:regular-batch-comparison}

Consider an online batch $J_0$ whose allowed offline neighbors are
partitioned into blocks $I_1,\ldots,I_k$. Let $n$ bound both the total
number of offline nodes and $|J_0|$. For $\varepsilon\in(0,1)$,
assume that every allowed block pair has the following property:
\begin{equation}
\label{eq:regular-no-empty-rectangle}
\begin{gathered}
A\subseteq J_0,\quad B\subseteq I_\ell,\quad
|A|\geq\varepsilon|J_0|,\quad |B|\geq\varepsilon|I_\ell|\\
\Longrightarrow\quad E(A,B)\neq\varnothing.
\end{gathered}
\end{equation}
The comparison batch is completely connected to every allowed block.
The online order within the batch is fixed and the same in both executions.

\begin{lemma}[One-batch comparison]
\label{lem:regular-one-batch-comparison}
Fix distinct ranks and any initially safe set $S$ in the allowed
blocks. Let $T$ be its $\min\{|J_0|,|S|\}$ lowest-ranked nodes,
and let $A_{\mathrm{sp}}$ be the set selected by the sparse batch.
Under~\eqref{eq:regular-no-empty-rectangle},
\[
|T\setminus A_{\mathrm{sp}}|\leq\varepsilon n,
\qquad
|A_{\mathrm{sp}}\mathbin{\triangle}T|\leq2\varepsilon n.
\]
\end{lemma}

\begin{proof}
Suppose more than $\varepsilon n$ members of $T$ are missed.
Since the allowed blocks contain at most $n$ nodes in total, some
block has more than an $\varepsilon$ fraction of missed nodes. Call
an arrival bad if it is unmatched or selects a node outside $T$.
There are
\[
|J_0|-|A_{\mathrm{sp}}\cap T|
\geq|T\setminus A_{\mathrm{sp}}|>\varepsilon n
\geq\varepsilon|J_0|
\]
bad arrivals. Every missed node stays safe throughout the batch.
A bad arrival has no edge to a missed node: otherwise it would either
match rather than remain unmatched, or choose that node in preference
to its selected node outside $T$. The bad arrivals and the missed
nodes in the chosen block therefore form an empty rectangle
contradicting~\eqref{eq:regular-no-empty-rectangle}.
Finally, $|A_{\mathrm{sp}}|\leq|T|$, so the symmetric difference is
at most twice the number of missed members of $T$.
\end{proof}

\begin{lemma}[Comparison across batches]
\label{lem:regular-many-batch-comparison}
Consider two instances on the same vertex sets, with at most $n$ nodes
on each side, the same $m$ ordered batches, and the same allowed block
pairs. Suppose one instance contains every allowed edge and the other
satisfies~\eqref{eq:regular-no-empty-rectangle} on each allowed pair.
For any fixed distinct ranks, their matching cardinalities differ by
at most $\varepsilon n(3^m-1)$.
\end{lemma}

\begin{proof}
Let $\delta_k$ be the size of the symmetric difference of their
matched offline sets after $k$ batches, with $\delta_0=0$.
Immediately before the next batch, their safe sets inside the allowed
blocks differ in at most $\delta_k$ nodes. For fixed ranks, the sets
of the first $q$ members of two sets, or all members when fewer than
$q$ are present, differ in at most twice the symmetric difference of
the input sets. This follows by inserting or deleting one member at
a time.

First compare the sparse batch with a complete batch started from its
own safe set, and then with the actual complete execution. The
one-batch lemma gives a difference of at most $2\varepsilon n$ in
the first comparison; the second contributes at most $2\delta_k$.
Including the previously matched sets,
\[
\delta_{k+1}\leq3\delta_k+2\varepsilon n.
\]
Iteration gives $\delta_m\leq\varepsilon n(3^m-1)$, which also
bounds the difference of the two matching cardinalities.
\end{proof}

\subsection{Exactly Biregular Blocks and Regular Triangular Instances}
\label{app:regular-triangular-construction}

We implement the block sizes in~\eqref{eq:regular-triangular-block-sizes}
using simple biregular graphs of density $1/p$ with no large empty
rectangle.

\begin{lemma}[Finite-field blocks and their blow-ups]
\label{lem:regular-finite-field-blocks}
For a prime $p$ and $h\geq1$, there is a simple bipartite graph with
$p^{h+1}$ nodes on each side, degree $p^h$, and second singular value
at most $p^{h/2}$. Replacing each left node by $u$ clones and each
right node by $v$ clones, with all edges between the clone classes of
adjacent nodes, preserves density $1/p$ and guarantees
\eqref{eq:regular-no-empty-rectangle} whenever
$\varepsilon>p^{-h/2}$.
\end{lemma}

\begin{proof}
Index the rows by $(x,z)\in\mathbb F_p^h\times\mathbb F_p$ and the
columns by $(a,b)\in\mathbb F_p^h\times\mathbb F_p$. Put an edge
between them exactly when $z=a\cdot x+b$. Each row and each column
has $p^h$ neighbors.

The row Gram matrix has diagonal entries $p^h$, off-diagonal entries
zero for two rows with the same $x$, and entries $p^{h-1}$ for rows
with different $x$. Its eigenvalues are $p^{2h}$ on the constant
vector, zero on vectors constant within each $x$-class and summing to
zero globally, and $p^h$ on vectors summing to zero within each
$x$-class. Taking square roots proves the singular-value bound.

Uniform left and right clone factors $u,v$ multiply the singular values
by $\sqrt{uv}$. The resulting graph is biregular with density $1/p$.
Subtracting the constant singular-vector contribution and applying
Cauchy--Schwarz gives, for subsets $A,B$ on the two sides,
\[
\left|e(A,B)-\frac{|A||B|}{p}\right|
\leq p^{h/2}\sqrt{uv}\sqrt{|A||B|}.
\]
If $A,B$ occupy fractions $\alpha,\beta$ of their respective sides
and $e(A,B)=0$, this inequality implies
$\sqrt{\alpha\beta}\leq p^{-h/2}$. Thus both fractions cannot be
at least $\varepsilon>p^{-h/2}$, proving the rectangle property.
\end{proof}

Fix $m,p,h$ and let $L=Tp^{h+1}$ for an integer $T\geq1$.
Every block size in~\eqref{eq:regular-triangular-block-sizes} is then
a multiple of $p^{h+1}$. Use the preceding blow-up for each strict
forward pair $J_j,I_i$, $i>j$, with clone factors given by the two
block sizes. Connect diagonal pairs completely and put no backward
edges. Every online node has degree
$|I_j|+p^{-1}\sum_{i>j}|I_i|$, and every offline node has degree
$|J_i|+p^{-1}\sum_{j<i}|J_j|$. The geometric sums in
\eqref{eq:regular-triangular-degree} therefore give a simple
$d_0$-regular instance, with
\begin{equation}
\label{eq:regular-triangular-size}
d_0=Tp^{h+m},
\qquad
n=L\bigl(p^m-(p-1)^m\bigr),
\qquad
\frac{n}{d_0}=p\left[1-\left(1-\frac1p\right)^m\right].
\end{equation}
The diagonal pairs automatically satisfy the rectangle property, and
all strict forward pairs satisfy it with
$\varepsilon=2p^{-h/2}$, provided $h$ is chosen large enough that
this number is less than one. Increasing $T$ changes the scale, not
these density or rectangle guarantees.

\subsection{The Complete-Batch Limit}
\label{app:regular-complete-batch-limit}

Complete all allowed forward block pairs of the regular construction.
This comparison instance need not be regular. For fixed $m,p,h$,
let $T\to\infty$ and normalize block and matching sizes by $d_0$.
The normalized block sizes are
\[
a_i:=\frac{|I_i|}{d_0}=\left(1-\frac1p\right)^{m-i},
\qquad
b_j:=\frac{|J_j|}{d_0}=\left(1-\frac1p\right)^{j-1}.
\]
The total normalized sizes on the two sides agree.

Define $x_0:=0$ and, for $j\in[m]$,
\begin{equation}
\label{eq:regular-complete-batch-recursion}
v_j:=\min\left\{b_j,(1-x_{j-1})\sum_{i=j}^m a_i\right\},
\qquad
x_j:=x_{j-1}+\frac{v_j}{\sum_{i=j}^m a_i}.
\end{equation}
Here $v_j$ is the limiting matched mass in batch $j$, and $x_j$ is
the rank threshold above which its remaining eligible nodes are safe.
Write
\begin{equation}
\label{eq:regular-complete-batch-ratio}
R_{m,p}:=\frac{\sum_{j=1}^m v_j}{\sum_{i=1}^m a_i}.
\end{equation}

\begin{lemma}[Limit of the complete triangular batches]
\label{lem:regular-complete-batch-limit}
For fixed $m,p,h$, the expected normalized matching size of the
complete comparison instance converges to $R_{m,p}$ as $T\to\infty$.
For each fixed $m$, $R_{m,p}\to R_m$ as $p\to\infty$ through primes,
where $R_m$ is the value of~\eqref{eq:regular-complete-batch-recursion}
with $a_i=b_j=1$. Moreover, $R_m\to1-1/e$ as $m\to\infty$.
\end{lemma}

\begin{proof}
There are finitely many offline blocks, and each block size tends to
infinity with $T$. Its empirical rank CDF converges uniformly in
probability to the uniform CDF. For completeness, convergence on a
fixed grid follows from the variance bound for empirical proportions
and a union bound; monotonicity bounds the error between adjacent
grid points, and grid refinement gives uniform convergence.

A complete batch selects the smallest available ranks in its allowed
suffix. Before each batch, all safe nodes in that suffix therefore lie
above a common threshold. Uniform empirical-CDF convergence and
induction over the finitely many batches show that the available mass
is $(1-x_{j-1})\sum_{i=j}^m a_i+o(1)$, the selected mass converges
to $v_j$, and the next threshold converges to $x_j$. If the suffix is
exhausted, the threshold is set to one. This proves convergence in
probability to~\eqref{eq:regular-complete-batch-ratio}; boundedness of
the normalized matching size gives convergence in expectation.

For fixed $m$, the positive block masses $a_i,b_j$ tend to one as
$p\to\infty$. The finite recursion is continuous in these masses,
so $R_{m,p}\to R_m$. In the equal-block case, each full batch
increases the common threshold by $1/(m-j+1)$. Let $k_m$ be the
largest $k\in\{0,\ldots,m\}$ for which
\[
\sum_{\ell=m-k+1}^m\frac1\ell\leq1,
\]
with the empty sum interpreted as zero. Exactly $k_m$ batches are
full, at most one additional batch is partial, and all later batches
are unmatched. Thus
\[
\frac{k_m}{m}\leq R_m\leq\frac{k_m+1}{m}.
\]
Integral bounds for the harmonic sum give
$k_m/m\to1-1/e$, proving the final assertion.
\end{proof}

By Lemma~\ref{lem:regular-many-batch-comparison}, the sparse regular
instance $G_T$ consequently satisfies
\begin{equation}
\label{eq:regular-sparse-core-ratio}
\frac{\mathbb E[\rank(G_T)]}{n}
\leq R_{m,p}+\varepsilon(3^m-1)+o_T(1),
\qquad \varepsilon=2p^{-h/2},
\end{equation}
for fixed $m,p,h$. Its optimum is $n$ because it is regular.
This establishes an upper approximation along the degrees
$d_0=Tp^{h+m}$. We next pass to every sufficiently large degree.

\subsection{Completing to an Arbitrary Target Degree}
\label{app:regular-degree-rounding}

\begin{lemma}[Degree completion with bounded cardinality cost]
\label{lem:regular-degree-rounding}
Let $G$ be a simple $d_0$-regular instance on $n$ nodes per side.
If $d=d_0+\delta$ with $\delta\geq0$ and $\delta n\leq d^2$,
there is a simple $d$-regular instance $H$ on $n+d$ nodes per side
containing $G$ with its original online order such that
\begin{equation}
\label{eq:regular-degree-rounding-cost}
\mathbb E[\rank(H)]\leq\mathbb E[\rank(G)]+2d.
\end{equation}
\end{lemma}

\begin{proof}
Add $d$ nodes on each side. Give every old node $\delta$ neighbors
on the opposite new side by allocating consecutive slots cyclically
among the $d$ new nodes. Since $\delta\leq d$, each old node receives
distinct new neighbors. Each new node receives either
$\lfloor\delta n/d\rfloor$ or $\lceil\delta n/d\rceil$ old
neighbors, at most $d$ by hypothesis. Use identical allocations on
the two sides so that the new nodes' degree sequences agree
coordinatewise.

The desired degrees inside the new $d$-by-$d$ block are consequently
identical on the two sides, lie between zero and $d$, and differ by
at most one. If their minimum is $k<d$, take a $k$-regular circulant
avoiding the diagonal and add the diagonal edge at every position
requiring degree $k+1$. This realizes the desired degrees without
repeated edges. If the desired degree is constantly $d$, take the
complete block. Thus every vertex has degree exactly $d$.

Process all new online nodes last. For fixed ranks, insertion of one
offline node changes matching cardinality by at most one: the two
executions differ along a single chain of displaced selections, which
can create at most one additional match. Inserting the $d$ new offline
nodes therefore increases the cardinality on the old online sequence
by at most $d$. The $d$ final online arrivals add at most $d$ more
matches. This proves~\eqref{eq:regular-degree-rounding-cost} pathwise
under the coupling of the old ranks, and hence in expectation.
\end{proof}

\subsection{Completion of the Limiting Upper Bound}
\label{app:regular-limiting-upper-completion}

\begin{proof}[\textbf{Proof of Lemma~\ref{lem:regular-algorithmic-limiting-upper}}]
Fix $\zeta>0$. Choose $m$ sufficiently large that
\[
R_m<1-\frac1e+\frac\zeta5,
\qquad
\frac8m<\frac\zeta5.
\]
Next choose a prime $p$ sufficiently large that
\[
|R_{m,p}-R_m|<\frac\zeta5,
\qquad
p\left[1-\left(1-\frac1p\right)^m\right]\geq\frac m2.
\]
Choose $h$ so large that $\varepsilon:=2p^{-h/2}<1$ and
$\varepsilon(3^m-1)<\zeta/5$. These parameters are now fixed.

For a large target degree $d$, take
\[
T:=\left\lfloor\frac{d}{p^{h+m}}\right\rfloor,
\qquad d_0:=Tp^{h+m},
\qquad \delta:=d-d_0<p^{h+m}.
\]
The corresponding sparse core has degree $d_0$ and satisfies
\eqref{eq:regular-triangular-size}. Since $n/d_0$ is fixed and
$\delta$ is bounded independently of $d$, the condition
$\delta n\leq d^2$ holds for all sufficiently large $d$.
Lemma~\ref{lem:regular-degree-rounding} gives $H\in\mathcal B_d$
with
\[
\frac{\mathbb E[\rank(H)]}{n+d}
\leq\frac{\mathbb E[\rank(G_T)]+2d}{n+d}
\leq\frac{\mathbb E[\rank(G_T)]}{n}+\frac{2d}{n}.
\]
For sufficiently large $d$, $d_0/d\geq1/2$, so
\[
\frac nd=\frac n{d_0}\frac{d_0}d\geq\frac m4,
\qquad \frac{2d}{n}\leq\frac8m<\frac\zeta5.
\]
Also $T\to\infty$ with $d$, so the error $o_T(1)$ in
\eqref{eq:regular-sparse-core-ratio} is smaller than $\zeta/5$ for
all sufficiently large $d$. Combining the five error allowances gives
\[
\frac{\mathbb E[\rank(H)]}{\operatorname{OPT}(H)}
\leq1-\frac1e+\zeta.
\]
This proves the upper bound for every sufficiently large integer $d$,
not merely for a subsequence of regular degrees.
\end{proof}


\section{The Role of the Gain-Sharing Function Class}
\label{app:gain-sharing-function-classes}

This appendix proves
Proposition~\ref{prop:intro-degree-two-incompleteness} and justifies
the function-class choices in
Subsection~\ref{subsec:intro-instance-and-function-classes}. We first
establish non-tightness over the full measurable class $\cA$, then
separate the effects of continuity and monotonicity. We use the exact
certifying values and competitiveness established in
Theorems~\ref{thm:Gamma-infinity}, \ref{thm:Gamma-two}, and
\ref{coro:Gamma-two-regular}.

The degree-two edge reduction and the power-filter score formula in
Lemmas~\ref{lem:exact-degree-two-edge-reduction} and
\ref{lem:power-filter-gadget-score} remain valid for $f\in\cA$:
their proofs use measurability, $0\leq f\leq1$, and $f(1)=1$,
but neither monotonicity nor continuity. We use this previously
established extension without repeating its proof.

\subsection{Proof of Proposition~\ref{prop:intro-degree-two-incompleteness}}
\label{subsec:function-class-full-barrier}

\begin{proof}
\xhdr{Online degree at most two.} For every $f\in\cA$, the
singleton and power-filter constructions bound $\chi(f;2)$ from
above by each of $H_f(1)$, $\Psi_{0,0}(f)$, $\Psi_{1,1}(f)$,
and $\Psi_{2,1}(f)$. Taking their convex combination with weights
$194/1000$, $512/1000$, $63/1000$, and $231/1000$, respectively,
gives
\begin{equation}
\label{eq:function-class-four-gadget-certificate}
\chi(f;2)\leq\frac{2853}{4000}+\int_0^1k(y)f(y)\,dy
\leq\frac{2853}{4000}+\int_0^1(k(y))_+\,dy,
\end{equation}
where
\[
k(y):=\frac{154y^4+63y^3-462y^2+310y-48}{2000}.
\]
The second inequality uses only $0\leq f\leq1$.

We certify the last integral by exact rational bounds. The numerator
of $k$ changes sign across each of the four disjoint intervals
\[
(-3,-2),\quad
\left(\frac{229499}{10^6},\frac{229500}{10^6}\right),\quad
\left(\frac{746005}{10^6},\frac{746006}{10^6}\right),\quad
\left(\frac{824213}{10^6},\frac{824214}{10^6}\right).
\]
Since its degree is four, each interval contains exactly one root,
and there are no other roots. On $[0,1]$, the polynomial is positive
between its first two positive roots and above its third positive
root. An antiderivative is
\[
J(y):=\frac{154}{10000}y^5+\frac{63}{8000}y^4
-\frac{77}{1000}y^3+\frac{31}{400}y^2-\frac3{125}y.
\]
Moreover, $|dk(y)/dy|\leq2039/2000$ on $[0,1]$. Replacing a
positive root by the lower endpoint of its bracket changes the
corresponding integral by at most $(2039/2000)10^{-12}$, because
$k$ vanishes at that root and the bracket has width $10^{-6}$.
Consequently, exact rational evaluation gives
\[
\begin{aligned}
\frac{2853}{4000}+\int_0^1(k(y))_+\,dy
&\leq\frac{2853}{4000}
+J\!\left(\frac{746005}{10^6}\right)
-J\!\left(\frac{229499}{10^6}\right)
+J(1)-J\!\left(\frac{824213}{10^6}\right)\\
&\qquad+\frac{6117}{2000\cdot10^{12}}<\frac{715361}{10^6}.
\end{aligned}
\]
This bound is independent of $f$, so
$\Gamma_{\cA}(2)<715361/10^6$. The competitiveness established in
Theorem~\ref{thm:Gamma-two}, together with
\eqref{eq:degree-two-competitiveness-rational-bracket}, gives
\[
\Gamma_{\cA}(2)<\frac{715361}{10^6}<\frac{71777}{100000}<\gamma(2).
\]

\xhdr{Two-regular instances.} The long-cycle construction proving
\eqref{eq:regular-two-H-upper} uses only $0\leq f\leq1$ and
$f(1)=1$. It therefore gives
$\tilde\chi(f;2)\leq H_f(1)$ for every $f\in\cA$.
We combine this bound with two scores from the six-cycle $C_6$
whose online arrival order is $j_1\prec j_2\prec j_3$ and whose
neighborhoods are
\[
N_{C_6}(j_1)=\{i_1,i_2\},\qquad
N_{C_6}(j_2)=\{i_1,i_3\},\qquad
N_{C_6}(j_3)=\{i_2,i_3\}.
\]
All scores below are evaluated after the complete execution.
Conditional on $Y_{i_1}=y$, the vertex $i_1$ is unmatched exactly
when the other two ranks are below $y$. The rank selected by $j_1$
is the minimum of two independent uniform ranks. Thus
\begin{equation}
\label{eq:function-class-six-cycle-early}
\Phi_{C_6}(i_1,j_1;f)
=\int_0^1(1-y^2)f(y)\,dy+1-2\int_0^1(1-y)f(y)\,dy
=1-\int_0^1(1-y)^2f(y)\,dy.
\end{equation}
The first two arrivals form the $(1,0)$ power-filter gadget. Since
the last arrival is not incident to $i_1$, it does not affect the
score of $(i_1,j_2)$. Lemma~\ref{lem:power-filter-gadget-score}
therefore gives
\begin{equation}
\label{eq:function-class-six-cycle-late}
\Phi_{C_6}(i_1,j_2;f)=\Psi_{1,0}(f)
=1-\int_0^1\left(y-\frac{y^2}{2}\right)f(y)\,dy.
\end{equation}
These three upper constraints satisfy the identity
\begin{equation}
\label{eq:function-class-regular-equalizing-mixture}
\frac14H_f(1)+\frac14\Phi_{C_6}(i_1,j_1;f)
+\frac12\Phi_{C_6}(i_1,j_2;f)=\frac34,
\end{equation}
since the coefficient of $f(y)$ is identically
$\frac14-\frac14(1-y)^2-\frac12(y-y^2/2)=0$.
It follows that $\tilde\chi(f;2)\leq3/4$ for every $f\in\cA$.
Conversely, Theorem~\ref{coro:Gamma-two-regular} supplies the profile
$\tilde f_2^\star\in\cF\subseteq\cA$ with certifying value
$3/4$. Hence
\begin{equation}
\label{eq:intro-regular-two-function-class-chain}
\tilde\Gamma_{\cF}(2)=\tilde\Gamma_{\cA}(2)=\frac34
<\frac{119}{144}=\tilde\gamma(2).
\end{equation}
Both gaps are strictly positive, proving non-tightness over $\cA$.
\end{proof}

\subsection{Continuity, Right-Continuity, and Attainment}
\label{subsec:function-class-continuity-proofs}

For the comparisons in this subsection, write
\[
\begin{aligned}
\cA&=\{f:[0,1]\to[0,1]:f\text{ is Borel measurable},\ f(1)=1\},\\
\cA_{\mathrm{cont}}&:=\cA\cap C([0,1]),\\
\cF&=\{f\in\cA:f\text{ is nondecreasing and right-continuous on }[0,1)\},\\
\cF_{\mathrm{cont}}&:=\cF\cap C([0,1]).
\end{aligned}
\]
We first establish the continuity estimate and approximation facts
needed to compare their certifying capacities.

\begin{lemma}[Uniform $L^1$ stability at bounded online degree]
\label{lem:function-class-L1-stability}
For every finite integer $d\geq1$, every
$G\in\mathcal B_{\leq d}^{\mathrm{on}}$, every $(i,j)\in E(G)$,
and every $f,h\in\cA$,
\begin{equation}
\label{eq:function-class-edge-stability}
|\Phi_G(i,j;f)-\Phi_G(i,j;h)|\leq(d+1)\|f-h\|_1.
\end{equation}
Consequently, for any $\cG\subseteq\mathcal B_{\leq d}^{\mathrm{on}}$
containing an instance with positive optimum,
\begin{equation}
\label{eq:function-class-inner-stability}
|\chi(f,\cG)-\chi(h,\cG)|\leq(d+1)\|f-h\|_1.
\end{equation}
\end{lemma}

\begin{proof}
The offline contribution satisfies
$\mathbb E[|f(Y_i)-h(Y_i)|X_i]\leq\|f-h\|_1$.
Whenever $j$ matches, its partner belongs to $N_G(j)$, so
\[
\mathbb E[|f(Y_j)-h(Y_j)|X_j]
\leq\sum_{\hat i\in N_G(j)}\mathbb E[|f(Y_{\hat i})-h(Y_{\hat i})|]
\leq d\|f-h\|_1.
\]
Adding the two bounds proves
\eqref{eq:function-class-edge-stability}. The bound is uniform
over graph-edge pairs, so taking their infima proves
\eqref{eq:function-class-inner-stability}.
\end{proof}

\begin{lemma}[Density of continuous profiles]
\label{lem:function-class-continuous-density}
The classes $\cA_{\mathrm{cont}}$ and $\cF_{\mathrm{cont}}$ are
$L^1$-dense in $\cA$ and $\cF$, respectively. In the monotone
case, for every $f\in\cF$ and integer $m\geq1$, there is
$f_m\in\cF_{\mathrm{cont}}$ satisfying
\begin{equation}
\label{eq:function-class-monotone-interpolation-error}
\|f_m-f\|_1\leq\frac{f(1)-f(0)}m\leq\frac1m.
\end{equation}
\end{lemma}

\begin{proof}
\xhdr{Measurable profiles.} Fix $f\in\cA$ and $\varepsilon>0$.
By density of continuous functions in $L^1([0,1])$, choose a
continuous real-valued approximation with error less than
$\varepsilon/2$. Projecting its values onto $[0,1]$ cannot increase
the error. This gives a continuous $h:[0,1]\to[0,1]$ with
$\|h-f\|_1<\varepsilon/2$. Choose
$0<\delta<\min\{1,\varepsilon/2\}$ and replace $h$ on
$[1-\delta,1]$ by the linear interpolation from $h(1-\delta)$
to $1$. The resulting $\hat h$ belongs to $\cA_{\mathrm{cont}}$
and satisfies
$\|\hat h-f\|_1\leq\|h-f\|_1+\delta<\varepsilon$.

\xhdr{Nondecreasing profiles.} For $f\in\cF$, interpolate
its values at $k/m$, $k=0,\ldots,m$, linearly to obtain
$f_m\in\cF_{\mathrm{cont}}$. On each grid interval, both
$f$ and $f_m$ lie between the same two endpoint values. Hence
\[
\|f_m-f\|_1\leq\frac1m\sum_{k=0}^{m-1}
\bigl(f((k+1)/m)-f(k/m)\bigr)
=\frac{f(1)-f(0)}m\leq\frac1m.
\]
\end{proof}

\begin{proposition}[Continuity preserves certifying capacity]
\label{prop:continuity-preserves-certifying-values}
For every $d\in\mathbb Z_{\geq1}\cup\{\infty\}$,
\begin{equation}
\label{eq:intro-untilded-continuity-equivalence}
\Gamma_{\cA}(d)=\Gamma_{\cA_{\mathrm{cont}}}(d),\qquad
\Gamma_{\cF}(d)=\Gamma_{\cF_{\mathrm{cont}}}(d).
\end{equation}
For every finite integer $d\geq1$,
\begin{equation}
\label{eq:intro-regular-continuity-equivalence}
\tilde\Gamma_{\cA}(d)=\tilde\Gamma_{\cA_{\mathrm{cont}}}(d),\qquad
\tilde\Gamma_{\cF}(d)=\tilde\Gamma_{\cF_{\mathrm{cont}}}(d).
\end{equation}
These equalities concern supremum values and need not preserve
attainment.
\end{proposition}

\begin{proof}
Fix a finite $d\geq1$. For any $f\in\cA$, choose
$f_m\in\cA_{\mathrm{cont}}$ with $\|f_m-f\|_1\to0$.
Lemma~\ref{lem:function-class-L1-stability} gives
\[
\chi(f_m;d)\longrightarrow\chi(f;d),\qquad
\tilde\chi(f_m;d)\longrightarrow\tilde\chi(f;d).
\]
Thus each value certified by a measurable profile is a limit of
values certified by continuous profiles. Taking suprema proves the
inequalities from $\cA$ to $\cA_{\mathrm{cont}}$; the reverse
inequalities follow from class inclusion. The same argument, using
the monotone approximations in
Lemma~\ref{lem:function-class-continuous-density}, proves the
identities for $\cF$ and $\cF_{\mathrm{cont}}$.

For $d=\infty$, the degree-dependent stability estimate does not
supply a uniform bound. Instead, use the continuous nondecreasing
optimizer $f_\infty^\star(y)=e^{y-1}$ from
Theorem~\ref{thm:Gamma-infinity} and soundness over $\cA$:
\[
1-\frac1e\leq\Gamma_{\cF_{\mathrm{cont}}}(\infty)
\leq\Gamma_{\cF}(\infty)\leq\Gamma_{\cA}(\infty)
\leq\gamma(\infty)=1-\frac1e.
\]
Since $\cF_{\mathrm{cont}}\subseteq\cA_{\mathrm{cont}}\subseteq\cA$,
all four unrestricted-degree suprema equal $1-1/e$.
\end{proof}

\xhdr{Right-continuity as a representative convention.} If
$f\in\cA$ is nondecreasing, define $f^+(y):=\lim_{t\downarrow y}f(t)$
for $y<1$ and $f^+(1):=1$. Then $f^+\in\cF$ and $f^+=f$
outside a countable set. Every finite graph-edge score is unchanged
by this modification, by
Lemma~\ref{lem:function-class-L1-stability}. Thus right-continuity
can be imposed on nondecreasing profiles without changing even their
individual scores. Full continuity, in contrast, is imposed through
approximation and need not preserve an individual maximizer.

\begin{lemma}[Compactness and attainment in the monotone class]
\label{lem:function-class-monotone-compactness}
The class $\cF$, identified with its image in $L^1([0,1])$, is
compact. For every fixed finite $d\geq1$, both $\Gamma_{\cF}(d)$
and $\tilde\Gamma_{\cF}(d)$ are attained.
\end{lemma}

\begin{proof}
The correspondence $f(y)=\pi_f([0,y])$ is an affine bijection
between $\cF$ and $\cP([0,1])$. Any sequence of probability
measures on the compact interval $[0,1]$ has a weakly convergent
subsequence, say $\pi_{f_{n_k}}\Rightarrow\pi$. For every
$y\in(0,1)$ with $\pi(\{y\})=0$, weak convergence gives
$f_{n_k}(y)\to f(y):=\pi([0,y])$. There are at most countably
many exceptional points, so dominated convergence yields
$\|f_{n_k}-f\|_1\to0$, where $f\in\cF$. This proves
compactness in $L^1$. Right-continuity and the common normalization
at $1$ ensure that two elements of $\cF$ agreeing almost everywhere
agree everywhere. Finally, the inner certifying values are
continuous by Lemma~\ref{lem:function-class-L1-stability}, and
therefore attain their maxima on $\cF$.
\end{proof}

The equal suprema over $\cF_{\mathrm{cont}}$ at online degree two
and regular degree two are not attained: their unique optimizers in
$\cF$, established in Theorems~\ref{thm:Gamma-two} and
\ref{coro:Gamma-two-regular}, are discontinuous. Thus preservation of certifying capacity does not imply preservation
of an optimizer.

\subsection{Monotonicity Can Reduce Certifying Capacity}
\label{subsec:function-class-degree-two-separation}

\begin{proposition}[Monotonicity is a substantive restriction]
\label{prop:monotonicity-is-substantive}
For online degree at most two,
\begin{equation}
\label{eq:intro-degree-two-function-class-chain}
\Gamma_{\cF}(2)=\frac{2}{3-s^\star}<\frac{143}{200}
<\frac{17877}{25000}\leq\Gamma_{\cA}(2)<\frac{715361}{10^6}<\gamma(2).
\end{equation}
Some continuous nonmonotone profile also certifies a value strictly
larger than $\Gamma_{\cF}(2)$. On two-regular instances, by contrast,
$\tilde\Gamma_{\cF}(2)=\tilde\Gamma_{\cA}(2)=3/4$.
\end{proposition}

\begin{proof}
The measurable upper bound and the regular-class equality follow
from Proposition~\ref{prop:intro-degree-two-incompleteness} and
\eqref{eq:intro-regular-two-function-class-chain}. For the strict
monotonicity loss, define
\begin{equation}
\label{eq:function-class-nonmonotone-profile}
g(y):=
\begin{cases}
0,&0\leq y<\frac15,\\[1mm]
\frac{2071}{2500},&\frac15\leq y<\frac25,\\[1mm]
\frac{19}{20},&\frac25\leq y<\frac45,\\[1mm]
\frac{847}{1000},&\frac45\leq y<1,\\[1mm]
1,&y=1.
\end{cases}
\end{equation}
This profile belongs to $\cA$ and has a downward jump at $4/5$.
Direct integration gives
\[
H_g(1)=\frac15\frac{2071}{2500}+\frac25\frac{19}{20}
+\frac15\frac{847}{1000}=\frac{17877}{25000}.
\]
For every integer $r\geq0$, its moment is
\begin{equation}
\label{eq:function-class-step-moments}
M_r(g):=\int_0^1y^r g(y)\,dy
=\frac{4235\cdot5^{r+1}+515\cdot4^{r+1}-608\cdot2^{r+1}-4142}
{5000(r+1)5^{r+1}}.
\end{equation}
By~\eqref{eq:power-filter-gadget-score},
\begin{equation}
\label{eq:function-class-step-margin}
\Psi_{p,q}(g)-H_g(1)
=\frac{p+q+1}{(p+1)(q+1)}-M_p(g)-M_q(g)
+\frac{M_{p+q+1}(g)}{p+1}.
\end{equation}

\xhdr{The infinite tail.} Since $g\leq19/20$ almost everywhere,
$M_r(g)\leq19/(20(r+1))$. For $p+q\geq18$, dropping the last,
nonnegative term in~\eqref{eq:function-class-step-margin} yields
\[
\Psi_{p,q}(g)-H_g(1)
\geq\frac{p+q+1}{(p+1)(q+1)}
-\frac{19}{20(p+1)}-\frac{19}{20(q+1)}
=\frac{(p+q+2)/20-1}{(p+1)(q+1)}\geq0.
\]

\xhdr{The finite cases.} For the 171 ordered pairs with
$p+q\leq17$, substituting the rational moments into
\eqref{eq:function-class-step-margin} gives
\[
\begin{array}{c|ccccc}
p+q&0&1&2&3&4,\ldots,17\\ \hline
\text{lower bound}
&\frac1{250000}&\frac{361}{156250}&\frac{1329}{5000000}
&\frac{4649}{468750000}&\frac1{100}.
\end{array}
\]
The first four bounds are the exact minima for the respective sums;
the last is a uniform bound over the stated range. The following
exact-arithmetic calculation verifies all entries:
\begin{verbatim}
from fractions import Fraction as Q

def M(r):
    return Q(4235*5**(r+1) + 515*4**(r+1)
             - 608*2**(r+1) - 4142,
             5000*(r+1)*5**(r+1))

bounds = [Q(1,250000), Q(361,156250),
          Q(1329,5000000), Q(4649,468750000)]
assert M(0) == Q(17877,25000)
for total in range(18):
    margins = []
    for p in range(total+1):
        q = total-p
        margins.append(Q(p+q+1,(p+1)*(q+1))
                       - M(p) - M(q) + M(p+q+1)/Q(p+1))
    if total < 4:
        assert min(margins) == bounds[total]
    else:
        assert min(margins) >= Q(1,100)
\end{verbatim}
Together with the tail estimate, this proves
$\Psi_{p,q}(g)\geq H_g(1)$ for every $p,q\geq0$.
Lemma~\ref{lem:exact-degree-two-edge-reduction}, in its measurable
version, now gives
\[
\chi(g;2)=H_g(1)=\frac{17877}{25000}\leq\Gamma_{\cA}(2).
\]

\xhdr{Comparison with the monotone optimum.} The polynomial in
\eqref{eq:s-two-star-polynomial} is strictly decreasing on $[0,1]$,
since its derivative satisfies
$8z^3+6z^2-16z-33\leq-2z-33<0$. Its value at $29/143$ is
$-525948/418161601<0$, so $s^\star<29/143$.
Theorem~\ref{thm:Gamma-two} therefore implies
\begin{equation}
\label{eq:monotonicity-strict-separation}
\Gamma_{\cF}(2)=\frac2{3-s^\star}
<\frac{143}{200}<\chi(g;2)=\frac{17877}{25000},
\end{equation}
where $17877/25000-143/200=1/12500$.

Finally, choose $h\in\cA_{\mathrm{cont}}$ with
$\|h-g\|_1<1/37500$, using
Lemma~\ref{lem:function-class-continuous-density}. Then
\[
\chi(h;2)\geq\chi(g;2)-3\|h-g\|_1
>\frac{143}{200}>\Gamma_{\cF}(2).
\]
Such an $h$ cannot be nondecreasing. Hence the strict improvement
survives full continuity and results from relaxing monotonicity.
\end{proof}

The two sources of loss are therefore distinct:
\[
\Del(2)=\bigl(\Gamma_{\cA}(2)-\Gamma_{\cF}(2)\bigr)
+\bigl(\gamma(2)-\Gamma_{\cA}(2)\bigr).
\]
Both terms are strictly positive. For two-regular graphs the first
term in the corresponding decomposition is zero, while the second
is exactly $11/144$. The measurable-class inequalities establish
strict non-tightness of the fixed protocol; they do not claim an
exact value for $\Gamma_{\cA}(2)$.

\section*{Acknowledgement}

\noindent\textbf{AI methodology.}
The author used ChatGPT (OpenAI) for assistance with mathematical
exploration, developing and refining proof arguments, checking
symbolic and numerical calculations, and drafting and revising
the manuscript. This assistance materially influenced parts of
the degree-two and large-degree analyses and the comparisons
between gain-sharing function classes, including their supporting
proofs and computational verifications, as well as the organization
and exposition throughout the paper. The author takes full
responsibility for the correctness, originality, and integrity
of the paper, including all mathematical arguments, computational
results, and references.
\end{document}